\documentclass[12pt]{article}

\usepackage{fancyhdr,titlesec,geometry,ragged2e}
\usepackage[dvipsnames]{xcolor}
\usepackage[many]{tcolorbox}
\usepackage{layout}
\usepackage{lipsum}
\usepackage{hyperref}
\hypersetup{
  colorlinks,                
  citecolor=red,
  linkcolor=blue,
  linktoc=all
}

\usepackage[T1]{fontenc}                            
\usepackage{lmodern,mathrsfs}

\usepackage[shortlabels]{enumitem}
\usepackage{mathtools,amssymb,amsfonts,amsthm,bm,mathrsfs}
\usepackage{amsmath} 
\usepackage{array,tabularx,booktabs}                
\usepackage{graphicx,wrapfig,float,caption,epsfig}         
\usepackage{setspace,multicol}                      
\usepackage{tikz,tikz-cd,physics,cancel}            
\usepackage{circuitikz}
\usepackage{esint}
\allowdisplaybreaks

\usepackage{parskip}
\DeclareGraphicsExtensions{.pdf,.png,.jpg}
\usepackage{geometry}
\renewcommand{\thefootnote}{\arabic{footnote}}

\titleformat{\section}[hang]{\normalfont\large\bfseries}{\thesection.}{3mm}{}
\titlespacing{\section}{0mm}{15mm}{4mm}

\titleformat{\subsection}[hang]{\normalfont\bfseries}{\thesubsection}{2mm}{}
\titlespacing{\subsection}{0mm}{10mm}{4mm}

\titleformat{\subsubsection}[hang]{\normalfont\bfseries}{\thesubsubsection}{5mm}{}
\titlespacing{\subsubsection}{0mm}{5mm}{1mm}

\newtheoremstyle{thmstyle}
{5pt} 
{3pt} 
{\itshape} 
{} 
{\bfseries} 
{.} 
{0.5em} 
{} 

\theoremstyle{remark}{

}

\theoremstyle{thmstyle}{

\newtheorem{result}{Result}[section]
}

\numberwithin{equation}{section}

\newlength{\extraspace}
\newlength{\extraspaces}
\newcommand{\be}{\begin{equation}
\addtolength{\abovedisplayskip}{\extraspaces}
\addtolength{\belowdisplayskip}{\extraspaces}
\addtolength{\abovedisplayshortskip}{\extraspace}
\addtolength{\belowdisplayshortskip}{\extraspace}}
\newcommand{\ee}{\end{equation}}
\newcommand{\ba}{\begin{eqnarray}
\addtolength{\abovedisplayskip}{\extraspaces}
\addtolength{\belowdisplayskip}{\extraspaces}
\addtolength{\abovedisplayshortskip}{\extraspace}
\addtolength{\belowdisplayshortskip}{\extraspace}}
\newcommand{\ea}{\end{eqnarray}}
\newcommand{\bas}{\begin{eqnarray*}
\addtolength{\abovedisplayskip}{\extraspaces}
\addtolength{\belowdisplayskip}{\extraspaces}
\addtolength{\abovedisplayshortskip}{\extraspace}
\addtolength{\belowdisplayshortskip}{\extraspace}}
\newcommand{\eas}{\end{eqnarray*}}

\renewcommand{\dd}{\partial} 
\renewcommand{\bra}{\langle}
\renewcommand{\ket}{\rangle}
\newcommand{\ra}{\rightarrow}

\newcommand{\sspace}{\makebox[1cm]{ }}

\newcommand{\nspace}{\!\!\!\!\!\!\!\!\!\!}

\newcommand{\nonum}{\nonumber \\[1.5mm]}
\newcommand{\is }{&\!\!=\!\!&} 

\DeclarePairedDelimiterX\set[1]\lbrace\rbrace{#1}

\newcommand{\1}{\mbox{1\hspace{-.8ex}1}}

\newcommand{\lb}{\lambda}

\newcommand{\om}{\omega}

\renewcommand{\th}{\theta}
\newcommand{\vp}{\varphi}
\newcommand{\eps}{\epsilon}

\newcommand{\cE}{{\cal E}}

\newcommand{\cH}{{\cal H}}

\newcommand{\cL}{{\cal L}}

\newcommand{\cO}{{\cal O}}

\newcommand{\cR}{{\cal R}}

\newcommand{\cW}{{\cal W}}

\newcommand{\C}{\mathbb{C}}
\newcommand{\N}{\mathbb{N}}
\newcommand{\R}{\mathbb{R}}
\newcommand{\Z}{\mathbb{Z}}

\def\fnum@figure{\textbf{\figurename\nobreakspace\thefigure}}
\def\fnum@table{\textbf{\tablename\nobreakspace\thetable}}

\newcommand{\smallcap}[1]{{\normalfont\textsc{#1}}}
\renewcommand{\l}{\smallcap{l}}
\newcommand{\h}{\smallcap{h}}

\newcommand{\lbn}{\lambda_{\smallcap{n}}}

\begin{document}

\begin{titlepage}

\renewcommand{\thefootnote}{\fnsymbol{footnote}}
\makebox[1cm]{}
\vspace{1cm}

\begin{center}
\mbox{{\Large \bf  Asymptotic Velocity Domination}}\\[3mm] 
\mbox{{ \Large \bf in quantized
polarized Gowdy Cosmologies}}
\vspace{1.8cm}

{\sc Max Niedermaier}\footnote{email: {\tt mnie@pitt.edu}},
{\sc Mahdi Sedighi Jafari}\footnote{email: {\tt mas1371@pitt.edu}}
\\[8mm]
{\small\sl Department of Physics and Astronomy}\\
{\small\sl University of Pittsburgh, 100 Allen Hall}\\
{\small\sl Pittsburgh, PA 15260, USA}
\vspace{10mm}


\vspace{10mm} 

{\bf Abstract} \\[5mm]

\begin{quote}
Asymptotic velocity domination (AVD) posits that when back-propagated to
the Big Bang generic cosmological spacetimes solve a drastically
simplified version of the Einstein field equations, where all 
dynamical spatial gradients are absent (similar as in the
Belinski-Khalatnikov-Lifshitz scenario). Conversely, a solution can in
principle be reconstructed from its behavior near the Big Bang. 
This property has been rigorously proven for the Gowdy class of cosmologies,
both polarized and unpolarized. Here we establish for the polarized case a
quantum version of the AVD property formulated in terms of two-point functions
of (the integrands of) Dirac observables: these correlators approach their
much simpler velocity dominated counterparts when the time support is
back-propagated to the Big Bang. Conversely, the full correlators can be
expressed as a uniformly convergent series in averaged spatial gradients
of the velocity dominated ones. 
\end{quote}  
\end{center}

\vfill

\setcounter{footnote}{0}
\end{titlepage}


\thispagestyle{empty}
\makebox[1cm]{}

\vspace{-23mm}
\begin{samepage}
{\hypersetup{linkcolor=black}
\tableofcontents}
\end{samepage}

\nopagebreak

\newpage

\section{Introduction}

The Big Bang singularity is one of the prime happenstances where
quantum gravitational effects are expected to be important,
both in imprinting observable signatures and as a quest for
ultimate origins. So far theoretical efforts to develop a quantum
framework specific to the situation have mostly been
limited to mini-super space models with a few degrees of freedom.
Gowdy cosmologies offer an interesting
alternative:~(i) they have field theoretical degrees of freedom;
(ii) mirror the structure of full Einstein gravity; 
(iii) are amenable to standard quantization techniques; and  
(iv) they exhibit classically a remarkable behavior near the Big Bang
known as Asymptotic Velocity Domination ({\bf AVD}).  AVD posits that,
when generic
cosmological solutions are evolved backward towards the Big Bang,
they enter a velocity dominated regime in which temporal derivatives
overwhelm spatial ones (similar as in the Belinski-Khalatnikov-Lifshitz
scenario \cite{BelHenbook}) and in which they are governed by a much simpler
system of equations (called ``Velocity Dominated'' {\bf VD}) without
dynamical spatial gradients. Conversely, the full solution can
in principle be reconstructed from the limiting one and
eventually from data specified on the Big Bang ``boundary”;
see \cite{RingstroemCR} for an overview. For Gowdy cosmologies,
AVD has been rigorously proven \cite{IsenMonc,Ringstroemproof},
as reviewed in \cite{RingstroemLR}.   

Gowdy cosmologies \cite{Gowdy} are exact, spatially inhomogeneous,
vacuum solutions of the Einstein equations with two commuting spacelike
Killing vectors. They describe nonlinear colliding gravitational waves
emerging from a Big Bang singularity. Prominent among them are soliton-like
solutions reviewed in \cite{Gravisolitons}. The system can be quantized
and for generic
polarizations leads to an effectively $1\!+\!1$ dimensional quantum field
theory with exponential self-interactions and a noncompact internal
symmetry. It is found to be quasi-renormalizable to all orders of
perturbation theory in a way compatible with the broader Asymptotic
Safety Scenario \cite{2Kren1,2Kren2,reducedquant}. These results do not
address aspects related to AVD. Precisely those aspects, however,
have the potential to generalize beyond the two-Killing vector subsector
and thus should contain pointers towards a `quantum theory of the Big Bang'. 
With this motivation, we investigate here whether a quantum
version of AVD holds in polarized Gowdy cosmologies. The polarized
case shares most qualitative features with the general case, but
is technically much simpler in that the action is quadratic in
the main dynamical field $\phi$. In a reduced phase space formulation
the  quantum theory can therefore be explored
using methods familiar from free quantum field theories on
curved backgrounds; see \cite{Berger,Torre,QGowdy0,TorreSchroed}.
We shall do so here as well, but focus on aspects pertaining to AVD. 

An important property of polarized Gowdy cosmologies is that they admit
an infinite set of Dirac observables \cite{TorreObs}. Dirac
observables are thought to capture the intrinsic `gauge invariant'
content of a gravity theory, but are normally elusive, see e.g.~\cite{Pons}.
In Appendix B we construct a new one-parameter family of the form 
\begin{equation}
\label{i1} 
\cO(\th) = \int_0^{2\pi} \! \frac{dx^1}{2\pi} \,
\jmath_0^{\smallcap{p}}(x^0,x^1;\th) =
\cO_0(\th) + O(\rho^2/\lbn^2) \,, \quad {\rm Im} \th \neq 0\,.
\end{equation}
Here $\rho>0$ is a temporal function whose $\rho=0$ level set can be
identified with the Big Bang. Further, 
$(x^0,x^1)$ are the non-Killing
coordinates of the Gowdy spacetime and $\lbn >0$ is the dimensionless
reduced Newton constant. These quantities strongly Poisson commute with the
constraints {\it without using equations of motion}. The leading term
$\cO_0(\th)$ is in one-to-one correspondence to a family
of Dirac observables
in the VD system, while the subleading terms are organized alternatively
according to powers of $\rho^2/\lbn^2$ or by the number
of dynamical {\it spatial} derivatives. The integrand $\jmath_0^{\smallcap{p}}$
is linear in the Gowdy scalar $\phi$ (defining $M = {\rm diag}(e^{\phi}, e^{-\phi})$ in (\ref{2Kmetric}) below) and its momentum $\pi^{\phi}$, and
in a gauge fixed reduced phase space formulation
only $(\phi,\pi^{\phi})$ is canonically quantized, while the
field $\tilde{\sigma}$ in (\ref{2Kmetric}) is treated as a renormalized
composite operator. Suitable gauge specializations
single out coordinates
$(\tau,\zeta) \in \R^2$, such that the Big Bang is located at
$\tau \ra - \infty$. Then $\cO(\th)$ becomes an on-shell conserved charge,
$\dd_{\tau} \cO^{\rm on}(\th) =0$, whose integrand $q(\tau,\zeta;\th)$
differs from $\jmath_0^{\smallcap{p}}(e^{\tau}, \zeta;\th)$ by a total
$\dd_{\zeta}$ derivative, and stays finite as $\tau \ra - \infty$.
The two-point function of the on-shell $q$ can be taken as the central
object in the quantum theory, $\bra 0_T| q(\tau,\zeta;\th)
q(\tau',\zeta';\th) |0_T\ket$, where $|0_T\ket$ is some (non-unique)
Fock vacuum defining a `state'. Only its symmetric part 
is state dependent and on account of the linearity in $\phi$ its
properties are coded by those of 
$\bra 0_T| \phi(\tau,\zeta) \phi(\tau',\zeta') + \phi(\tau',\zeta')
\phi(\tau,\zeta) |0_T\ket$ and its $\dd_{\tau},\dd_{\tau'},\dd_{\tau}\dd_{\tau'}$
derivatives. These are conveniently combined into
a $2\times 2$ matrix two-point function $\cW^s$.
As described in Appendix A, one can view the velocity dominated
Gowdy system as a Carroll-type gravity theory in its own right and 
follow analogous steps through VD Dirac observables and
their integrand's two-point function to find its properties coded by
a matrix two-point function $\mathfrak{W}^s$ in the VD system
analogous to $\cW^s$. The question of {\bf quantum AVD} can then
be posed as follows: does $\cW^s$ have a $\tau, \tau' \ra - \infty$
asymptotics that relates to its VD counterpart $\mathfrak{W}^s$? If so,
can $\cW^s$ be reconstructed from $\mathfrak{W}^s$? We find that the
answers are affirmative for a large class of states we dub
``time consistent''. Specifically, we obtain the  

{\bf Result.} {\it For any time consistent state the matrix two-point
function $\cW^s$ admits upon averaging with $f,g \in C^{\infty}(S^1)$
test functions a series expansion of the form  }
\begin{small} 
\begin{eqnarray}
\label{i2} 
\!\!\!\!&\nspace & 
\int\!\!\frac{d\zeta}{2\pi}\frac{d\zeta'}{2\pi}
f(\zeta) g(\zeta') \cW^s(\tau,\tau',\zeta\! - \!\zeta')=
\int\!\!\frac{d\zeta}{2\pi}\frac{d\zeta'}{2\pi}
f(\zeta) g(\zeta') \,\mathfrak{W}^s(\tau,\tau',\zeta\! - \!\zeta')
\nonum
\!\!\!\!&\nspace & \quad +  
\sum_{l \geq 1} \sum_{k=0}^l e^{2 k \tau} e^{ 2(l-k) \tau'} 
\int\!\!\frac{d\zeta}{2\pi}\frac{d\zeta'}{2\pi}
f(\zeta) g(\zeta') (\dd_{\zeta} \dd_{\zeta'})^l
I_k \,\mathfrak{W}^s(\tau,\tau',\zeta\! - \!\zeta') \,I_{l-k}^T\,.
\end{eqnarray}
\end{small}
{\it $\!$Here $I_k$ are numerical $2\times 2$ matrices with a known
generating functional.
The expansion is fully determined by the spatial gradients of the two-point
function in the VD system,
$\mathfrak{W}^s(\tau,\tau',\zeta - \zeta')$, which is linear in $\tau,\tau'$.  
The series (\ref{i2}) is uniformly convergent for all $\tau,\tau'
< - \delta$, $\delta >0$ and $\tau - \tau'$ bounded. }

The paper is organized as follows. After introducing the polarized Gowdy
system in relation to its AVD property in Subsections \ref{section2.1}
and \ref{section2.2}, we
outline the reduced phase space quantization and define the
$qq$ two-point functions as well as $\cW^s$. In Section \ref{section3}
the properties of
the various two-point functions are explored, leading to
the notion of time dependent states and the above spatial gradient
expansion. The construction of the $\sigma$ field solving the
quantum constraints is done in Section \ref{section4}.  The Hamiltonian actions
and their symmetries are detailed in Appendix \ref{appendixA}, which
are prerequisites for the construction of Dirac observables (\ref{i1})
in Appendix \ref{appendixB}. 

\section{Polarized Gowdy cosmologies and AVD}
\label{section2} 

Gowdy cosmologies are exact solutions of the 4-dimensional Einstein
equations with two commuting spacelike Killing
vectors $K_1,K_2$. In adapted coordinates $X= (x^0,x^1,y^1,y^2)$
the line element can be brought into the form
\be
\label{2Kmetric}
g^{\rm 2K}_{IJ}(X) dX^I dX^J= \gamma_{\mu\nu}(x) dx^{\mu} dx^{\nu} +
\rho(x) M_{ab}(x) dy^a dy^b\,.
\ee
Here, all fields only depend on $(x^0,x^1)$ while $(y^1,y^2)$ are
Killing coordinates, i.e.~are such that locally $K_a = K_a^I \dd_I =
\dd/\dd y^a$, $a =1,2$. Further, $\rho(x) > 0$, and  $M(x)$
is a real-valued $2\times 2$ matrix with unit determinant.
A selfcontained derivation of (\ref{2Kmetric}) (for
the case of one timelike and one spacelike Killing vector) can be
found in \cite{Straumannbook}, Appendix C. For the Lorentzian metric
$\gamma$ we adopt the following lapse-shift
type parameterization
\begin{equation}
\label{2Dmetric}
\gamma_{\mu\nu}(x) dx^{\mu} dx^{\nu} =
e^{\tilde{\sigma}}[ - n^2 (dx^0)^2 + (dx^1 + s \,dx^0)^2]\,,
\end{equation}
i.e.~$\gamma_{00} = e^{\tilde{\sigma}}(-n^2 + s^2)$,
$\gamma_{01} = e^{\tilde{\sigma}}s = \gamma_{10}$,
$\gamma_{11} = e^{\tilde{\sigma}}$. By a slight abuse of terminology
we shall refer to $n,s$ as lapse, shift, respectively.%
\footnote{ For a two-dimensional Lorentzian metric the
  lapse $N$, shift $N^1$, spatial metric $\gamma_{11}$ proper would be
  $N = e^{\tilde{\sigma}/2} n$, $N^1 = s$, $\gamma_{11} = e^{ \tilde{\sigma}}$.
  Note that $e^{\tilde{\sigma}}, n$ are spatial densities of weight $2,-1$,
  respectively, and that $s$, being one-dimensional spatial vector,
  can also be viewed as a spatial density of weight $-1$.
}
Cosmological spacetimes arise in this setting if $\rho$ is assumed
to have everywhere timelike gradient,
$\gamma^{\mu\nu}(x) \dd_{\mu} \rho \dd_{\nu} \rho <0$. This evaluates to
$e_0(\rho)^2 > n^2 (\dd_1 \rho)^2$, where $e_0(\rho) = (\dd_0 - s \dd_1)\rho$. The field equations are presented in (\ref{evolpt}) below.
The solutions generically have a Big Bang singularity, which
in the coordinates singled out by $s=0,n=1, \rho =t$ occurs at
the $t \ra 0^+$ hypersurface. The evolution equations can be
coded by a Lax pair and some of the integrable systems methodology can
be adapted to this situation. In particular, there is a rich
set of gravitational soliton solutions \cite{Gravisolitons}. 

The Killing coordinates $y^1,y^2$ are assumed to be periodic,
corresponding to orbits with the topology of a 2-torus $T^2$. 
The $x^1$ spatial sections are normally taken to be
circular as well, in which case all fields in (\ref{2Kmetric}) are 
periodic functions of $x^1 \in \R$. The overall topology of the manifold
then is $\R \times T^3$ and the solutions are referred to as $T^3$ Gowdy
cosmologies. Technically, it is often simpler to allow the
fields to be be nonperiodic, with suitable fall-off conditions. The
topology then is $\R\times \R \times T^2$. Other possible topologies
are $\R \times S^3$ or $\R \times S^1 \times S^2$ \cite{Gowdy}. 

Here we shall focus on the $\R \times T^3$ case, arguably the most
natural one in that all spatial dimensions are treated on the same footing.
Moreover, AVD has been rigorously proven for this class \cite{RingstroemLR}.
Specifically, we consider the polarized $T^3$ Gowdy cosmologies, where the
matrix $M$ is diagonal, $M(x) = {\rm diag}(e^{\phi(x)}, e^{- \phi(x)})$. 
AVD for this case has been established much earlier in \cite{IsenMonc};
moreover, the key field equations are linear, and on the
reduced phase space the system is amenable to a straightforward
canonical quantization \cite{Berger,Torre,QGowdy0,Husain}.
For the construction of Dirac observables on the full phase
space a  Hamiltonian action principle and its gauge symmetries 
is the appropriate starting point. These aspects are prepared in
Appendices \ref{appendixA} and \ref{appendixB}. Below we outline how
the action leads to the reduced
phase space convenient for quantization.

\subsection{Action, proper time gauge, and reduced phase space}
\label{section2.1}
Inserting (\ref{2Kmetric}) with $M(x) = {\rm diag}(e^{\phi(x)}, e^{- \phi(x)})$
into the Gibbons-Hawking action results in a valid action
principle for the polarized Gowdy cosmologies:
\be
\label{SL}
S^\l = \frac{1}{\lbn} \int_{t_i}^{t_f} \!\! dx^0 \!\!\int_0^{2\pi} \!\! dx^1
\Big\{\! - \frac{1}{n} e_0(\rho) e_0(\sigma) + n \big( \dd_1 \rho \dd_1 \sigma
\!-\! 2 \dd_1^2 \rho \big) + \frac{\rho}{2 n} e_0(\phi)^2 - \frac{\rho n}{2}
(\dd_1 \phi)^2\Big\}.
\ee
This action can be inferred from \cite{BM,QGowdy0,TorreObs} or directly verified using computer algebra. Here, $\lbn>0$ is the dimensionless reduced Newton constant,
$e_0 = \dd_0 - \cL_s$ (with $\cL_s$ the one-dimensional Lie-derivative
acting on spatial tensor densities) is the derivative transversal
to the leaves of the $x^0 = {\rm const}.$ foliation,
and $\sigma = \tilde{\sigma} + \frac{1}{2}\ln \rho$. We use 
$x^1 \in \R$ but all fields in (\ref{SL}) are assumed to be spatially
periodic with period $2\pi$ (and thus having a well-defined projection
onto $S^1 = \R/2\pi \Z$.)
This form of
the action is tailored towards holding the fields $\rho,\sigma,\phi$
fixed at the boundaries $x^0=t_i,t_f$. Its Hamiltonian
version (\ref{SH}) allows one to identify the canonically induced
gauge symmetries (\ref{gtrans1}). The subset (\ref{gtrans1},a,b)
with the velocity-momentum relations inserted on the right hand
side of (\ref{gtrans1}b) yields the gauge variations $\delta_{\eps}^\l$
of (\ref{SL}).   
Variation of $S^\l$ with respect to $n,s$ gives the
Lagrangian constraints
\be
\label{constrL}
\frac{\delta S^\l}{\delta n} = - \cH_0^\l\,, \quad
\frac{\delta S^\l}{\delta s} = - \cH_1^\l\,. 
\ee
Here $\cH_0^\l, \cH_1^\l$ coincide with $\cH_0,\cH_1$ from
(\ref{SH}) upon insertion of the velocity-momentum relations.
Importantly, the vanishing conditions $\cH_0^\l = 0 = \cH_1^\l$ can
be solved algebraically for $e_0(\sigma) = F_0(\rho,\phi)$
and $n \dd_1 \sigma = F_1(\rho,\phi)$, where
\ba
\label{Fdef}
F_0(\rho,\phi) \is \frac{e_0(\rho)}{ e_0(\rho)^2 \!- \!(n \dd_1\rho)^2}
  \Big( 2 n^2 \dd_1^2 \rho + \frac{\rho}{2}
  [ e_0(\phi)^2 + (n \dd_1\phi)^2] \Big) 
\nonum
  &-& \frac{n \dd_1 \rho}{ e_0(\rho)^2 \!- \!(n \dd_1 \rho)^2}
  \big( 2 n^2 \dd_1 ( e_0(\rho)/n)  + \rho
  e_0(\phi) n \dd_1  \phi \big)\,, 
\nonum
F_1(\rho,\phi) \is
\frac{e_0(\rho)}{ e_0(\rho)^2 \!-\! (n \dd_1 \rho)^2}
  \big( 2 n^2 \dd_1 ( e_0(\rho)/n)  + \rho
  e_0(\phi) n \dd_1  \phi \big) 
\nonum
&-&  \frac{n \dd_1\rho}{ e_0(\rho)^2 \!- \!(n \dd_1 \rho)^2}
  \Big( 2 n^2 \dd_1^2 \rho + \frac{\rho}{2}
  [ e_0(\phi)^2 + (n \dd_1\phi)^2] \Big)\,. 
  \ea
The fact that the constraints in 2-Killing vector reductions 
can be solved for one of the fields is well-known
\cite{Gravisolitons,Straumannbook};
we present (\ref{Fdef}) mostly to highlight that this can be
done without gauge fixing or coordinate choices.  
By assumption,  $e_0(\rho)^2 \!-\! (n \dd_1 \rho)^2 >0$ for Gowdy
cosmologies. Since $\dd_1 e_0(\sigma + 2 \ln n) =
e_0 \dd_1 (\sigma + 2 \ln n )$  the integrability condition
$\dd_1 [F_0 + 2 e_0(\ln n)]= e_0[ F_1/n + 2 \dd_1 \ln n]$
must be satisfied. This is indeed the case subject to the
evolution equations for $\rho,\phi$, and re-expresses the preservation
(\ref{evol2}) of the constraints under time evolution. The two first order
relations can therefore locally be integrated to yield $\sigma$ as a
functional of the on-shell $\rho$ and $\phi$,
i.e.~$\sigma = \sigma[\rho^{\rm on},\phi^{\rm on}]$.
Further, $e_0 (n^{-1} e_0(\sigma)) - \dd_1( n \dd_1 \sigma)=
e_0(n^{-1} F_0) - \dd_1 F_1 = 2 \dd_1^2 n - e_0(\phi)^2/(2n) +
n (\dd_1 \phi)^2/2$ reproduces the equation of motion for $\sigma$,
i.e.~$\delta S^\l/\delta \rho =0$. That is, the functional
$\sigma = \sigma[\rho^{\rm on}, \phi^{\rm on}]$ is automatically on-shell
as well. Moreover, if $\rho^{\rm on},\phi^{\rm on}$ are parameterized
by their boundary values at $x^0 = t_i,t_f$, so will be
$\sigma[\rho^{\rm on}, \phi^{\rm on}]$, rendering $\sigma|_{t_i},
\sigma|_{t_f}$ a functional of $\rho|_{t_i}, \rho|_{t_f}, \phi|_{t_i},
\phi|_{t_f}$. Together, this allows one to take the fields $\rho,\phi$
as basic and to treat $\sigma$ as a composite field in them.

{\bf Proper time gauge.} So far, no gauge fixing entered.
Both, for setting up a functional integral and for
ensuring a unique classical time evolution one needs to fix a gauge.  
The action principle (\ref{SL}) with fixed boundary fields
would enter the functional integral for the propagation kernel,
in which case the proper time gauge is the appropriate choice
\cite{Teitelpropertime}. The constraints are then imposed
on the Schr\"{o}dinger picture wave functionals and the propagation
kernel is evaluated in the gauge 
\begin{equation}
\label{gf1} 
s = 0 = \dd_0 n\,.
\end{equation}
In this gauge the proper time interval $\tau(x^1) = n(x^1) \int_{t_i}^{t_f}
dx^0 (e^{\tilde{\sigma}/2} )(x^0, x^1)$, between points on the
initial and final spacelike hypersurfaces is fixed. Since $\tau$ is a spatial
scalar, the elapsed proper time does not depend on the choice of spatial
coordinates used (the same though for all level surfaces).  
The only gauge transformations $\delta^\l_{\eps}$ (with $\eps|_{t_i} = 0 =
\eps|_{t_f}$) preserving these conditions have parameters of the form
$\eps =0, \eps^1 = \eps^1(x^1)$. Writing $\delta_{\eps^1}$ for the restricted
gauge variations one has $\delta_{\eps^1} \sigma = \eps^1 \dd_1 \sigma +
2 \dd_1 \eps^1$, $\delta_{\eps^1} n = \eps^1 \dd_1 n - n \dd_1 \eps^1$,
$\delta_{\eps^1} \rho = \eps^1 \dd_1 \rho$, 
$\delta_{\eps^1} \phi = \eps^1 \dd_1 \phi$. 
The residual purely spatial gauge invariance is useful for
bookkeeping purposes. 

As seen above, the solution of the constraints determines the
$\sigma$ field in terms of on-shell $\rho,\phi$ satisfying (\ref{F1int}). 
Consistent therewith, the periodic $\sigma$ field can in the functional
integral for the propagation kernel in proper time gauge be integrated
out to yield a $\delta$ distribution insertion with argument
$\dd_0^2 \rho - (n \dd_1)^2 \rho$. Here we allude to the fact
(not elaborated here) that the Faddeev-Popov factors for the gauge
(\ref{gf1}) do not contain $\sigma$ in differentiated form and can
naturally be interpreted as $\sigma$ independent. The $\delta$ function
insertion produced implements, of course, the evolution equation
for $\rho$, so the Schr\"{o}dinger picture wave functionals
should be projected onto the joint solution set of
$\cH_0^\l =0, \cH_1^\l =0$, $\dd_0^2 \rho - (n \dd_1)^2 \rho =0$. 
This leads to the reduced phase space quantization: 
The basic quantum fields are $\phi, \rho$ constrained by
(\ref{F1int}) below and $\dd_0^2 \rho - (n \dd_1)^2 \rho =0$.
The dynamics of the $\phi$ field is governed by
the propagation kernel computed from the second part of the
action $S^\l$ evaluated in proper time gauge and with prescribed
boundary values $\phi|_{t_i}, \phi|_{t_f}$. The $\rho$ field at
this point is nondynamical and just sets a background geometry of sorts.  
The propagation kernel could be evaluated explicitly in terms of
the Hamilton-Jacobi principal function, but will not be needed. 
It acts on spatially reparameterization invariant Schr\"{o}dinger
picture wave functionals $\Psi[\phi|_{t_i}]$ by convolution. The
non-unique vacuum wave functionals will be Gaussians, analogous to
the situation for free quantum field theories on spatially homogeneous
time dependent backgrounds, see e.g.~\cite{TorreSchroed,BonusSLE}. 
By construction, the $\phi$ two-point functions in the
so-defined Schr\"{o}dinger picture will coincide with the ones
evaluated in the Heisenberg picture with an underlying (non-unique)
Fock vacuum, see Section \ref{section3} for the latter. The only remnants of
the gravitational origin of the system are: (i) the
constraint (\ref{F1int}) needs to be taken into account, and
(ii) the $\sigma$ field needs to be constructed as a renormalized 
composite operator in $\rho,\phi$, see Section \ref{section4}.

So far, we maintained spatial reparameterization invariance
by carrying $n = n(x^1)$ along. This residual gauge invariance
could be gauge fixed as well. Technically, it is simpler to absorb
$n$ into a pseudo-scalar
\begin{equation}
\label{gf2}
\zeta(x^1) = \int_{y^1}^{x^1}\! \frac{dx}{n(x)}\,.
\end{equation}
Note that with a spatially periodic $n$ the quantity
(\ref{gf2}) is quasi-periodic, $\zeta(x^1 + 2\pi) = \zeta(x^1) + \zeta_0$,
$\zeta_0 = \int_0^{2\pi} dx/n(x)$. By suitable constant rescalings
we may assume that $\zeta_0 =2\pi$, so that the periods in $x^1$
and $\zeta$ are the same. Since $\dd_1 \zeta = 1/n>0$ the 
map $\zeta: \R \ra \R$ is invertible and all quantities can be
regarded as functions of $\zeta$ rather than $x^1$. For 
simplicity we retain the original function symbols and write
$\phi(x^0,\zeta)$, $n(\zeta)$, etc., viewed as $2\pi$ periodic functions
of $\zeta$.  The transcription of the action and field equations
is straightforward as mostly $n \dd_1 = \dd_{\zeta}$ enters, while
the exceptional $n \dd_1^2 n$ term transcribes into $\dd_{\zeta}^2 \ln n$.
In particular, the constraints are solved by the transcription of
(\ref{Fdef}) and yield $\dd_0 \sigma$ and $\dd_{\zeta} (\sigma + 2 \ln n)$
as a function of the $\dd_0$ and $\dd_{\zeta}$ derivatives of $\phi,\rho$,
while $n$ no longer occurs.
For later reference we note the independent field equations 
in this version of the proper time gauge 
\ba
\label{evolpt} 
&& \dd_0( \rho \dd_0 \phi) = \dd_{\zeta}(\rho \dd_{\zeta} \phi)\,, \quad
\dd_0^2 \rho = \dd_{\zeta}^2 \rho\,, \quad
\nonum
&& \dd_0 \sigma = F_0^{\rm pt}(\rho,\phi) \,, \quad
\dd_{\zeta} (\sigma +2 \ln n)= F_1^{\rm pt}(\rho,\phi) \,, 
\ea
where $F_0^{\rm pt}$, $F_1^{\rm pt}$ are obtained from $F_0, F_1$ in
(\ref{Fdef}) by the specialization to (\ref{gf1}), (\ref{gf2}).
Explicitly,
\begin{eqnarray}
\label{Fdefpt}   
F_0^{\rm pt}(\rho,\sigma)\is \frac{\dd_0 \rho}{ (\dd_0 \rho)^2 \!- \!(\dd_{\zeta} \rho)^2}
  \Big( 2 \dd_{\zeta}^2 \rho + \frac{\rho}{2}
  [ (\dd_0 \phi)^2 + (\dd_{\zeta} \phi)^2] \Big) 
\nonum
  &-& \frac{\dd_{\zeta} \rho}{ (\dd_0 \rho)^2 \!- \!(\dd_{\zeta} \rho)^2}
  \big( 2 \dd_{\zeta}\dd_0 \rho + \rho
  \dd_0 \phi \dd_{\zeta} \phi \big) 
\nonum 
F_1^{\rm pt}(\rho,\sigma) \is
\frac{\dd_0 \rho}{ (\dd_0 \rho)^2 \!-\! (\dd_{\zeta} \rho)^2}
\big( 2 \dd_{\zeta}\dd_0 \rho + \rho
  \dd_0 \phi \dd_{\zeta} \phi \big) 
\nonum
&-&  \frac{\dd_{\zeta} \rho}{ (\dd_0 \rho)^2 \!- \!(\dd_{\zeta} \rho)^2}
 \Big( 2 \dd_{\zeta}^2 \rho + \frac{\rho}{2}
  [ (\dd_0 \phi)^2 + (\dd_{\zeta} \phi)^2] \Big) \,.
\end{eqnarray}
The integrability condition mentioned after Eq.(2.5) specializes to
$\dd_{\zeta} F_0^{\rm pt} = \dd_0 F_1^{\rm pt}$, and can be checked to
hold on account of the first two equations in (2.8). Note that the
$\phi$ field equation is through $\rho$ explicitly $(x_0,\zeta)$-dependent;
in particular, it is not conformally invariant.
The other vestige of the gravitational origin of the system then is
that in order for $\sigma$ to be spatially periodic, one must have
\be
\label{F1int} 
\int_0^{2\pi} \frac{d\zeta}{2\pi} F_1^{\rm pt}(\rho,\phi)
\stackrel{\displaystyle{!}}{=}0\,.
\ee

{\bf VD Gowdy system.} The velocity dominated (VD) Gowdy system is
normally introduced only on the level of the gauge fixed
field equations \cite{IsenMonc,RingstroemLR}. It can, however, be
viewed as a (electric) Carroll-type gravity theory \cite{HenneauxC,GRstrongI} 
in its own right with Lagrangian action
\be
\label{SVDL}
S^{\smallcap{lvd}} =
\frac{1}{\lbn} \int_{t_i}^{t_f} \!\! dx^0 \!\!\int_0^{2\pi} \!\! dx^1
\Big\{\! - \frac{1}{n} e_0(\varrho) e_0(\varsigma) +
\frac{\rho}{2 n} e_0(\vp)^2\Big\},
\ee
where $\varrho, \varsigma, \vp$ are the counterparts
$\rho,\sigma,\phi$ in the VD system. Strictly speaking different
symbols should be used also for lapse and shift, for readability's
sake we retain the original notation. Observe that in (\ref{SVDL})
spatial points are only coupled through the Lie derivative term
in $e_0$. Correspondingly, there is a Diffeomorphism constraint
$\delta S^{\smallcap{lvd}}/\delta s = - \cH_1^{\smallcap{lvd}}$
of the same form as in the full Gowdy system, while the Hamiltonian
constraint $\delta S^{\smallcap{lvd}}/\delta n = - \cH_0^{\smallcap{lvd}}$
simplifies. The constraints can be solved for $e_0(\varsigma)$ and
$\dd_1 \varsigma$ and determine $\varsigma$ as a functional
of the on-shell $\varrho, \vp$, i.e.~$\varsigma =
\varsigma[\varrho^{\rm on},\vp^{\rm on}]$. A more detailed
exposition of the VD Hamiltonian action and its gauge symmetries
is relegated to Appendix \ref{appendixA}. The transition to a reduced phase
space formulation proceeds in parallel to the previous discussion.  

For comparison's sake, we note the independent field equations
in the proper time gauge (\ref{gf1}), (\ref{gf2}):
\ba
\label{VDevolpt}
&& \dd_0(\varrho \dd_0 \vp) =0\,, \quad 
\dd_0^2 \varrho =0\,,
\\[2mm] 
&& 
\dd_0 \varsigma = \frac{\varrho}{2 \dd_0 \varrho} (\dd_0 \vp)^2\,,
\quad 
\dd_{\zeta} ( \varsigma + 2 \ln n) = \frac{1}{\dd_0 \varrho}
\big( \dd_0 \dd_{\zeta} \varrho + \varrho \dd_0 \vp \dd_{\zeta} \vp \big) -
\frac{ \varrho\dd_{\zeta} \varrho}{ 2 (\dd_0 \varrho)^2} (\dd_0 \vp)^2\,. 
\nonumber
\ea
Here the counterpart of (\ref{F1int}) needs to be imposed in order
to ensure that $\varsigma$ is spatially periodic.

{\bf Matched foliations.} So far we kept $\rho$ as a field,
which in the reduced phase space formulation sets a background
geometry of sorts for the dynamical field $\phi$ and the induced
$\sigma$. It is worth noting that $\rho$ meets the criteria for  a
temporal function in Lorentzian geometry: it is a strictly
positive scalar function which has an everywhere timelike
gradient $d\rho = \dd_{\mu} \rho dx^{\mu}$ such that the associated
vector field $\gamma^{\mu\nu} \dd_{\nu} \rho$ is past pointing. 
This means, the level surfaces $\rho = {\rm const}.$ define
a foliation of the two-dimensional Lorentzian manifold
$(\R \times S^1, \gamma)$. Specifically, in proper time gauge
one can use the equation of motion $(\dd_0 - \dd_{\zeta})(\dd_0 + \dd_{\zeta})
\rho = 0$ and the spatial periodicity requirement to conclude that
$\rho$ must be quasi-periodic in $x^0$, i.e.~$\rho(x^0 + n 2\pi, \zeta)
= \rho(x^0, \zeta) + n \rho_0$, with $\rho_0 = \int_0^{2\pi} dx
(\dd_0 \rho)$. Positivity then requires either $x^0 >0, \rho_0>0$
or $x^0<0, \rho_0<0$. Choosing the former and taking into account
$(\dd_0 \rho)^2 > (\dd_{\zeta} \rho)^2 >0$ one sees that 
$\rho(x^0,\zeta)$ is strictly increasing in $x^0$, pointwise in $\zeta$.
Its level surfaces $\rho(x^0,\zeta) = {\rm const}.$ define
a foliation of the half cylinder $\R_+ \times S^1$. 

On the other hand, the defining relation for the lapse in terms
of an ADM temporal function $T$ specializes to
$\sqrt{\gamma} \gamma^{\mu\nu} \dd_{\mu} T \dd_{\nu} T = - 1/n$ for 
the metric (\ref{2Dmetric}). This is satisfied for the choice
$T = x^0$, and the level surfaces $x^0 = {\rm const}$ set the
standard ADM foliation to which the coordinatization of the fields
in terms of $(x^0,\zeta)$ refers. In most of the Gowdy literature
the $\rho$-foliation and the standard ADM foliation are identified
by stipulating\footnote{As a metric component $\rho$ is dimensionless.
For simplicity, we suppress a dimensionful conversion factor in this
and subsequent identifications.}
\begin{equation}
\label{folimatch} 
\rho = x^0 =:t>0\,.
\end{equation}
This is especially natural in the reduced phase space formulation,
where $\rho$ merely sets a background for the $\phi,\sigma$
evolution. Technically, (\ref{folimatch}) should not be regarded as
a gauge choice complementing the proper time gauge. Indeed,
for an off shell $\rho$ the condition $n\dd_1 \rho =0$ is
compatible with the residual gauge symmetry of the proper time gauge:
$\delta_{\eps^1} \dd_1 \rho|_{\dd_1 \rho =0} = 
\dd_1[\delta_{\eps^1} \rho]|_{\dd_1 \rho =0} = 0$.
One can still take $\rho = \rho(x^0)$ to be an arbitrary
(positive, strictly increasing) function of $x^0$. The
identification (\ref{folimatch}) picks one of them. 
We stress this because in the discussion of Dirac observables,
the full phase space must be used, including the 
$(\rho, \pi^{\rho})$ canonical pair as off-shell dynamical degree
of freedom, see Appendix \ref{appendixB}. Once Dirac observables $\cO(\th)$ have
successfully constructed the choice (\ref{folimatch}) eliminates 
the $(\rho, \pi^{\rho})$ canonical pair and reduces them to
(semilocal) charges which are conserved on-shell, $\dd_t \cO(\th) =0$.

A similar discussion applies to the VD Gowdy system. To avoid
a detour into Carroll geometry, we consider directly the
properties of $\varrho$ in proper time gauge. The conditions  
$\dd_0^2 \varrho =0$, $\varrho >0$, and the spatial periodicity
fix $\varrho(x^0,\zeta) = \varrho_1(\zeta) x^0 + \varrho_0(\zeta)$.
Taking $x^0>0$ one needs $\varrho_0,\varrho_1$ to be periodic and
positive.  Again, $\varrho$ plays the role of a local time function,
now linear in $x^0$. The counterpart of (\ref{folimatch}) is
the choice $\varrho = x^0>0$, which also here it is not a gauge choice
on top of the proper time gauge. Together, the matched foliations  
condition identifies three temporal functions, $\rho$, $\varrho$,
and the ADM one:
\begin{equation}
\label{AVDtime} 
\rho = x^0 = t = \varrho>0\,.
\end{equation}
This sets a shared coordinate time $t$ in which the
solutions of the full and the VD Gowdy system evolve.

\subsection{The AVD property} 
\label{section2.2}

In the shared coordinate time  $\rho = \varrho = t>0$ both sets
of field equations (\ref{evolpt}) and (\ref{VDevolpt}) can essentially
be solved in closed form. For (\ref{VDevolpt}) the general solution is
\ba
\label{VDsol} 
&& \vp(t, \zeta) = v(\zeta) \ln(t/t_0) + \vp_0(\zeta)\,, \quad
\varsigma(t, \zeta) = \ln(t/t_0) \frac{1}{2} v(\zeta)^2
+ \varsigma_0(\zeta)\,,
\nonum
&& \varsigma_0(\zeta) = - 2 \ln (\frac{n(\zeta)}{n(\zeta_0)})
+ \int_{\zeta_0}^{\zeta} \! d\zeta'
v(\zeta') \dd_{\zeta'} \vp_0(\zeta')\,.
\ea 
Note that this carries a largely arbitrary $\zeta$ dependence
merely through the choice of initial conditions at $t = t_0$. 
Further, $\varsigma$ is determined by $\vp$ up to a constant. 
The solution of (\ref{evolpt}) in $\rho = t$ time
proceeds via spatial Fourier decomposition of $\dd_t ( t \dd_t \phi) =
t \dd_{\zeta}^2 \phi$ and leads to Bessel functions with index zero.  
Upon Fourier synthesis, one can insert the solution into
the constraint equations and integrate them to find $\sigma$.
Needless to say, this is only possible in the polarized case.
In the non-polarized Gowdy system the evolution equations are
nonlinear partial differential equations and despite the existence of
a Lax pair they cannot be integrated in closed form. On the other
hand, the VD Gowdy system admits a simple explicit solution even
in the non-polarized case. This pattern underlines why the
AVD property is interesting and nontrivial.  

Foreshadowing much of the subsequent developments the AVD 
property for polarized Gowdy cosmologies has been proven by
Isenberg and Moncrief \cite{IsenMonc}, see Thm.~III.5 for the $T^3$
topology considered here.%
\footnote{The notational correspondence to \cite{IsenMonc} is:
$\phi \mapsto W, \,\tilde{\sigma} +2 \ln n = \sigma + \ln(n^2\sqrt{t})
\mapsto 2 a, \,\ln(t/t_0) = \tau \mapsto - (\tau\!- \!\tau_0)$.}
Schematically, the result asserts the following:
Given initial data $\phi_0(\zeta), \pi_0(\zeta), \sigma_0(\zeta)$
at some $t_0 >0$ there exists a solution (\ref{VDsol})
of the VD system and correction terms $\Phi(t,\zeta), \Sigma(t,\zeta)$
such that
\ba
\label{AVD} 
\!\!& \nspace & \phi(t,\zeta) = \vp(t,\zeta) + \Phi(t,\zeta), \quad
\sigma(t,\zeta) = \varsigma(t,\zeta) + \Sigma(t,\zeta)\,,
\nonum
\!\!&\nspace &
|\dd_{\zeta}^k \Phi(t,\zeta)|, |t \dd_t \dd_{\zeta}^k \Phi(t,\zeta)|,
|\dd_{\zeta}^k \Sigma(t,\zeta)| \leq
c (1 \!+\! \ln^2 (t/t_0) )t^2\,,
\ea
(with $c>0$ and some finite differentiation order $k$) is a solution of
the full Gowdy field equations (\ref{evolpt})
with initial data $\phi(t_0,\zeta) = \phi_0(\zeta)$,
$(t \dd_t \phi)(t_0,\zeta) = \pi_0(\zeta)$,
$\sigma(t,\zeta) = \sigma_0(\zeta)$. The same holds if instead
of initial data a non-exceptional solution (\ref{VDsol}) for
the VD system is prescribed.  

In other words, all solutions of the full Gowdy system approach a
VD solution near the Big Bang. Conversely, almost every solution
of the VD system can be lifted to a solution of the full
Gowdy system via (\ref{AVD}).  

The origin of the exceptional points can be understood from
the blow-up behavior of the curvature scalars. Considering
$\cR_2 := R_{\mu\nu\rho\sigma}R^{\mu\nu\rho\sigma}$, the result
of \cite{IsenMonc}, Thm.~IV.1, is that the rate of blow up for
$t \ra 0^+$ is determined solely by the coefficient $v(\zeta)$
in the velocity dominated solution (\ref{VDsol}). Specifically,  
\ba
\label{Kretsch} 
\cR_2(t, \zeta_0) =
\left\{ \begin{array}{cc}  O(t^{-3 - v(\zeta_0)^2})  & \mbox{if} \;\;
  v(\zeta_0)^2 \neq 1\,,\\[2mm]
  O(t^{-2}) & \mbox{if} \;\; v(\zeta_0)^2 =1\,,\;
  (\dd_{\zeta}v)(\zeta_0) \neq 0\,,\\[2mm]
  O(\ln^2 t/t_0) & \mbox{if} \;\; v(\zeta_0)^2 =1\,,\;
  (\dd_{\zeta}v)(\zeta_0) =0\,, \;\;(\dd_{\zeta}^2v)(\zeta_0) \neq 0\,.
\end{array}
\right. 
\ea  
This means, a form of ``cosmic censorship'' holds: except
for exceptional cases with `fine tuned' data, the polarized Gowdy
cosmologies have a  Big Bang singularity. Moreover, they do
not admit an analytic extension beyond the singularity to
include non-globally hyperbolic (`acausal') regions.


\subsection{Fourier decomposition and canonical quantization}

Returning to (\ref{AVD}), since $\sigma$ and $\varsigma$
are determined by $\phi$ and $\vp$, respectively, the key aspect
is the one-to-one correspondence between $\phi(t,\zeta)$ and $\vp(t, \zeta)$.
As their defining wave equations are linear, the problem can naturally
be recast in terms of the spatial Fourier transforms. Our conventions
for the Fourier transform are
\ba
f(\tau, \zeta) \is \sum_{n \in \Z} f_n(\tau) e^{i n \zeta}\,,
\quad \;\;f_n(\tau) = \int_{0}^{2 \pi} \! \frac{d\zeta}{2\pi} \,
e^{ - i n \zeta} f(\tau,\zeta)\,. 
\ea
For the scalar field, we use the mode expansion
\be
\label{ModePP}
\phi(\tau,\zeta) = T_0(\tau) a_0 + T_0(\tau)^* a_0^* +
\sum_{ n\neq 0} \big\{ T_{n}(\tau) e^{i n \zeta} a_n +
  T_{n}(\tau)^* e^{-i  n \zeta} a_n^* \big\}\,,
\ee
where the time variable is taken to be $\tau = \ln t/t_0$,
with $t \in \R_+$ transcribing to $\tau \in \R$, and the Big Bang
located at $\tau = - \infty$. Classically, the $a_n, a_n^*$ are
just complex parameters; later, they will be promoted to Fock space operators, see \cite{Berger, Torre, QGowdy0, TorreSchroed}. 
Applying a spatial Fourier transform
to the wave equation for $\phi$ in Eq.~(\ref{evolpt})
for $\rho = t$ yields 
\be
\label{Tsol1} 
[ \dd_{\tau}^2 + e^{2 \tau} (t_0 n)^2 ] T_n(\tau) =0\,,
\quad (\dd_{\tau} T_n) {T_n}^* - (\dd_{\tau} T_n)^* T_n = - i\,.  
\ee
When using the dimensionless $\tau$ time variable only the
dimensionless combination $t_0 n$ occurs. For notational 
simplicity we will just write $n$ for this, with $t_0$ tacit
whenever demanded for dimensionality reasons. 
In the second equation we imposed a Wronskian normalization condition
based on the fact that the quantity on the left hand side is constant
in $\tau$. The general solution can be written in terms of Hankel
functions in the form
\ba
\label{Tsol2} 
T_n(\tau) \is
\frac{\sqrt{\pi}}{2} \lb_n H_0^{(2)}(|n| e^{\tau}) + 
  \frac{\sqrt{\pi}}{2}\mu_n H_0^{(1)}(|n| e^{\tau}) \,,
    \quad \,|\lb_n|^2 - |\mu_n|^2 =1\,,\quad n \neq 0\,,
\nonum
T_0(\tau) \is - \frac{i}{\sqrt{\pi}}(\tilde{\lb}_0 -\tilde{\mu}_0)
\tau 
    + \frac{\sqrt{\pi}}{2}(\tilde{\lb}_0 + \tilde{\mu}_0)\,, \quad
    |\tilde{\lb}_0|^2 - |\tilde{\mu}_0|^2 =1\,.
    \ea
We assume throughout that $\lb_n = \lb_{-n}, \mu_n= \mu_{-n}$ are
bounded in $|n|$ and take this as part of the definition of a `state'
via the two-point function (\ref{Wsdef}) below. 
An alternative parameterization is in terms of initial data 
\be
\label{Initialdata}
z_n(\tau_0):= T_n(\tau_0), \quad w_n(\tau_0) := (\dd_\tau T_n)(\tau_0).
\ee 
Displaying also a potential $\tau_0$-dependence of the Bogoliubov parameters
$\lb_n = \lb_n(\tau_0)$ and $\mu_n = \mu_n (\tau_0)$, both parameterizations
are for $n \neq 0$ related by 
\be
\label{FHad11}
\begin{pmatrix} z_n(\tau_0) \\ w_n(\tau_0) \end{pmatrix}
= H(|n|e^{\tau_0})
\begin{pmatrix} \lb_n(\tau_0) \\ \mu_n (\tau_0) \end{pmatrix}\,, \quad
H(x) := \frac{\sqrt{\pi}}{2} \begin{pmatrix}
  H_0^{(2)}(x) &  H_0^{(1)}(x) \\[1.5mm]
  -x H_1^{(2)}(x) &  -x H_1^{(1)}(x)
  \end{pmatrix},
\ee
with $\det H(x) = i$.

Upon canonical quantization the $a_{n}$ and $a_{n}^*$
are promoted to annihilation and creation operators
normalized according to $[a_n, a_m^*] = \delta_{n,m}$. Each
choice of Wronskian normalized solution $T_n(\tau)$, corresponds to
a different decomposition (\ref{ModePP}), i.e.~to different
Fock space operators $a_n = a_n^T$, $a_n^* = (a_n^T)^*$,
and we associate a Fock vacuum to some such choice via
\begin{equation}
\label{Tvac}
a_n | 0_T\ket = 0\,, \quad n \in \mathbb{Z}.
\end{equation}

For the velocity dominated system,  we adapt (\ref{ModePP}) to
\begin{equation}
\label{ModeVPP}
\vp(\tau,\zeta) = T_0(\tau) a_0 + T_0(\tau)^* a_0^* +
\sum_{ n\neq 0} \big\{ \mathfrak{t}_{n}(\tau)
e^{i  n \zeta} a_n +
\mathfrak{t}_{n }(\tau)^* e^{-i  n \zeta}
a_n^* \big\}\,,
\end{equation}
where the defining relation for $\mathfrak{t}_{n}(\tau)$ is simply
$\dd_{\tau}^2 \mathfrak{t}_{n}=0$. We parameterize the
general Wronskian normalized solution as 
\begin{equation}
\label{gpol6}
\mathfrak{t}_n(\tau) = - \frac{i}{\sqrt{\pi}}(\tilde{\lb}_n -\tilde{\mu}_n)
\tau + \frac{\sqrt{\pi}}{2}(\tilde{\lb}_n + \tilde{\mu}_n)
= \tilde{w}_n(\tau_0) (\tau -\tau_0) + \tilde{z}_n(\tau_0)\,.
\end{equation}
Guided by (\ref{AVD}) one will describe the quantized velocity dominated
scalar $\vp(\tau, \zeta)$
by the {\it same} set of creation/annihilation operators in Fourier
space as $\phi(\tau,\zeta)$, but use as the classical coefficient functions
those $\mathfrak{t}_n$ of the velocity dominated system. We highlight
the concomitant identification of the vacua
\begin{equation}
\label{Ttvac} 
|0_T\ket = |0_{\mathfrak{t}}\ket\,.
\end{equation}
For $n=0$, we identify $T_0(\tau) = \mathfrak{t}_0(\tau)$.
The Wronskian normalization condition amounts respectively to 
$|\tilde{\lb}_n|^2 - |\tilde{\mu}_n|^2 =1$ and
$(\tilde{w}_n\tilde{z}_n^* - \tilde{w}_n^* \tilde{z}_n)(\tau_0) = -i$.
For later use, we express this in parallel to (\ref{FHad11}) as
\ba
\label{H0def} 
\begin{pmatrix} \tilde{z}_n(\tau_0) \\ \tilde{w}_n(\tau_0) \end{pmatrix}
= \tilde{H}(\tau_0)
\begin{pmatrix} \tilde{\lb}_n(\tau_0) \\ \tilde{\mu}_n(\tau_0) \end{pmatrix}
\,,\quad
\tilde{H}(\tau_0) := \begin{pmatrix}
  h(\tau_0) & h(\tau_0)^* \\[2mm]
 - i/\sqrt{\pi} &  i/\sqrt{\pi}
\end{pmatrix}\,,
\ea 
where $h(\tau) := \sqrt{\pi}/2 - i \tau/\sqrt{\pi}$.

\subsection{Two-point functions of gauge-fixed Dirac observables} 

In Appendix \ref{appendixB}, a one-parameter family of off-shell Dirac
observables for the polarized $T^3$ Gowdy cosmologies is constructed,
which strongly Poisson commutes with the constraints. 
In proper time gauge, with $\lbn =1$, and $\rho = t$, $\tilde{\rho}
= \zeta$ coordinates the on-shell Dirac observables (\ref{diracon2}) can after
integration-by-parts be written as
\ba
\label{diractwopt1} \cO(\th) \!\is \!\int_0^{2 \pi}
\frac{d\zeta}{2\pi} q(\tau,\zeta;\th)\,, \quad {\rm Im} \th \neq 0\,,
\\[2mm] 
q(\tau,\zeta;\th) \!&\!\!:=\!\!& \!
c_0(\zeta\! + \!\th, e^{\tau}) \dd_{\tau} \phi +
e^{2 \tau} \dd_{\zeta} c_1(\zeta\! +\! \th, e^{\tau}) \phi
= \jmath_0^{\smallcap{p}}(e^{\tau},\zeta;\th) + \dd_{\zeta}
[ e^{2 \tau} c_1(\zeta \!+ \!\th, e^{\tau}) \phi]\,,
\nonumber
\ea
where we use $\tau = \ln t$ as time variable and omit the superscript
``on''. By construction,
the $c_0,c_1$ are $2\pi$-periodic functions in $\zeta$ and are such that 
$\dd_{\tau} \cO(\th) =0$ holds, subject to the evolution equations in
the same specialization. This conservation is a consequence of the stronger
gauge invariance of the $\cO(\th)$ in (\ref{jaDirac}), but in the
fully gauge-fixed setting convenient for (canonical) quantization  
it is the main indicator for the observable property. 
In relativistic quantum field theories conserved charges normally
have to be regularized in order to give rise to well-defined (unbounded)
operators. Remarkably, this is not necessary here. Inserting the
mode expansion (\ref{ModePP}) for $\phi$ a well-defined unbounded
operator on Fock space arises as long as ${\rm Im} \th \neq 0$. It is given
by (\ref{diracon3}) with the $a_n^*,a_n$ read as creation, annihilation
operators. The two-point function in the vacuum (\ref{Tvac}) is
simply
\begin{equation}
\label{Otwopt}
\bra 0_T| \cO(\th) \cO(\th') |0_T \ket =
\frac{1}{\pi} \sum_{n \geq 1} |\lb_n\!-\!\mu_n|^2 \,n^2
e^{ \pm i n (\th - {\th'}^*)}\,, \quad \pm {\rm Im} \th >0,\,
\pm {\rm Im} \th' >0\,,
\end{equation}
where $\lb_n,\mu_n$ are the Bogoliubov parameters entering via
(\ref{Tsol2}). Clearly, the sum in (\ref{Otwopt}) is rapidly convergent
for all $\lb_n,\mu_n$ bounded in $|n|$. In Section 4 we shall
argue that physically viable states have in fact parameters obeying
$\lb_n \ra 1$, $\mu_n \ra 0$, as $n \ra \infty$. The two-point
function (\ref{Otwopt}) is manifestly $\tau$ independent and turns
to coincide exactly with its counterpart computed in the velocity
dominated system, see (\ref{diracon4}). While conceptually satisfying
this renders (\ref{Otwopt}) unsuited to study AVD. 

Better suited for this purpose is the two-point functions of the integrands
$q(\tau,\zeta;\th)$ in the Fock vacuum (\ref{Tvac}). As often, it is
useful to consider the symmetric and anti-symmetric parts
separately. These can be expressed in matrix form as follows
\ba
\label{diractwopt2} 
\!\!&\nspace & \bra 0_T| q(\tau,\zeta;\th)q(\tau',\zeta';\th') +
q(\tau',\zeta';\th')q(\tau,\zeta;\th) |0_T\ket
\\[2mm] 
\!\!&\nspace & \quad =
\big( e^{2 \tau} \dd_{\zeta} c_1(\zeta\!+\!\th, e^{\tau}),
c_0(\zeta\!+\!\th, e^{\tau}) \big) 
\cW^s(\tau,\tau', \zeta - \zeta') \begin{pmatrix}
  e^{2 \tau'} \dd_{\zeta'} c_1(\zeta'\!+\!\th', e^{\tau'})
  \\[2mm] c_0(\zeta'\!+\!\th', e^{\tau'})
\end{pmatrix},  
\nonumber
\ea
where the following symmetric matrix two-point function enters
\ba
\label{Wsmatdef} 
\cW^s(\tau,\tau', \zeta - \zeta') &:=& 
\begin{pmatrix} 1 & \dd_{\tau'} \\[2mm] \dd_{\tau} & \dd_{\tau} \dd_{\tau'} 
\end{pmatrix} W^s(\tau, \tau', \zeta - \zeta') \,,
\nonum
W^s(\tau, \tau', \zeta - \zeta') &:=& 
\bra 0_T| \phi(\tau,\zeta) \phi(\tau',\zeta') +
\phi(\tau',\zeta') \phi(\tau,\zeta) |0_T\ket \,. 
\ea 
In the parameterization (\ref{Tsol2}) this gives
\ba
\label{Wsdef}
&& W^s(\tau, \tau', \zeta - \zeta') = W_0(\tau,\tau') + 
2\sum_{n \geq 1} \cos n (\zeta\!- \!\zeta') W_n(\tau,\tau') ,
\\[2 mm]
&& W_0(\tau,\tau') = 
\frac{2}{\pi} |\tilde{\lb}_0 \!- \!\tilde{\mu}_0|^2
\tau \tau'
-  i (\tilde{\lb}_0\tilde{\mu}_0^* \!- \!\tilde{\lb}_0^*\tilde{\mu}_0)
(\tau\! +\!\tau' )
+ \frac{\pi}{2} |\tilde{\lb}_0 + \tilde{\mu}_0|^2 \,,
\nonum
&& W_n(\tau,\tau') = \frac{\pi}{2}
\big( |\lb_n|^2 + |\mu_n|^2 \big) 
\big( J_0(|n|e^{\tau}) J_0( |n| e^{\tau'}) + Y_0(|n| e^{\tau'}) Y_0(|n|e^{\tau}) \big) 
\nonum
&& \quad + \!\frac{\pi}{2}  (\lb_n \mu_n^* + \lb_n^* \mu_n)
\big( J_0(|n|e^{\tau}) J_0( |n| e^{\tau'}) - Y_0(|n| e^{\tau'}) Y_0(|n|e^{\tau}) \big)
\nonum
&& \quad - \!\frac{\pi}{2}  i(\lb_n \mu_n^* - \lb_n^* \mu_n)
\big( J_0(|n|e^{\tau}) Y_0( |n| e^{\tau'}) + Y_0(|n| e^{\tau}) J_0(|n|e^{\tau'}) \big) \,,
\quad n \neq 0\,.
\nonumber
\ea
The anti-symmetric part of $q$'s two-point function is independent
of the choice of vacuum state $|0_T\ket$ and is just given by the
commutator, i.e.~$\bra 0_T| q(\tau,\zeta;\th)q(\tau',\zeta';\th) -
q(\tau',\zeta';\th)q(\tau,\zeta;\th) |0_T\ket =
[q(\tau,\zeta;\th), q(\tau',\zeta';\th)]$. 
In matrix form 
\ba 
\label{diractwopt3} 
&\nspace & i[ q(\tau,\zeta;\th), q(\tau',\zeta';\th')]
\\[2mm] 
& \nspace & \quad =
\big(e^{2 \tau} \dd_{\zeta} c_1(\zeta\!+\!\th, e^{\tau}),
c_0(\zeta\!+\!\th, e^{\tau}) \big) 
D(\tau,\tau', \zeta - \zeta')
\begin{pmatrix} 0 & -1
\\[2mm] 1 & 0 
\end{pmatrix}   
\begin{pmatrix} e^{2 \tau'} \dd_{\zeta'} c_1(\zeta'\!+\!\th', e^{\tau'})
\\[2mm] c_0(\zeta'\!+\!\th', e^{\tau'})
\end{pmatrix}. 
\nonumber
\ea 
The matrix $D$ is defined in parallel to (\ref{Wsmatdef}) but with
a `symplectic unit' matrix taken out for later convenience.
Explicitly
\ba
\label{Dmatdef} 
\nspace D(\tau,\tau', \zeta - \zeta')
\begin{pmatrix} 0 & -1
\\[2mm] 1 & 0 
\end{pmatrix}   
&:=& 
\begin{pmatrix} 1 & \dd_{\tau'} \\[2mm] \dd_{\tau} & \dd_{\tau} \dd_{\tau'} 
\end{pmatrix} \Delta(\tau, \tau', \zeta - \zeta') \,,
\nonum
\nspace \Delta(\tau, \tau', \zeta - \zeta') &:=& 
i \bra 0_T| \phi(\tau,\zeta) \phi(\tau',\zeta') -
\phi(\tau',\zeta') \phi(\tau,\zeta) |0_T\ket \,. 
\ea 
We recall from Appendix B that $c_0(\zeta\!+\!\th, e^{\tau})$,
$c_1(\zeta\!+\!\th, e^{\tau})$
are explicitly known and given by convergent power series in
$e^{2\tau}$, see (\ref{csums4}). The computation of the two-point
functions (\ref{diractwopt2}), (\ref{diractwopt3}) therefore amounts to the
evaluation of the matrix two-point functions $\cW^s$ and $D$.  
Our primary interest will lie in their dependence on $\tau,\tau'$.

\newpage 
\section{Time evolution of two-point functions} 
\label{section3} 

The time evolution of the Heisenberg picture field operators
and Fock space operators can be described by convolution with a
symplectic matrix kernel that is defined in terms of the commutator
function; see (\ref{opevol}). This in turn gives rise to an initial value
parameterization for the symmetric part of the two-point functions;
see (\ref{twopmat2}), (\ref{twopmat3}). The bonus property 
that the time evolution of (slightly modified) Heisenberg picture field
operators can be unitarily implemented on Fock space \cite{QGowdy0}
will not be  needed. For simplicity,
we set $\lbn =1$ throughout this section.

\subsection{Time evolution of field operators}

The solution of the initial value problem of Heisenberg picture field operators can concisely be expressed in terms of the commutator function
$\Delta(\tau,\tau',\zeta-\zeta')$ defined in (\ref{Dmatdef}). 
Its Fourier components $\Delta_n(\tau,\tau')$ are real valued and
can be characterized by the following relations:
$[\dd_{\tau}^2 + n^2 e^{2 \tau}] \Delta_n(\tau,\tau') =0$,
$\Delta_n(\tau,\tau') = - \Delta_n(\tau',\tau)$,
$\dd_{\tau} \Delta_n(\tau,\tau')|_{\tau'  = \tau} =~1$. Merely on account
of these properties one has
\be
\label{initial3}
\!\!\begin{pmatrix} \phi_n(\tau) \\[1.5mm] \pi_n(\tau) \end{pmatrix}
= D_n(\tau, \tau_0)
\begin{pmatrix} \phi_n(\tau_0) \\[1.5mm] \pi_n(\tau_0) \end{pmatrix},
\quad
D_n(\tau,\tau_0) := 
\begin{pmatrix} - \dd_{\tau_0} \Delta_n(\tau,\tau_0)
  & \Delta_n(\tau,\tau_0) \\[1.5mm]
  -\dd_{\tau} \dd_{\tau_0} \Delta_n(\tau,\tau_0)
  & \dd_{\tau}\Delta_n(\tau,\tau_0) \end{pmatrix},
\ee
for all $n \in \Z$, where $\pi_n(\tau) := \dd_\tau \phi_n (\tau)$. We take
$\Delta_0(\tau, \tau_0) := \tau - \tau_0$. If the data are propagated in
two time steps, from $\tau_0$ to $\tau_1$ and then from $\tau_1$
to $\tau$, uniqueness of the solution implies consistency conditions
on the evolution matrix in (\ref{initial3}). The four conditions
arising consist of a basic one (the 1,2 matrix component) and
derivatives thereof. The basic identity reads
\be
\label{initial2}
-\dd_{\tau_1} \Delta_n(\tau,\tau_1) \Delta_n(\tau_1, \tau_0) 
+ \Delta_n(\tau, \tau_1) \dd_{\tau_1} \Delta_n(\tau_1,\tau_0)
= \Delta_n(\tau,\tau_0)\,,
\ee
and again holds only on account of the above properties of $\Delta_n$. 
The matrix $D_n$ obeys $D_n(\tau, \tau_1) D_n(\tau_1, \tau_0) =
D_n(\tau, \tau_0)$ by (\ref{initial2}). In particular,
$D_n(\tau,\tau_0)^{-1} = D_n(\tau_0, \tau)$ and by interpreting
the adjoint in terms of the inverse it is also symplectic.

Upon Fourier synthesis the matrix kernel
$D(\tau,\zeta;\tau',\zeta')$ governs the time evolution
of the Heisenberg picture field operators,
see (\ref{opevol}) below. It is manifestly built from
the time derivatives of the basic commutator function
$\Delta(\tau,\zeta;\tau',\zeta')$. However, the inverse
Fourier transform of $\Delta_n$ can not be evaluated
in closed form and even certain qualitative properties are 
masked. Explicitly, we seek to evaluate 
\ba
\label{Gowdycomm}
\Delta(t,t',\zeta - \zeta') \is \ln(t/t') + 2\sum_{n \geq 1}
\cos n (\zeta - \zeta') \Delta_n(t,t')\,,
\nonum
\Delta_n(t,t') \is \frac{\pi}{2}\big\{Y_0(|n|t)J_0(|n|t') - 
J_0(|n|t)Y_0(|n|t') \big\}\,,
\ea
where for notational convenience we revert to  $t,t'>0$ as time variable.
When the spatial sections are diffeomorphic to $\R$ the
commutator function $\Delta_{\R}(t,t',\zeta - \zeta')$
can be shown analytically to vanish for spacelike distances,
$|\zeta - \zeta'| > |t-t'|$ (despite the time dependent background).
We state without derivation the following link to the
commutator function (\ref{Gowdycomm}) on the cylinder needed here:
\be
\label{CommAPvsPP}
2\pi \sum_{n \in \Z} \Delta_{\R}(t,t', \zeta - \zeta' +2 \pi n)
  = \Delta(t,t', \zeta - \zeta')\,,
  \ee
where the sum is finite for each fixed set of arguments.
This relation implies that within each periodicity
interval $0 < |\zeta - \zeta'| < 2\pi$ also the cylinder
commutator function vanishes exactly for $|\zeta - \zeta'| > |t-t'|$.

\begin{figure}[h]
    \centering
    \includegraphics[scale=0.8]{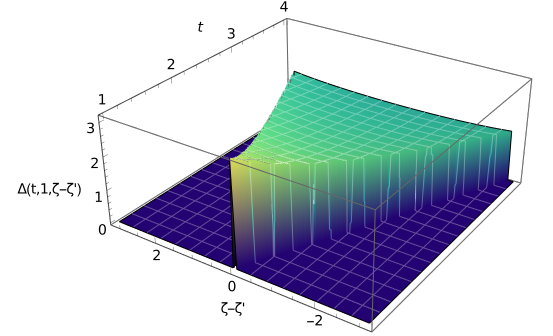}
    \caption{\label{Fig1} Commutator function $\Delta(t,1,\zeta-\zeta')$,
      $t \geq 1$, $\zeta - \zeta'\in [-\pi,\pi]$.} 
\end{figure}

A brute force implementation of the sum (\ref{Gowdycomm})
turns out to be numerically unstable. We therefore
perform the sums that govern its qualitative
behavior analytically and treat only the rest numerically. 
By the exact anti-symmetry in $t,t'$ we may restrict attention to
$t \geq t'>0$. For large $n >1$ the terms behave like 
\ba
\label{PPtwop3}
&\nspace & \frac{\pi}{2} \big( Y_0(nt) J_0(nt') - Y_0(nt') J_0(nt) \big)
\nonum
& \nspace & \quad =: \frac{1}{\sqrt{tt'}}
\frac{\sin n (t\!-\!t')}{n}
+ \frac{t\!-\!t'}{8(tt')^{3/2}} 
\frac{\cos n (t\!-\!t')}{n^2} + \Delta^{\rm sub2}_n(t,t')\,.
\ea 
Here, the `subtracted' summands $ \Delta_n^{\rm sub2}(t,t')$
are pointwise $O(1/n^3)$. Then (\ref{Gowdycomm}) can be rewritten as
\ba
\label{PPtwop4}
&\nspace & \Delta(t, t', \zeta - \zeta') = \ln(t/t') + 2 \sum_{n \geq 1}
\cos n(\zeta\! - \!\zeta') \Delta_n^{\rm sub2}(t,t')
\nonum
&\nspace & \quad + 2 \sum_{n \geq 1} \cos n(\zeta\! -\!\zeta')
\Big[ \frac{1}{\sqrt{tt'}}
\frac{\sin n (t\!-\!t')}{n}
+ \frac{t\!-\!t'}{8(tt')^{3/2}} 
\frac{\cos n (t\!-\!t')}{n^2} \Big]\,.
\ea 
The infinite sums in the second line can be performed exactly:
\ba
\label{PPtwop5}  
2\sum_{n =1}^{\infty} \cos n x \,\frac{\sin n T}{n} \is 
- T + \frac{\pi}{2} \sum_{n \in \Z}
\big\{ {\rm sign}(T \!+ \!x \!+\!2 \pi n) + 
 {\rm sign}(T \!- \!x \!- \!2 \pi n) \big\}\,,
\nonumber
\ea
\ba
&\nspace & {\rm C}_2(T, x) := 
\sum_{n =1}^{\infty} \cos n x \,\frac{\cos n T}{n^2}
\nonum
&\nspace & \quad = 
\frac{1}{4} \Big\{
{\rm Li}_2\big(e^{ i(T+x)} \big) + 
{\rm Li}_2\big(e^{ -i(T+x)} \big) +
{\rm Li}_2\big(e^{ i(T-x)} \big) +
{\rm Li}_2\big(e^{ -i(T-x)} \big) \Big\}\,.
\ea 
Here, ${\rm Li}_{\nu}(z) = \sum_{n \geq 1} z^n n^{-\nu}$, $|z| <1$, is the
polylogarithm function, analytically continued to $|z|=1$ via an integral
representation, and ${\rm C}_2$ is a continuous
periodic function of bounded variation of $O(1)$. Using
(\ref{PPtwop5}) in (\ref{PPtwop4}) one
obtains
\ba
\label{PPtwop7}
&\nspace & \Delta(t, t', \zeta - \zeta') = \ln(t/t') + 2 \sum_{n \geq 1}
\cos n(\zeta\! - \!\zeta') \Delta_n^{\rm sub2}(t,t')
\nonum
&\nspace & \quad 
- \frac{t\!-\!t'}{\sqrt{tt'}} + \frac{\pi}{2\sqrt{tt'}} \sum_{n \in \Z}
\big\{ {\rm sign}(t\!-\!t' \!+ \!(\zeta \!- \!\zeta') \!+\!2 \pi n) + 
 {\rm sign}(t\!-\!t' \!- \!(\zeta \!-\! \zeta') \!- \!2 \pi n) \big\}
\nonum
&\nspace & \quad  + \frac{t\!-\!t'}{4\pi(tt')^{3/2}}
{\rm C}_2(t\!-\!t', \zeta\! - \!\zeta')\,,
\ea 
The key aspect is that the remaining oscillatory sum is rapidly
converging due to the pointwise $O(1/n^3)$ decay of  $\Delta_n^{\rm sub2}$. 
In fact, truncating the $\sum_{n \geq 1}$ sums to $\sum_{n=1}^{15}$, say, 
only introduces an error of order $O(10^{-5})$. The so-truncated
expression (\ref{PPtwop7}) is then readily programmed and a plot is
shown in Fig.~\ref{Fig1}. Usually there are two discontinuities per periodicity
interval, located at $|\zeta - \zeta'| = t - t' \!\mod 2\pi$.
For $t- t' = \pi \! \mod 2\pi$, these merge to one. The vanishing
outside the lightcone in the zero-centered periodicity interval
$\zeta - \zeta' \in [-\pi,\pi]$ is clearly visible in the figure.

Returning to (\ref{initial3}) and $\tau = \ln t$ as time coordinate
we obtain its position space version as
\ba
\label{opevol}
&& \begin{pmatrix} \phi(\tau,\zeta) \\[1.5mm] \pi(\tau,\zeta) \end{pmatrix}
= \frac{1}{2\pi} \int_0^{2\pi}\! d\zeta_0 \,D(\tau,\zeta;\tau_0, \zeta_0)
\begin{pmatrix} \phi(\tau_0,\zeta_0) \\[1.5mm] \pi(\tau_0,\zeta_0) \end{pmatrix}
\nonum 
&& D(\tau,\zeta;\tau_0, \zeta_0) =
\begin{pmatrix} - \dd_{\tau_0} \Delta(\tau,\tau_0,\zeta-\zeta_0)
  & \Delta(\tau,\tau_0,\zeta-\zeta_0) \\[1.5mm]
  -\dd_{\tau} \dd_{\tau_0} \Delta(\tau,\tau_0, \zeta-\zeta_0)
  & \dd_{\tau}\Delta(\tau,\tau_0,\zeta-\zeta_0) \end{pmatrix}\,,
\ea 
where $\pi(\tau,\zeta):= \dd_\tau \phi(\tau,\zeta)$.

The main simplification in the velocity dominated system is that
the Hamiltonian is time independent and is determined by the
also time independent momentum operator.  
For the Fourier modes a precise counterpart of (\ref{initial3})
holds
\ba
\label{vdopevol} 
&& \begin{pmatrix} \vp_n(\tau) \\ \wp_n(\tau) \end{pmatrix}
= \mathfrak{D}_n(\tau,\tau_0)
\begin{pmatrix} \vp_n(\tau_0) \\ \wp_n(\tau_0) \end{pmatrix},
\nonum
&& \mathfrak{D}_n(\tau,\tau_0) = 
\begin{pmatrix} -\dd_{\tau_0} \mathfrak{d}_n(\tau,\tau_0) &
  \mathfrak{d}_n(\tau,\tau_0) 
\\
-\dd_{\tau} \dd_{\tau_0} \mathfrak{d}_n(\tau,\tau_0) &
\dd_{\tau} \mathfrak{d}_n(\tau,\tau_0) 
\end{pmatrix}
= \begin{pmatrix} 1 & \tau - \tau_0 \\ 0 & 1 \end{pmatrix},
\ea
where $\wp_n(\tau) := \dd_\tau \vp_n(\tau)$, and $\mathfrak{d}_n$ are the ($n$-independent) Fourier modes of the velocity dominated commutator
function.%
\footnote{Note that $\mathfrak{D}_n(\tau, \tau_0) = D_0(\tau,\tau_0)$,
$n \in \Z$, with $D_0$ the zero mode of the time evolution matrix
(\ref{initial3}) of the full system.} 
In parallel to (\ref{opevol}), the position space version of (\ref{vdopevol}) is 
\ba
\label{vdopevol2}
&& \begin{pmatrix} \vp(\tau,\zeta) \\[1.5mm] \wp(\tau,\zeta) \end{pmatrix}
= \frac{1}{2\pi} \int_0^{2\pi}\! d\zeta_0 \,
\mathfrak{D}(\tau,\zeta;\tau_0, \zeta_0)
\begin{pmatrix} \vp(\tau_0,\zeta_0) \\[1.5mm] \wp(\tau_0,\zeta_0) \end{pmatrix}
\nonum 
&& \mathfrak{D}(\tau,\zeta;\tau_0, \zeta_0) =
\begin{pmatrix} 1 & \tau - \tau_0 \\ 0 & 1 \end{pmatrix}
\delta_{2\pi}(\zeta - \zeta_0)\,,
\ea 
where $\wp(\tau,\zeta) := \dd_\tau \vp(\tau,\zeta)$.

\subsection{Initial value parameterization and time consistent states} 
\label{section3.2}

We return to the symmetric matrix two-point function $\cW^s$ in
(\ref{Wsmatdef}).  
Using (\ref{opevol}) in (\ref{Wsmatdef}) one finds
\ba
\label{twopmat2} 
&\nspace & \cW^s(\tau,\tau',\zeta - \zeta') =
\int_0^{2\pi} \! \frac{d\zeta_0}{2\pi} 
\int_0^{2\pi} \! \frac{d\zeta_0'}{2\pi}  
\\[2mm] 
& \nspace & \quad 
D(\tau,\zeta;\tau_0,\zeta_0) \,
\cW^s(\tau_0,\tau_0,\zeta_0 \!- \!\zeta'_0) \,
D(\tau',\zeta';\tau_0,\zeta'_0)^T \,,
\nonumber
\ea
where $D^T$ is the pointwise matrix transpose of the matrix $D$. The matrix
two-point function is therefore parameterized by its initial value kernel
$\cW^s(\tau_0,\tau_0,\zeta_0 \!- \!\zeta'_0)$ depending only on the single
reference time $\tau_0$. Below, we shall study the dependence on this
reference time. For now, we only note its Fourier representation in terms
of a positive definite matrix $Z_n(\tau_0) $,  
\ba
\label{twopmat3} 
&\nspace & \cW^s(\tau_0,\tau_0,\zeta\!-\!\zeta') =
\sum_{n \in \mathbb{Z}}  e^{ i n (\zeta - \zeta')} Z_n(\tau_0) \,,
\nonumber
\\
&\nspace & Z_n (\tau_0) \!\!:=\!\! \begin{pmatrix} 2|z_n|^2 &
  z_n w_n^* + z_n^* w_n
  \\[1.5mm] z_n w_n^* + z_n^* w_n & 2|w_n|^2 \end{pmatrix}(\tau_0) > 0 \,.
\ea 
where $z_n(\tau_0)$, $w_n(\tau_0)$ are the initial data from
(\ref{Initialdata}). The positive definiteness of $Z_n(\tau_0)$ is
linked to the fact that the underlying Fock vacuum (\ref{Tvac}) is a
state in the algebraic quantum field theory sense. It could be rendered
manifest by factorizing $Z_n(\tau_0)$ into a matrix and its adjoint.
For the basic two-point function $W^s$ an initial value parameterization
of this form is due to L\"{u}ders and Roberts \cite{LR}.
The matrix version (\ref{twopmat2}) is convenient in that the initial
values and the time evolved quantity have the same structure.

Mostly to set the notation we run through the same steps in the velocity
dominated system. Defining in parallel to (\ref{Wsmatdef})
\ba
\label{twopmat6} 
&\nspace & \mathfrak{W}^s(\tau,\tau',\zeta - \zeta') :=
\begin{pmatrix} 1 & \dd_{\tau'} \\ \dd_{\tau} & \dd_{\tau} \dd_{\tau'}
\end{pmatrix} \mathfrak{w}^s(\tau,\tau', \zeta - \zeta')  \,, 
\ea 
the counterpart of (\ref{twopmat2}) is spatially local
\be
\label{twopmat7} 
\nspace \mathfrak{W}^s(\tau,\tau',\zeta - \zeta') =
\begin{pmatrix} 1 & \tau \!-\!\tau_0 \\ 0 & 1 \end{pmatrix}  
\mathfrak{W}^s(\tau_0,\tau_0,\zeta \!- \!\zeta') \,
\begin{pmatrix} 1 & 0 \\\tau'\! -\!\tau_0 & 1 \end{pmatrix}  \,.
\ee
Again, the initial value kernel
$\mathfrak{W}^s(\tau_0,\tau_0,\zeta \!- \!\zeta')$ is parameterized
by a positive definite matrix in a Fourier space $\tilde{Z}_n(\tau_0)$.
\ba
\label{twopmat8} 
&\nspace & \mathfrak{W}^s(\tau_0,\tau_0,\zeta\!-\!\zeta') =
\sum_{n \in \mathbb{Z}}  e^{ i n (\zeta - \zeta')} \tilde{Z}_n(\tau_0) \,,
\nonumber
\\
&\nspace & \tilde{Z}_n (\tau_0) \!\!:=\!\!
\begin{pmatrix} 2|\tilde{z}_n|^2
  & \tilde{z}_n w_n^* + \tilde{z}_n^* \tilde{w}_n
  \\[1.5mm] \tilde{z}_n \tilde{w}_n^* + \tilde{z}_n^* \tilde{w}_n
  & 2|\tilde{w}_n|^2 \end{pmatrix}(\tau_0) > 0 \,,
\ea
where $\tilde{z}_n(\tau_0)$, $\tilde{w}_n(\tau_0)$ are the initial
data from (\ref{gpol6}).
\medskip

{\bf Definition.} A state in a linear quantum field theoretical system
on a spatially homogeneous background is called {\bf time consistent}
if in the realization of the (matrix-) two-point function in terms
of the commutator function and an initial data kernel at time $\tau_0$
the two-point function itself is independent of the choice of
$\tau_0$.

The definition draws on the fact that a homogeneous pure state in  a
linear quantum field theoretical system can be characterized by
the spatial Fourier transform of an inital data matrix \cite{LR}. 
The requirement that the two-point function associated with such a state
is independent of the time $\tau_0$ where the initial data are specified
is a very natural requirement, as the standard definition of a two-point
function in terms of the expectation value of a product of smeared
field operators does not refer to $\tau_0$. To the best of our knowledge
it has never been formulated as an explicit requirement, since in earlier
uses of the initial value formulation $\tau_0$ was always kept fixed
\cite{LR,Olbermann,BonusSLE}. In the present context often the initial
data of intermediary or derived quantities ought to be backpropagated
to the Big Bang as well, so $\tau_0$ cannot be kept fixed.
In the following we characterize the initial data that give rise
to a reconstructed two-point function independent of $\tau_0$. For
definiteness we do so directly for the Gowdy scalar at hand,
i.e.~(\ref{twopmat3}) versus (\ref{twopmat2}), (\ref{Wsmatdef}),
but the ensued Result \ref{4maps} would generalize to other linear
quantum field theories on spatially homogeneous backgrounds.
In Section \ref{sec4.2} we shall see that these initial data are
typically compatible with the Hadamard property.

We begin by spelling out the consistency condition imposed
on $Z_n(\tau_0)$ by time consistency. In view of 
(\ref{twopmat2}) this means that two initial value kernels
$\cW^s(\tau_0,\tau_0,\,\cdot\,)$ and
$\cW^s(\tau_1,\tau_1,\,\cdot\,)$ must be related by the
$\tau = \tau'=\tau_1$ specialization of (\ref{twopmat2}). In the
Fourier representation (\ref{twopmat3}) this amounts to 
\be
\label{initindep2}
\begin{pmatrix} z_n \\ w_n \end{pmatrix}\!(\tau_0)
= D_n(\tau_0, \tau_1)
\begin{pmatrix} z_n \\ w_n \end{pmatrix}\!(\tau_1)\,,
\quad 
Z_n(\tau_0) = 
D_n(\tau_0, \tau_1) Z_n(\tau_1) D_n(\tau_0, \tau_1)^T.
\ee
The same applies to the velocity dominated system where two initial
value kernels $\mathfrak{W}^s(\tau_0,\tau_0,\,\cdot\,)$ and
$\mathfrak{W}^s(\tau_1,\tau_1,\,\cdot\,)$ must be related by the
$\tau = \tau' = \tau_1$ specialization of (\ref{twopmat7}). In the
Fourier representation (\ref{twopmat8})
\be
\label{initindep4}
\begin{pmatrix} \tilde{z}_n \\ \tilde{w}_n \end{pmatrix}\!(\tau_0)
=
\begin{pmatrix} 1 & \tau_0\!-\! \tau_1 \\ 0 & 1 \end{pmatrix} 
\!\begin{pmatrix} \tilde{z}_n \\ \tilde{w}_n \end{pmatrix}\!(\tau_1)\,,
\quad 
\tilde{Z}_n(\tau_0) = 
\begin{pmatrix} 1 \!&\! \tau_0\!-\! \tau_1 \\ 0 \!&\! 1 \end{pmatrix} 
\tilde{Z}_n(\tau_1)
\begin{pmatrix} 1 \!&\! 0 \\ \!\tau_0\!-\! \tau_1 \!&\! 1 \end{pmatrix}.
\ee
We shall refer to initial data $\big(z_n(\tau_0), w_n(\tau_0)\big)$ satisfying
(\ref{initindep2}) as time consistent initial data; similarly, for
the velocity dominated system and (\ref{initindep4}). We now claim that
such time consistent initial data can be constructed by transitioning
to the corresponding Bogoliubov parameters. 

\begin{figure}[!ht]
  \centering
\resizebox{0.85\textwidth}{!}{%
\begin{circuitikz}
\tikzstyle{every node}=[font=\LARGE]


\draw[->, >=Stealth] (10.25,11.15) -- (4.25,11.15)
  node[midway, below=2pt, font=\normalsize] {$I^{\text{grad}}(|n|e^{\tau_0})$};

\node [font=\normalsize] at (11.75,11.75) {time-consistent};
\node [font=\normalsize] at (11.75,11.25) {$\tilde{z}_n(\tau_0)$ , $\tilde{w}_n(\tau_0)$};

\node [font=\normalsize, color={rgb,400:red,30; green,250; blue,100}] at (2.75,12.5) {Full Gowdy};
\node [font=\normalsize, color={rgb,400:red,30; green,250; blue,100}] at (11.75,12.5) {VD Gowdy};

\node [font=\normalsize] at (2.75,11.75) {time-consistent};
\node [font=\normalsize] at (2.75,11.25) {$z_n(\tau_0)$ , $w_n(\tau_0)$};

\draw[->, >=Stealth] (2.75,6.5) -- (2.75,10.5)
  node[midway, right=3pt, font=\normalsize] {$H(|n|e^{\tau_0})$};
\node [font=\normalsize] at (2.75,6) {($\lambda_n,\mu_n$)};

\draw[->, >=Stealth] (10.25,5.90) -- (4.25,5.90)
  node[midway, below=2pt, font=\normalsize] {$\ln \gamma_n(0)\ \text{map}$};

\draw[->, >=Stealth] (11.75,6.5) -- (11.75,10.5)
  node[midway, right=3pt, font=\normalsize] {$\tilde{H}(\tau_0)$};
\node [font=\normalsize] at (11.75,6) {($\tilde{\lambda}_n, \tilde{\mu}_n$)};

\node [font=\normalsize, color={rgb,300:red,20; green,40; blue,255}] at (0,11.75) {$\tau_0$-};
\node [font=\normalsize, color={rgb,300:red,20; green,40; blue,255}] at (0,11.25) {dependent};
\node [font=\normalsize, color={rgb,300:red,20; green,40; blue,255}] at (0,6) {$\tau_0$-};
\node [font=\normalsize, color={rgb,300:red,20; green,40; blue,255}] at (0,5.5) {independent};

\end{circuitikz}
}%
\caption{\label{Fig2} 
Commutative diagram. All four maps are invertible.}
\end{figure}
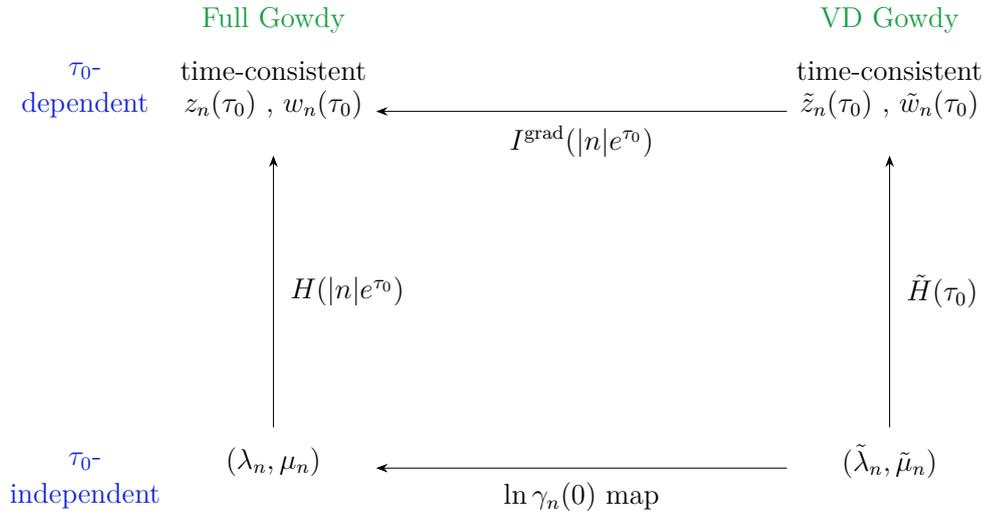

\begin{result} \label{4maps}
The transformations (\ref{FHad11}), (\ref{H0def})
trivialize the consistency conditions (\ref{initindep2}),
(\ref{initindep4}) for the initial data under a change of reference
time: the latter hold iff the Bogoliubov parameters
$\lb_n(\tau_0), \mu_n(\tau_0)$ and 
$\tilde{\lb}_n(\tau_0), \tilde{\mu}_n(\tau_0)$, $n \in \Z$,
are independent of $\tau_0$. Further, both sets of Bogoliubov parameters are  
consistently related by the time independent relation
(\ref{match8}) below. Both sets of time consistent initial data are in
turn related by $I^{\rm grad}(|n| e^{\tau_0})$ from (\ref{grad6}).
Overall, the four maps give rise to the commutative diagram in Figure
\ref{Fig2}.
\end{result}

The derivation of this result is based on the fact that the solutions of
the full Gowdy system are related to those of the velocity-dominated system
through the following {\bf gradient map}
\ba
\label{grad6} 
\begin{pmatrix} T_n(\tau) \\[1.5mm] \dd_{\tau} T_n(\tau) \end{pmatrix} \is
I^{\rm grad}(|n| e^{\tau})
\begin{pmatrix} \mathfrak{t}_n(\tau)
\\[1.5mm] \dd_{\tau} \mathfrak{t}_n(\tau) \end{pmatrix} \,, \quad n \in \Z\,,
\nonum
I^{\rm grad}(x) &:=& \begin{pmatrix} J_0(x) & U_0(x) \\[1.5mm]
  - x J_1(x) & - x U_1(x) \end{pmatrix} =: \1 + \sum_{k \geq 1} I_k \, x^{2k}\,, 
\ea
with $U_k(x) := (\pi/2) Y_k(x) -\ln(|x|e^{\gamma_E}/2) J_k(x)$, $k = 0,1$.
The matrix is sympletic, $\det I^{\rm grad}(x) =1$, which preserves the
Wronskian normalizations. 
We refer to it as the gradient map, because all matrix elements have
{\it absolutely convergent} Taylor expansions in powers of $x^2$, as
indicated. For $x^2 = n^2 e^{2\tau}$ the $n^2$ powers in turn correspond to
spatial gradients in position space. 

To see how (\ref{grad6}) comes about, one starts with 
\ba
\label{grad3} 
T_n(\tau) \is J_0(|n| e^{\tau}) 
\mathfrak{t}_n(\tau)  + U_0(|n| e^{\tau})
\dd_\tau \mathfrak{t}_n(\tau)
\nonum
\is J_0(|n| e^{\tau}) \big[\mathfrak{t}_n(\tau) 
  - \dd_\tau \mathfrak{t}_n(\tau) \ln(e^{\tau_0} e^{\gamma_E}|n|/2) \big] + \frac{\pi}{2}
 Y_0(|n| e^{\tau}) \dd_\tau \mathfrak{t}_n(\tau)\,.
\ea
The coefficients of $J_0(|n| e^{\tau})$ and $Y_0(|n| e^{\tau})$ are constants, and by rewriting the expression in terms of $H_0^{(1)}(|n| e^{\tau})$ and $H_0^{(2)}(|n| e^{\tau})$, one can match the result to (\ref{Tsol2}). Repeating these steps for $\dd_\tau T_n(\tau)$, one obtains another set of matching conditions. Both combine to the following time independent identification of Bogoliubov parameters.
\ba
\label{match8}
\begin{pmatrix} \lb_n(\tau_0) \\ \mu_n(\tau_0) \end{pmatrix}
\is 
\begin{pmatrix}
  1 + \frac{i}{\pi} \ln \gamma_n(0)
  &  - \frac{i}{\pi} \ln \gamma_n(0)  \\[1.5mm]
  \frac{i}{\pi} \ln \gamma_n(0)
  &  1 - \frac{i}{\pi} \ln \gamma_n(0) 
\end{pmatrix}
\begin{pmatrix} \tilde{\lb}_n(\tau_0) \\[1.5mm] \tilde{\mu}_n(\tau_0)
\end{pmatrix}, \quad n \neq 0\,,
\ea
where $\gamma_n(0) = e^{\gamma_{E}} |n|/2 $. Conversely, subject to the identification (\ref{match8}) the relation (\ref{grad6}) holds.  

Specializing (\ref{grad6}) to $\tau = \tau_0$ one has 
\ba
\label{init1} 
&& \begin{pmatrix} z_n(\tau_0) \\[1.5mm] w_n(\tau_0) \end{pmatrix} =
I^{\rm grad}(|n| e^{\tau_0})
\begin{pmatrix} \tilde{z}_n(\tau_0) 
\\[1.5mm] \tilde{w}_n(\tau_0) \end{pmatrix} \,, \quad n \in \Z\,.
\ea 
Since $I^{\rm grad}(0)$ is the unit matrix, the $n=0$ instance of (\ref{init1}) simply identifies $z_0(\tau_0) = \tilde{z}_0(\tau_0), \, w_0(\tau_0) = \tilde{w}_0(\tau_0)$.
The interrelation (\ref{init1}) must be compatible with the
respective consistency conditions (\ref{initindep2}), (\ref{initindep4})
on both sides. This amounts to the relation
\be
\label{init2} 
D_n(\tau_1,\tau_0) I^{\rm grad}(|n| e^{\tau_0})
\begin{pmatrix} 1& \tau_0\! -\!\tau_1 \\ 0 & 1 \end{pmatrix} 
=  I^{\rm grad}(|n| e^{\tau_1})\,,\quad n \in \Z\,,
\ee
which is indeed an identity. 
Inserting (\ref{FHad11}), (\ref{H0def}) into (\ref{init1})
one finds
\ba
\label{con1}
\begin{pmatrix} \lb_n(\tau_0) \\ \mu_n(\tau_0) \end{pmatrix}
\is
H(|n| e^{\tau_0})^{-1} I^{\rm grad}(|n|e^{\tau_0}) \tilde{H}(\tau_0)
\begin{pmatrix} \tilde{\lb}_n(\tau_0) \\[1.5mm] \tilde{\mu}_n(\tau_0)
\end{pmatrix} \,.
\ea
Consistency requires that the matrices in (\ref{con1}) and (\ref{match8}) are the same. This can be checked to be the case and yields the commutativity of the above diagram.

So far, we held the reference time $\tau_0$ fixed. Next we consider the transformation of the Bogoliubov parameters
$\lb_n(\tau_0), \mu_n(\tau_0)$ and 
$\tilde{\lb}_n(\tau_0), \tilde{\mu}_n(\tau_0)$, $n \in \Z$, under
a change of initial time. Note that the transformation law is uniquely determined
by (\ref{initindep2}), (\ref{FHad11}) and (\ref{initindep4}),
(\ref{H0def}), respectively. In the first case one finds for
$n \neq 0$
\be
\label{match10}
\begin{pmatrix} \lb_n(\tau_0) \\ \mu_n(\tau_0) \end{pmatrix}
= H(|n| e^{\tau_0})^{-1} D_n(\tau_0,\tau_1) H(|n| e^{\tau_1})
\begin{pmatrix} \lb_n(\tau_1 \\ \mu_n(\tau_1) \end{pmatrix}
=
\begin{pmatrix} \lb_n(\tau_1) \\ \mu_n(\tau_1) \end{pmatrix} \,,
\ee 
as the matrix reduces to the unit matrix. In the velocity dominated
system one has
\be
\label{match11}
\begin{pmatrix} \tilde{\lb}_n(\tau_0) \\ \tilde{\mu}_n(\tau_0) \end{pmatrix}
= \tilde{H}(\tau_0)^{-1}
\begin{pmatrix} 1 & \tau_0\! - \!\tau_1 \\ 0 & 1 \end{pmatrix} 
\tilde{H}(\tau_1)
\begin{pmatrix} \tilde{\lb}_n(\tau_1) \\ \tilde{\mu}_n(\tau_1) \end{pmatrix}
=
\begin{pmatrix} \tilde{\lb}_n(\tau_1) \\ \tilde{\mu}_n(\tau_1) \end{pmatrix} \,,
\ee 
where the matrix again reduces to the unit matrix. This shows the main assertion in the above result: the Bogoliubov parameters do not depend on the choice of reference time. The relation
(\ref{match11}) applies to all $n \in \Z$, while for (\ref{match10}) the zero modes need to be augmented. We do so by identifying 
\be
\label{match12}
\lb_0(\tau_0) = \tilde{\lb}_0(\tau_0)\,, \quad
\mu_0(\tau_0) = \tilde{\mu}_0(\tau_0)\,,
\ee
consistent with our initial identification $T_0(\tau) = \mathfrak{t}_0(\tau)$.

Conversely, re-expressing the first equality in (\ref{match10}) in terms of
initial data using (\ref{FHad11}), one obtains (\ref{initindep2}). That is,
time independent Bogoliubov parameters give via (\ref{FHad11}) automatically
rise to time consistent initial data. Similarly, re-expressing the first
equality in (\ref{match11}) in terms of initial data using (\ref{H0def}),
gives rise to (\ref{initindep4}). This completes the derivation of the above
result.

\subsection{Spatial gradient expansion of two-point functions}

As seen before, in Fourier space the basic mode functions
$(T_n, \dd_{\tau} T_n)$ and their velocity dominated counterparts
$(\mathfrak{t}_n, \dd_{\tau} \mathfrak{t}_n)$ are related by
the gradient map (\ref{grad6}). Since $I^{\rm grad}(x)$ is real
this extends to their complex conjugates and one has
\be
\label{grad14}
\begin{pmatrix}
  T_n(\tau) & T_n(\tau)^* \\[2mm]
  \dot{T}_n(\tau) & \dot{T}_n(\tau)^* 
\end{pmatrix} 
= I^{\rm grad}(|n|e^{\tau} )
\begin{pmatrix}
  \mathfrak{t}_n(\tau) & \mathfrak{t}_n(\tau)^* \\[2mm]
  \dot{\mathfrak{t}}_n(\tau) & \dot{\mathfrak{t}}_n(\tau)^* 
\end{pmatrix} \,,
\ee
where we write $\dot{f}(\tau) = (\dd_{\tau} f)(\tau)$. The
Fourier modes of the basic two-point functions can be written as
\ba 
\label{PPtwop}
W_n^s(\tau,\tau') &:= &
T_n(\tau) T_n(\tau')^* + T_n(\tau') T_n(\tau)^* \,,
\nonum
\Delta_n(\tau,\tau') &:= & i\big(T_n(\tau) T_n(\tau')^* -
T_n(\tau') T_n(\tau)^* \big) \,.
\ea
We seek to express their matrix versions in terms of
the left hand side in (\ref{grad14}). One has 
\ba
\label{grad15}
\cW_n^s(\tau,\tau') =
\begin{pmatrix} 1 & \dd_{\tau'} \\[2mm] \dd_{\tau} & \dd_{\tau} \dd_{\tau'}
\end{pmatrix} W_n^s(\tau,\tau') =
\begin{pmatrix}
  T_n(\tau) & T_n(\tau)^* \\[2mm]
  \dot{T}_n(\tau) & \dot{T}_n(\tau)^* 
\end{pmatrix} 
\begin{pmatrix}
  T_n(\tau') & T_n(\tau')^* \\[2mm]
  \dot{T}_n(\tau') & \dot{T}_n(\tau')^* 
\end{pmatrix}^{\dagger}\!.  
\ea 
The matrix commutator function enters naturally in the form
(\ref{initial3}) via the solution of the initial value problem.
This gives
\ba
\label{grad16}
&\nspace & D_n(\tau,\tau') \begin{pmatrix} 0 & -1 \\[2mm] 1 & 0 \end{pmatrix} = 
\begin{pmatrix} 1 & \dd_{\tau'} \\[2mm] \dd_{\tau} & \dd_{\tau} \dd_{\tau'}
\end{pmatrix} \Delta_n(\tau,\tau')
\nonum
& \nspace & \quad = i 
\begin{pmatrix}
  T_n(\tau) & T_n(\tau)^* \\[2mm]
  \dot{T}_n(\tau) & \dot{T}_n(\tau)^* 
\end{pmatrix} 
\begin{pmatrix} 1 & 0 \\[2mm] 0 & -1 \end{pmatrix}
\begin{pmatrix}
  T_n(\tau') & T_n(\tau')^* \\[2mm]
  \dot{T}_n(\tau') & \dot{T}_n(\tau')^* 
\end{pmatrix}^{\dagger}\,.
\ea 
The same relations apply to the velocity dominated matrix two-point
functions $\mathfrak{W}_n^s(\tau,\tau')$ and
$\mathfrak{D}_n(\tau,\tau')$, respectively, with the $T_n$'s replaced
by $\mathfrak{t}_n$'s. Inserting (\ref{grad14}) into (\ref{grad15})
one obtains 
\be
\label{grad17}
\cW_n^s(\tau,\tau') = I^{\rm grad}(|n| e^{\tau}) 
\mathfrak{W}_n^s(\tau,\tau') I^{\rm grad}(|n| e^{\tau'})^T\,, \quad n \neq 0. 
\ee
For the zero mode, we identified $T_0(\tau)$ with $\mathfrak{t}_0(\tau)$, so that $\cW_0^s(\tau,\tau') = \mathfrak{W}_0^s(\tau,\tau')$ holds by definition. For the commutator function one needs in addition
\begin{small}
\begin{equation}
\label{Igradinv}
\begin{pmatrix} 0 & 1 \\[2mm] -1 & 0 \end{pmatrix}^{\!\!-1}
I^{\rm grad}(x)^T 
\begin{pmatrix} 0 & 1 \\[2mm] -1 & 0 \end{pmatrix}
= I^{\rm grad}(x)^{-1}\,.
\end{equation}
\end{small}
This results in
\be
\label{grad18}
D_n(\tau,\tau') = I^{\rm grad}(|n| e^{\tau}) 
\mathfrak{D}_n(\tau,\tau') I^{\rm grad}(|n| e^{\tau'})^{-1}\,, \quad n\neq 0. 
\ee
For $n=0$ one has again trivially $D_0(\tau,\tau') = \mathfrak{D}_0(\tau,\tau') $.
Note that the inverse on the far right of (\ref{grad18}) ensures compatibility with
the composition law mentioned after (\ref{initial3}). An alternative,
slightly quicker, route to (\ref{grad18}) is by rearranging
(\ref{init2}) using $\mathfrak{D}_n(\tau',\tau) =
\mathfrak{D}_n(\tau,\tau')^{-1}$. 

Since $I^{\rm grad}(x)$ admits a convergent series expansion in
powers of $x^2$ the relations (\ref{grad17}), (\ref{grad18}) will
give rise to convergent expansions of the matrix two-point functions
in Fourier space around their velocity dominated counterparts.
Preparing
\ba
\label{Igradexp} 
I^{\rm grad}(x) \is \1 + x^2 I_1 + O(x^4) \,,
\quad
 I_1 = \begin{pmatrix} -1/4 & 1/4 \\ -1/2 & 1/4 \end{pmatrix}\,,
\ea
one finds to first order 
\ba
\label{grad19}
&\nspace& \cW^s_n(\tau,\tau') = \mathfrak{W}_n^s(\tau,\tau') +
n^2 [e^{2 \tau} I_1 \mathfrak{W}_n^s(\tau,\tau') +
  e^{2 \tau'} \mathfrak{W}_n^s(\tau,\tau') I_1^T] + O(n^4)\,,
\nonum
&\nspace& D_n(\tau,\tau') = \mathfrak{D}_n(\tau,\tau') +
n^2 [e^{2 \tau} I_1 \mathfrak{D}_n(\tau,\tau') -
  e^{2 \tau'} \mathfrak{D}_n(\tau,\tau') I_1] + O(n^4)\,.
\ea 
This illustrates a systematic way to reconstruct the
full two-point functions from the velocity dominated ones. The powers
of $n^2$ encountered amount to a spatial gradient expansion, so the
reconstruction does not rely on the time reversal in a 
unitary time evolution operator (supposed to undo the
$\tau, \tau' \ra -\infty$ asymptotics, as in the classical AVD results.)

In position space, the velocity dominated two-point functions will
in general contain distributional terms in $\zeta - \zeta'$ and will 
not be conventionally differentiable to all orders. However once
spatially averaged with real, smooth test functions
$f,g \in C^{\infty}(S^1)$, the differentiations can be defined
distributionally. That is, for any element 
$\mathfrak{F}$ in the dual%
\footnote{That is, $\mathfrak{F}(\zeta) = \sum_m \mathfrak{F}_m
e^{ i m \zeta}$, where there exists $c'>0,M\in\N$ such that $|\mathfrak{F}_m|
\leq c'(1+ |m|)^M$, for all $m \in \Z$. This is is dual to the
characterization of $f \in C^{\infty}(S^1)$,
i.e.~$f(\zeta)= \sum_n f_n e^{i n \zeta}$, with 
$|f_n| \leq c(1+ |n|)^{-N}$, for all $N\in \N$.}
of $C^{\infty}(S^1)$ and $l \in \N$ we set
\ba
\label{Distrdiff} 
\int\! \frac{d \zeta}{2\pi} f(\zeta)
\int\! \frac{d \zeta'}{2\pi} g(\zeta') (\dd_{\zeta} \dd_{\zeta'} )^l
\mathfrak{F}(\zeta\! - \!\zeta') &:=& 
\int\! \frac{d \zeta}{2\pi} \dd_{\zeta}^l f(\zeta)
\int\! \frac{d \zeta'}{2\pi} \dd_{\zeta'}^l g(\zeta') 
\mathfrak{F}(\zeta \!- \!\zeta')
\nonum
\quad &=& \sum_{n \in \Z} n^{2 l} f_{-n} g_{n} \,\mathfrak{F}_n\,. 
\ea
Convergence of the sum is ensured by the rapid decay of
$f_n, g_{-n}$ in $n$. This can be used to show the 
\begin{result} \label{AVDexpansion}
The matrix two-point function $\cW^s(\tau,\tau',\zeta - \zeta')$
admits upon averaging with $f,g \in C^{\infty}(S^1)$ test functions
a series expansion of the form  
\begin{small} 
\ba
\label{grad20} 
\!\!\!\!&\nspace & 
\int\!\!\frac{d\zeta}{2\pi}\frac{d\zeta'}{2\pi}
f(\zeta) g(\zeta') \,\cW^s(\tau,\tau',\zeta\! - \!\zeta')=
\int\!\!\frac{d\zeta}{2\pi}\frac{d\zeta'}{2\pi}
f(\zeta) g(\zeta') \mathfrak{W}^s(\tau,\tau',\zeta\! - \!\zeta')
\nonum
\!\!\!\!&\nspace & +  
\sum_{l \geq 1} \sum_{k=0}^l e^{2 k \tau} e^{ 2(l-k) \tau'} 
\int\!\!\frac{d\zeta}{2\pi}\frac{d\zeta'}{2\pi}
f(\zeta) g(\zeta') (\dd_{\zeta} \dd_{\zeta'})^l
I_k \,\mathfrak{W}^s(\tau,\tau',\zeta\! - \!\zeta') I^T_{l-k}\,,
\ea
\end{small}
Here, $I_k$ are the numerical $2\times 2$ matrices in the
expansion of $I^{\rm grad}(x)$ in powers of $x^2$. All terms in
the expansion are spatial gradients of the two-point
function $\mathfrak{W}^s(\tau,\tau',\zeta - \zeta')$ of
the velocity dominated system, which is linear in $\tau,\tau'$.  
The series (\ref{grad20}) is uniformly convergent for all $\tau,\tau'
< - \delta$, $\delta >0$ and $|\tau - \tau'|$ bounded. 
\end{result}

Only the convergence part needs to be shown. We start from (\ref{grad17}) 
and insert the Taylor series $I^{\rm grad}(x) =
\sum_{k =0}^K I_k x^{2k} + C_K x^{2 K}$ with Peano remainder,
i.e.~$C_K$ goes to zero as $K \ra \infty$ in one (and hence any)
$2 \times 2$ matrix norm. A brief computation gives
\ba
&& \cW^s_n(\tau,\tau') - \mathfrak{W}_n^s(\tau,\tau') -
\sum_{l=1}^{2 K} n^{2l} \Big( \sum_{k=0}^l I_k
\mathfrak{W}^s_n(\tau,\tau') I_{l-k}^T e^{2 k \tau} e^{2 (l-k) \tau'} \Big)
\nonum
&& \quad = \sum_{k=0}^K n^{2 (k+K)}
\Big( e^{2 k \tau + 2 K \tau'} I_k \mathfrak{W}^s_n(\tau,\tau') C_K^T
+ e^{2 k \tau' + 2 K \tau} C_K \mathfrak{W}^s_n(\tau,\tau') I_k^T \Big)
\nonum
&& \quad + n^{4 K} e^{2 K(\tau + \tau')} C_K \mathfrak{W}^s_n(\tau,\tau') C_K^T\,.
\ea 
In view of (\ref{Distrdiff}) we multiply this by $f_n g_{-n} + f_{-n} g_n$
and sum over $n \in \N$. The resulting right hand side is denoted by
RHS and we seek to bound it in some $2\times 2$ matrix norm. For definiteness
we use the $\Vert \cdot \Vert_{\infty}$ norm from (\ref{mnorm}) below.  
We only sketch the rest of the argument, deferring some
of the details to Section \ref{section3.4}:
For large enough $K$ we may assume $\Vert C_K\Vert_{\infty} \leq 1$.
By inspection of (\ref{grad6}) one sees that $\sum_{k=0}^K
\Vert I_k \Vert_{\infty} \leq 2$, for all $K$. This readily gives
$\Vert {\rm RHS} \Vert_{\infty} \leq
2(2 e^{2 K \tau} + 2 e^{2 K \tau'} + e^{2 K(\tau + \tau'}) \sum_{n\geq 1}
|f_n| |g_n| n^{4 K} \Vert \mathfrak{W}^s_n(\tau,\tau')\Vert_{\infty}$.
A bound for $\Vert \mathfrak{W}^s_n(\tau,\tau')\Vert_{\infty}$ will be
provided in (\ref{Boundw}), (\ref{avdpr7a}). Since $|\tilde{\lb}_n|,
|\tilde{\mu}_n|$ grow like $\log n$ for the relevant cases, see
Section \ref{sec4.2}, we may safely bound the sum by $\sum_{n \geq 1}
|f_n| |g_n| n^{4 K + 2 \nu}$, for a small power $0< \nu <1$. 
From footnote 5 we know $|f_n|, |g_n| \leq c (1 + |n|)^{-N}$,
for all $N \geq 1$. Taking $N = 2 \nu + 5K$, the previous sum can
be bounded by $c \sum_{n \geq 1} n^{-K} = c\, \zeta(K)$. For large $K$
the $\zeta$ function approaches $1$, and for real $K \geq 2$ it is 
bounded by $2$. It remains to take into account the
$\tau,\tau'$ powers from (\ref{Boundw}). They lead to terms of the form
$e^{2 K \tau},e^{2 K \tau'}$, $\tau e^{2 K \tau},\tau' e^{2 K \tau'}$,
and $\tau' e^{2 K \tau},\tau e^{2 K \tau'}$, $K \geq 2$. All of them are readily
seen to be uniformly bounded for $|\tau\! - \!\tau'| < d$
and $\tau,\tau' < -\delta$, for some (small) $\delta >0$ and (large) $d>0$. 
This establishes Result \ref{AVDexpansion}.

\subsection{Asymptotic velocity domination for two-point functions}
\label{section3.4}

Here we derive an explicit bound on the second line of (\ref{grad20}) 
that pins down the leading rate of decay. 
This could be done for each of the $2\times 2$ matrix components
separately, 
but it is technically convenient to use a matrix norm. A matrix
norm on the space of $n \times n$ matrices obeys:
$\Vert A \Vert \geq 0$;
$\Vert A \Vert =0$ iff $A =0$; 
$\Vert \alpha A \Vert = |\alpha| \Vert A \Vert$, $\alpha \in \C$;  
$\Vert A + B \Vert \leq \Vert A \Vert + \Vert B \Vert$;
$\Vert A B \Vert \leq \Vert A \Vert \Vert B \Vert$.
For finite dimensional matrices any two of such norms are equivalent
and we use the sup-norm for convenience
\begin{equation}
\label{mnorm} 
\Vert A \Vert_{\infty} = \max_{1 \leq j \leq n} \sum_{i=1}^n |a_{ij}|\,,
\quad (\mbox{row sums}) 
\end{equation} 
where $a_{ij}$, $1\leq i,j \leq n$, are the matrix elements of $A$. 
In this norm we consider
\begin{small} 
\ba
\label{avdpr1} 
\!\!\!\!&\nspace & \bigg\Vert
\int\!\!\frac{d\zeta}{2\pi}\frac{d\zeta'}{2\pi}
f(\zeta) g(\zeta')
\Big\{ \cW^s(\tau,\tau',\zeta\! - \!\zeta') -  
\mathfrak{W}^s(\tau,\tau',\zeta \!- \!\zeta') \,  
\Big\}
\bigg\Vert_{\infty}
\nonum
\!\!\!\!&\nspace & \leq \sum_{n \neq 0}
\Big\Vert f_n g_n
{\scriptsize
\Big\{ \cW_n^s(\tau,\tau') - \mathfrak{W}_n^s(\tau,\tau')
\Big\} \Big\Vert_{\infty} 
}\,.
\ea
\end{small} 
The $n=0$ term drops out on account of identification of zero modes. Next,
we insert (\ref{grad17}) and split off the unit matrix part from
$I^{\rm{grad}}$, defining $\hat{I}(x) = I^{\rm{grad}}(x) - \1$. Termwise
this gives
\begin{small}
\ba
\label{Ibound0}
\!\!\!\!&\nspace &  
\Big\Vert f_n g_n
\Big\{ \cW_n^s(\tau,\tau') - \mathfrak{W}_n^s(\tau,\tau')
\Big\} \Big\Vert_{\infty} 
\nonum
&\nspace & = 
\Big\Vert f_n g_n
\Big\{ \hat{I}(|n| e^{\tau}) \mathfrak{W}_n^s(\tau,\tau') + \mathfrak{W}_n^s(\tau,\tau') \hat{I}(|n| e^{\tau'})^T + \hat{I}(|n| e^{\tau}) \mathfrak{W}_n^s(\tau,\tau') \hat{I}(|n| e^{\tau'})^T
\Big\} \Big\Vert_{\infty} 
\nonum
&\nspace & \leq |f_n| |g_n| \Big\{ \Vert\hat{I}(|n| e^{\tau})\Vert_{\infty} \Vert\mathfrak{W}_n^s(\tau,\tau')\Vert_{\infty} + \Vert\mathfrak{W}_n^s(\tau,\tau')\Vert_{\infty} \Vert\hat{I}(|n| e^{\tau'})^T\Vert_{\infty}
\nonum
&\nspace &
\quad + \Vert\hat{I}(|n| e^\tau) \Vert_{\infty} \Vert\mathfrak{W}_n^s(\tau,\tau')\Vert_{\infty} \Vert\hat{I}(|n| e^{\tau'})^T\Vert_{\infty}
\Big\} .
\ea
\end{small}
Clearly only bounds on $\hat{I}$ and $\mathfrak{W}^s_n$ are needed. 
To obtain a bound on $\hat{I}(x)$, we recall the definition of
$I^{\rm grad}(x)$ from
(\ref{grad6}) and the redefinitions $J_0(x) = 1 - (x^2/4)
\hat{\jmath}_0(x)$, $J_1(x) = (x/2) \hat{\jmath}_1(x)$,
$U_0(x) = (x^2/4) \hat{u}_0(x)$, $U'_0(x) = (x/2) \hat{u}_1(x)$.  
For the lower right entry of $I^{\rm grad}(x)$ we need $-x U_1(x)$,
where $U_1(x) := (\pi/2) Y_1(x) - \ln(|x|e^{\gamma_E}/2)J_1(x)$.
Differentiating $U_0(x) := (\pi/2) Y_0(x) - \ln(|x|e^{\gamma_E}/2)
J_0(x)$ one finds $U'_0(x) = - U_1(x) - J_0(x)/x$, so that
$- x U_1(x) = 1 + (x^2/4)(2 \hat{u}_1(x) - \hat{\jmath}_0(x))$. 
Inserted into (\ref{grad6}) one gets
\begin{small}
\be
\label{Ibound1} 
\hat{I}(x) = I^{\rm grad}(x) - \begin{pmatrix} 1 &0\\ 0 & 1 \end{pmatrix}
= \frac{x^2}{4}
\begin{pmatrix} - \hat{\jmath}_0(x) & \hat{u}_0(x) \\
- 2 \hat{\jmath}_1(x) & 2 \hat{u}_1(x) - \hat{\jmath}_0(x) \end{pmatrix}. 
\ee
\end{small}
Since the modulus of the `hatted' functions is globally bounded
by $1$, this gives straightforwardly
\be
\label{Ibound2} 
\Vert \hat{I}(x) \Vert_{\infty} \leq \frac{5}{4} x^2\,,\quad
\Vert \hat{I}(x)^T \Vert_{\infty} \leq x^2\,. 
\ee 
Next, for a bound on $\mathfrak{W}^s_n(\tau,\tau')$ we combine equations
(\ref{twopmat7}) and (\ref{twopmat8}) to obtain 
\ba
\label{Boundw}
\Vert \mathfrak{W}^s_n(\tau,\tau') \Vert_{\infty} \leq ( 1 + |\tau - \tau_0|)( 1 + |\tau'-\tau_0|)\Vert \tilde{Z}_n(\tau_0) \Vert_{\infty}\,.
\ea
For the $\tilde{Z}_n$ norm we use
\begin{eqnarray}
\label{avdpr7}
\Vert \tilde{Z}_n(\tau_0) \Vert_{\infty} \is \max\big\{
2|\tilde{z}_n|^2 \!+\!
| \tilde{w}_n \tilde{z}_n^* \!+\! \tilde{w}_n^* \tilde{z}_n|,  
2|\tilde{w}_n|^2 \!+\!
|\tilde{w}_n \tilde{z}_n^* \!+\! \tilde{w}_n^* \tilde{z}_n|  
\big\}(\tau_0) 
\nonum
&\leq &
2 \big( |\tilde{z}_n(\tau_0)| \!+\!
|\tilde{w}_n(\tau_0)| \big)^2\,.
\end{eqnarray}
This is still $\tau_0$ dependent, but for time consistent initial
data we know that $\tilde{z}_n(\tau_0), \tilde{w}_n(\tau_0)$ arise
from $\tau_0$-independent Bogoliubov parameters
$\tilde{\lb}_n, \tilde{\mu}_n$ via (\ref{H0def}). This gives
\be
\label{avdpr7a}
\Vert \tilde{Z}_n(\tau_0) \Vert_{\infty} \leq \mathfrak{z}_n(\tau_0)\,,
\quad
\mathfrak{z}_n(\tau_0) = \frac{2}{\pi}\big(| \tilde{\lb}_n| +
| \tilde{\mu}_n| \big)^2 \big(\tau_0 + 1 + \frac{\pi}{2} \big)^2\,.
\ee
With these bounds in place we return to (\ref{Ibound0}) to estimate
\ba
\label{Iboundd}
\!\!\!\!&\nspace &  
\Big\Vert f_n g_n
{\scriptsize
\Big\{ \cW_n^s(\tau,\tau') - \mathfrak{W}_n^s(\tau,\tau')
\Big\} \Big\Vert_{\infty} 
}
\\ [2mm]
&\nspace & \leq |f_n||g_n| ( 1 + |\tau - \tau_0|)( 1 + |\tau'-\tau_0|)\mathfrak{z}_n(\tau_0)\Big\{ \frac{5}{4}n^2 e^{2\tau} + n^2 e^{2\tau'} + \frac{5}{4}n^4 e^{2(\tau + \tau')} \Big\}
\nonumber
\ea
Defining 
\be
m_{2l}(\tau_0) = \frac{5}{4} \sum_{n \neq 0}  |f_n| |g_n|\, \mathfrak{z}_n(\tau_0)
\,n^{2l} \,, \quad l \in \N\,,
\ee
we can return to the mode sum (\ref{avdpr1}) and arrive at the

\begin{result} \label{AVDbounds} 
The difference between the spatially averaged matrix
two-point function $\cW^s(\tau,\tau',\zeta \!- \!\zeta')$ in the full
Gowdy system and its counterpart $\mathfrak{W}^s(\tau,\tau',\zeta\!-\!\zeta')$
in the velocity dominated system can for $f,g \in C^{\infty}(S^1)$
be bounded as follows:
\ba
\label{avdresult} 
\!\!\!\!&\nspace & \bigg\Vert
{\scriptsize
\int\!\!\frac{d\zeta}{2\pi}\frac{d\zeta'}{2\pi}
f(\zeta) g(\zeta')
\Big\{ \cW^s(\tau,\tau',\zeta\! - \!\zeta') -  
\mathfrak{W}^s(\tau,\tau',\zeta \!- \!\zeta') \,  
\Big\}
}
\bigg\Vert_{\infty}
\nonum
\!\!\!\!&\nspace & \leq ( 1 + |\tau - \tau_0|)( 1 + |\tau'-\tau_0|)
\Big\{ e^{2\tau + 2\tau'} m_4(\tau_0) + (e^{2\tau}
+ e^{2\tau'}) m_2(\tau_0) \Big\}\,.
\ea
Here, both $W^s$ and $\mathfrak{W}^s$ are taken to derive
from time consistent initial data, so that the left hand side
(though not the bound) is independent of $\tau_0$. 
\end{result}

Remarks.

(i) This bound entails the desired result: if $\tau,\tau',\tau_0$ are
back-propagated towards the Big Bang via  $\tau = \tilde{\tau} + \eta$,
$\tau' = \tilde{\tau}' + \eta$, $\tau_0 = \tilde{\tau}_0 + \eta$,
$\eta \ra -\infty$, the right hand side vanishes like $O(\eta^2 e^{2 \eta})$. 
In particular $\tau - \tau' = \tilde{\tau} - \tilde{\tau}'$ stays
fixed in the limit, mirroring the condition for uniform convergence
in Result \ref{AVDexpansion}. In the latter case the time consistency of
the underlying state does not directly enter, as it is tacit on
both sides of (\ref{grad17}). 

(ii) The assumption $f,g \in C^{\infty}(S^1)$ is clearly not
necessary for the proof to go through. Even with the crude
Bessel function bounds leading to (\ref{Ibound2}) only the  
moments up to $m_4(\tau_0)$ need to be finite. With
$\tilde{\lb}_n, \tilde{\mu}_n$ to be at most of $\ln |n|$ type
growth this requires the sequence $|f_n| |g_n| n^{4} \ln |n|$ to be summable.
This means one could replace $C^{\infty}(S^1)$ by the space
of twice differentiable functions $C^2(S^1)$, equipped with the
usual norm $\Vert f \Vert_{C^2(S^1)} = \sum_{j=0}^2 \sup_{\zeta \in S^1}
|\dd_{\zeta}^j f|$.

(iii) A similar bound can be obtained for the difference of the commutator
functions. Starting from (\ref{grad18}) and proceeding as above, one finds
\ba
\label{avdresult2} 
\!\!\!\!&\nspace & \bigg\Vert
{\scriptsize
\int\!\!\frac{d\zeta}{2\pi}\frac{d\zeta'}{2\pi}
f(\zeta) g(\zeta')
\Big\{ D(\tau,\tau',\zeta\! - \!\zeta') -  
\mathfrak{D}(\tau,\tau',\zeta \!- \!\zeta') \,  
\Big\}
}
\bigg\Vert_{\infty}
\nonum
\!\!\!\!&\nspace & \leq ( 1 + |\tau - \tau'|)
\big\{ e^{2\tau + 2\tau'} k_4 + (e^{2\tau} + e^{2\tau'}) k_2 \big\}\,,
\ea
where we defined
\be
k_{2l} = \frac{5}{4} \sum_{n \neq 0}  |f_n| |g_n|\, 
\,n^{2l} \,, \quad l \in \N\,.
\ee

For illustration, we show in Fig.~3 the actual behavior of the difference
for the two-point functions in the Bunch-Davies vacuum, as detailed
in Section 4.   
\begin{figure}[h]
    \centering
    \includegraphics[scale=0.7]{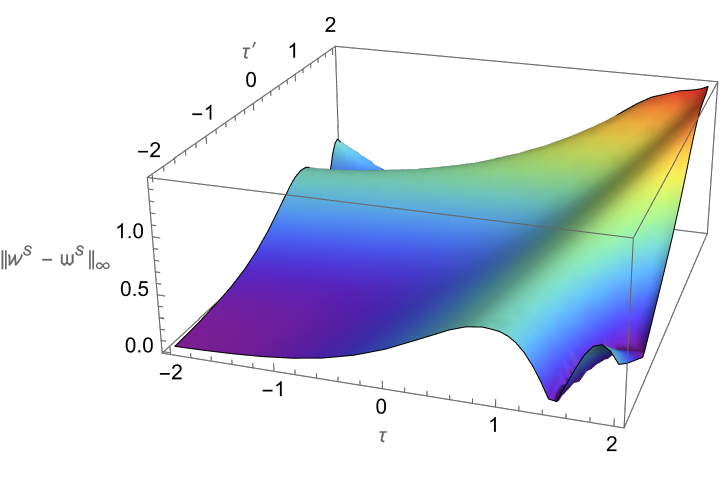}
    \caption{Decay of the difference $\Vert\mathcal{W}^s -  
      \mathfrak{W}^s\Vert_{\infty}$ in the Bunch–Davies vacuum state,
      plotted as a function of the two time variables $\tau$ and $\tau'$.}
\end{figure}

\newpage 
\section{Hadamard states and composite operators}
\label{section4} 

The concept of Hadamard states normally only applies to (free) quantum
field theories on a curved, non-dynamical background. In the present
context, the metric components of the Gowdy cosmologies themselves
are treated as quantum fields, so the notion of a Hadamard state is
not directly applicable. Nevertheless, the basic wave equation for
the Gowdy scalar $\phi$ in (\ref{evolpt}) "looks as if" it lived on a
$1+1$ dimensional time-dependent background set by $\rho$. 
In this sense, the notion of a Hadamard state {\it is} applicable to the
quantized $\phi$ fields. In the following we argue that the
Fock vacuum entering in (\ref{Tvac}) should in fact be 
associated with a Hadamard state.


\subsection{Hadamard condition in Fourier space}

Hadamard states on a globally hyperbolic background admit a characterization
in terms of a local parametrix defined in terms of the Synge function,
see e.g.~\cite{KMreview,Hackbook}, but constructing them is difficult.
On spatially homogeneous backgrounds 
(with spatial sections diffeomorphic to~$\R^d$), one can alternatively
characterize {\it and} construct Hadamard states in Fourier space.

{\bf Result\cite{BianchiSLE}.}
{\it Let $T_p$ be a Wronskian‑normalized solution of
$\bigl[\partial_\tau^2 + \nu(\tau) + |p|^2 w(\tau)\bigr]T_p(\tau) = 0$, 
with $\nu,w\in C^\infty(\R)$, $w>0$, and $p=(p_1,\ldots,p_d)$
the spatial momentum. Define the two‑point function}
\begin{equation}
\label{FHad0}
\varpi(\tau,x;\tau',x')
=\int\!\frac{d^dp}{(2\pi)^d}\,T_p(\tau)\,T_p(\tau')^*\,e^{ip\cdot(x-x')}.
\end{equation}
{\it Then $\varpi$ is associated with a Hadamard state if and only if
$|T_p(\tau)|^2$ admits an asymptotic expansion of the form} 
\begin{equation}
\label{FHad1}
|T_p(\tau)|^2 \asymp \frac{1}{2 \sqrt{w} |p|}\Big\{
1 + \sum_{n \geq 1} (-)^n G_n(\tau) |p|^{-2n} \Big\}\,,
\end{equation}
{\it 
where the $G_n$ are known functionals of $\nu,w$.  In particular,}
\begin{eqnarray}
\label{FHad2} 
G_1 &\! =\! & \frac{\nu}{2 w} + \frac{5}{32} \frac{{w'}^2}{ w^3} 
- \frac{1}{8} \frac{w''}{w^2}\,,
\nonumber \\[1.5mm]
G_2 &\! =\! & \frac{3}{8w^2} \Big( \nu^2 + \frac{1}{3} \nu'' \Big) 
- \frac{5}{16 w^3} \Big( \nu w'' + \nu'w' - \nu\frac{7 {w'}^2}{4w} \Big)     
\nonumber \\[1.5mm]
&+&\frac{1}{32 w^3} \Big(\! - w^{(4)} +\frac{21 {w''}^2}{4 w}
+\frac{7 w^{(3)} w'}{w} -\frac{231 {w'}^2w''}{8 w^2} 
+\frac{1155 {w'}^4}{64 w^3} \Big)\,.
\end{eqnarray}

The asymptotics (\ref{FHad1}) is consistent with the one induced by the
adiabatic iteration but much streamlined, see \cite{PT,PlaWinst} and
Appendix A of \cite{BianchiSLE}. Heuristically, Hadamard states are
adiabatic vacua of infinite order. Mathematically, this is expressed
by Lemma III.2 of \cite{BianchiSLE} which we assume to extend
to generic $d \geq 1$. Importantly, the Fourier space Hadamard condition
(\ref{FHad1}) only
prescribes the large $|p|$ asymptotics of $|T_p(\tau)|^2$. Different
Hadamard states will differ by their non-universal small $|p|$ behavior.
It is commonly assumed that a state is also infrared finite,
which requires $|T_p(\tau)|^2 = o(|p|^{-\delta})$, $\delta < d$,
for the integral to be infrared finite. For example, the States
of Low Energy are Hadamard states with $\delta =1$ for all $d \geq 1$,
so they are infrared finite for $d \geq 2$ \cite{BianchiSLE}. 
\smallskip

\noindent\textbf{Two prominent Hadamard states for the Gowdy scalar.}
For $d=1$, $\nu=0$, $w=e^{2\tau}$ the above result entails
that $T_n(\tau)$ is associated with a Hadamard state if and only if
$|T_n(\tau)|^2$ has the a large $|n|$ asymptotic expansion of the form
\begin{equation}
\label{FHad3} 
|T_n(\tau)|^2 \asymp \frac{1}{2 e^{\tau} |n|}\Big\{
1 -\frac{1}{8} \frac{e^{-2 \tau}}{n^2} +
\frac{27}{128} \frac{e^{-4 \tau}}{n^4} + O(n^{-6}) 
\Big\}\,, 
\end{equation}
where the coefficients turn out to be those of $(\pi/4)
|H_0^{(2)}(|n| e^{\tau})|^2$. Here we transitioned to discrete momenta
and circular spatial sections. This is expected to be legitimate on
general grounds, but in the situation at hand it follows more directly from
an identity of the form
\begin{equation}
\label{FHad4} 
\sum_{n = - N}^N 2\pi W^s_{\R}(\tau,\tau', \zeta - \zeta' +2 \pi n) -
[W^s(\tau,\tau', \zeta - \zeta') - W_0(\tau,\tau')] =
c_N(\tau,\tau')\,. 
\end{equation}
Here $W^s_{\R}$ is the symmetric two-point function for noncompact
spatial sections and the term in square brackets is
the previous two-point function $W^s$ from (\ref{Wsdef})
defined on $S^1$, with the zero mode term $W_0(\tau,\tau')$ removed.
The function $c_N(\tau,\tau')$ is explicitly computable
and is regular at $\tau = \tau'$. It diverges for $N \ra \infty$ in a
way that depends on the infrared properties of $W^s_\R$'s Fourier
transform. In $d=1$ one needs this Fourier transform to scale
as $p^{-\delta},0 \leq \delta <1$, for the integral to be infrared finite.
In this case the leading divergence is $O(N^{\delta} \ln^2 N)$.
For $\delta=1$, relevant for States of Low Energy, the
Fourier integral needs to be cut off at some small $0 < \mu/(2N+1)
\ll 1$. In this case the leading divergence is $O(N \ln^2 \mu)$
or $O(N \ln^2 N \ln \mu)$. Note that the periodic two-point function
is always infrared finite as $W_0(\tau,\tau')$ of the form in
(\ref{Wsdef}) can be chosen at will. Based on (\ref{FHad4}) one can
recover $W^s(\tau,\tau',\zeta\! - \!\zeta') - W_0(\tau,\tau')$ as
a subtracted limit of $2\pi W_\R^s$'s periodic extension.
The notion of a Hadamard state thereby carries over from the
noncompact to the compact spatial sections without having to 
construct the parametrix in the latter case. We now discuss
two important instances of so-understood Hadamard states for
the $T^3$-Gowdy system.
\smallskip

{ \bf (Gowdy) Bunch–Davies vacuum.} The choice $\lb_n =1, \mu_n=0$,
$n \neq 0$, in (\ref{Tsol2}) gives
\begin{equation}
\label{FHad6} 
T_n^{\rm BD}(\tau) = \frac{\sqrt{\pi}}{2}H_0^{(2)}(|n|e^\tau)\,,
\quad n \neq 0\,,
\end{equation} 
which satisfies (\ref{FHad3}). Augmented by the zero-mode choice   
$\tilde{\lb}_0 =1, \tilde{\mu}_0 =0$ it defines a spatially
periodic two-point function via (\ref{Wsdef}). The Fock vacuum
associated with the decomposition (\ref{ModePP}) for this choice will be referred
to as the ($T^3$-Gowdy) Bunch-Davies vacuum. The rationale for this
terminology is as follows. Starting from the basic wave equation for
$\phi$ in (\ref{evolpt}) with $\rho = t$ and redefining $\phi$ according
to $\Phi(t,\zeta) := \sqrt{t} \phi(t,\zeta)$ leads to \cite{QGowdy0}
\be
\label{dSwaves}
\Big[ \dd_t^2 - \dd_{\zeta}^2 + \frac{1}{4t^2}\Big] \Phi =0\,,
\quad \Phi(t, \zeta) := \sqrt{t} \phi(t,\zeta)\,.  
\ee
This can be recognized as the wave equation of a free massive field on
$1+1$ dimensional de Sitter space in the Poincar\'{e} patch and conformal
time $t$, see e.g.~\cite{dSQFTreview}. Except for the late time limit $t\rightarrow\infty$ it is not conformally invariant. The mass parameter here is
$m = 1/2$, which is at the unitarity threshold. For noncompact
spatial sections the two-point function $W_{\R}(t,t',\zeta\!-\!\zeta')$
can be computed explicitly and indeed coincides with the known 
$W_{{\rm dS}_2}(t,t',\zeta\!-\!\zeta')/\sqrt{tt'}$, expressible
in terms of an ${}_2F_1$ hypergeometric function for $m=1/2$ and
with an appropriate $i\eps$-prescription. The symmetric part can
be expressed as follows
\begin{equation}
\label{WsBD}
W^s_\R(t,t',\zeta - \zeta') 
= \frac{1}{ \pi \sqrt{tt'}} \Big\{ \th(-\xi) \frac{1}{\sqrt{1 - \xi}}
{\bf K}\Big(\frac{1}{1-\xi}\Big) + \th(\xi) {\bf K}(1-\xi)\Big\}\,.
\end{equation}
Here ${\bf K}(z) = \int_0^{\pi/2} \! ds (1- z \sin s)^{-1/2}$ is
the complete elliptic integral and
$\xi = [(\zeta - \zeta')^2 - (t-t')^2]/(4 t t')$ is the de Sitter
embedding distance. Near the lightcone $|t-t'| = |\zeta-\zeta'|$ the function $W^s_\R$ has a logarithmic singularity, at $t+t'=|\zeta-\zeta'|$ it is regular, and for $|t-t'| \rightarrow \infty$ it behaves like $1/|t-t'|$. The relation (\ref{FHad4}) can be directly
verified in this case and used to (re-)construct the periodic
two-point function (\ref{Wsdef}) with the above Bogoliubov
parameters. 
\smallskip

{\bf States of Low Energy (SLE).} SLE are a class of Hadamard states
that are explicitly constructable on any spatially homogeneous background.
The original construction \cite{Olbermann} was for Friedmann-Lema\^{i}tre 
cosmologies; further properties and the extension to Bianchi I
spacetimes can be found in \cite{BonusSLE,BianchiSLE}. The term SLE stems from
the fact that in Fourier space the temporarily averaged Hamiltonian
is minimized modewise. The averaging is done with a
$f \in C_c^{\infty}(\R)$ `window' function and the minimization
produces a set of Bogoliubov parameters that depend on $f$ and
the momenta. This construction can be applied to the 
Hamiltonian governing the dynamics of the Gowdy scalar in
the reduced phase space formulation, see (\ref{Hamdef}) below.
Specifically, taking the wave equation (\ref{Tsol1}) as basic
the formulas in \cite{BonusSLE,BianchiSLE} apply with the specialization to
$d=1$ and $\om_n(\tau) = |n| e^{\tau}$. The SLE solution can be
expressed solely in terms of the commutator function, see Thm.II.2
of \cite{BianchiSLE}. Using the expression (\ref{Gowdycomm})
for the commutator function of the Gowdy system SLE solution
for $G_n^{\rm SLE}(\tau) = |T_n^{\rm SLE}(\tau)|^2$ can be evaluated. 
The result is of the following form
\ba
\label{GDGowdysol}
&& G_n(\tau) = \frac{\pi}{2} c_{1,n} J_0(|n|e^{\tau})^2 +   
\frac{\pi}{2} c_{2,n} J_0(|n|e^{\tau}) Y_0(|n|e^{\tau}) +     
\frac{\pi}{2} c_{3,n} Y_0(|n|e^{\tau})^2 \,,
\nonum
&& 
c_{1,n} = \frac{1}{2} |\lb_n + \mu_n|^2\,, \quad
c_{3,n} = \frac{1}{2} |\lb_n - \mu_n|^2\,, \quad
c_{2,n} = -i (\lb_n \mu_n^* - \lb_n^* \mu_n) \,,
\nonum
&&  4 c_{1,n} c_{3,n} - c_{2,n}^2 = (|\lb_n|^2 - |\mu_n|^2)^2 = 1\,,
\quad n \neq 0\,.
\ea
For SLE the $n \neq 0$ coefficients are
\ba
\label{FHad10}
c_{1,n}^f \is \frac{\pi n^2}{8 \cE_n^{\rm SLE}}
\int\! d\tau f(\tau)^2 e^{2 \tau} \big[
Y_0(|n|e^{\tau})^2 + Y_1(|n|e^{\tau})^2 \big] \,, 
\nonum
c_{2,n}^f \is -\frac{\pi n^2}{4 \cE_n^{\rm SLE}}
\int\! d\tau f(\tau)^2 e^{2 \tau} \big[
  J_0(|n|e^{\tau}) Y_0(|n|e^{\tau}) + J_1(|n|e^{\tau})
  Y_1(|n|e^{\tau}) \big] \,, 
\nonum
c_{3,n}^f \is \frac{\pi n^2}{8 \cE_n^{\rm SLE}}
\int\! d\tau f(\tau)^2 e^{2 \tau} \big[
J_0(|n|e^{\tau})^2 + J_1(|n|e^{\tau})^2 \big] \,. 
\ea 
Here $\cE_n^{\rm SLE}$ is the energy of the time averaged
Hamiltonian in the SLE. By inverting the $c_{1,n}, c_{2,n},c_{3,n}$
versus $\lb_n,\mu_n$ relations in (\ref{GDGowdysol}) one obtains
expressions for
the $f$-dependent Bogoliubov parameters $\lb_n^f,\mu_n^f$ that
define the SLE solution $T_n^{\rm SLE}(\tau)$ in (\ref{Tsol2}).
The induced $n$-dependence is rather complicated and it is not
obvious that  $|T_n^{\rm SLE}(\tau)|^2$ will satisfy (\ref{FHad3}).  
Actually, it does, as one can show 
\be
\label{FHad15}
c_{1,n}^f \asymp 1/2, \quad c_{3,n}^f \asymp 1/2, \quad c_{2,n}^f \asymp 0,
\ee
where `$\asymp$' means ``equality in an asymptotic expansion in
powers of $1/|n|$''. For $\lb_n^f, \mu_n^f$ this means
$\mu_n^f \asymp 0$, $|\lb_n^f| \asymp 1$, while the
phase of $\lb_n^f$ would require separate examination.  
This phase drops out in the asymptotics of $G_n^{\rm SLE}(\tau) =
|T_n^{\rm SLE}(\tau)|^2$, which therefore has the same asymptotics
as (\ref{FHad6}), i.e. (\ref{FHad3}).

In summary, both the Bunch-Davies state and the SLE define
Hadamard states for the Gowdy scalar by specifying a set
of Bogoliubov parameters in (\ref{Tsol2}). Via (\ref{FHad11})
the Bogoliubov parameters define initial data, which enter
the initial value parameterization of the two-point function
(\ref{twopmat2}), (\ref{twopmat3}). The Result \ref{4maps}
delineates conditions under which these states
are also time consistent.


\subsection{Hadamard and time consistent states}
\label{sec4.2}

Both the Bunch–Davies vacuum (BD) and State of Low Energy (SLE)
are also time consistent. Technically, this is because their
Bogoliubov parameters
$(\lambda_n,\mu_n)$ are independent of the reference time~$\tau_0$,
albeit for distinct reasons.

The Bunch-Davies state corresponds in the parameterization
(\ref{Tsol2}) simply to $\lb_n^{\rm BD} =1$, $\mu_n^{\rm BD} =0$, $n \in \Z$.
Then (\ref{match8}), (\ref{H0def}) gives for $n \neq 0$ 
\ba
\label{match17}
&& \tilde{\lb}_n^{\rm BD}(\tau_0) = 1 - \frac{i}{\pi} \ln \gamma_n(0)\,,
\sspace \;\; \tilde{\mu}_n^{\rm BD}(\tau_0) =
- \frac{i}{\pi} \ln \gamma_n(0)\,,
\nonum
&& \tilde{z}_n^{\rm BD}(\tau_0) = \frac{\sqrt{\pi}}{2} - \frac{i}{\sqrt{\pi}}
\ln \gamma_n(\tau_0) \,, \quad
\tilde{w}_n^{\rm BD}(\tau_0) = - \frac{i}{\sqrt{\pi}}\,.
\ea
 For $n=0$ one has from
(\ref{match12}) $\tilde{\lb}_0 =1, \tilde{\mu}_0 =0$ and (\ref{H0def})
then gives
\begin{equation}
\label{match18}
\tilde{z}_0(\tau_0) = \frac{\sqrt{\pi}}{2} - i \frac{\tau_0}{\sqrt{\pi}}\,,
\quad \tilde{w}_0(\tau_0) = - \frac{i}{\sqrt{\pi}}\,.
\end{equation}

The SLE are likewise time consistent. Indeed, according to the result
in Section \ref{section3.2} this should hold because the parameters
$c_{1,n}^f, c_{2,n}^f, c_{3,n}^f$  defined in (\ref{FHad10}) are
independent of $\tau_0$. Returning to (\ref{GDGowdysol}), (\ref{FHad10}) and
considering the $\tau \ra - \infty$ behavior of the solution
$G_n^{\rm SLE}(\tau)$ one finds for $n \neq 0$
\ba
\label{match20} 
G_n^{\rm SLE}(\tau) \is \tilde{c}_{3,n} \tau^2 + \tilde{c}_{2,n} \tau
+ \tilde{c}_{1,n} + O(\tau e^{2\tau})\,,
\nonum
\tilde{c}_{3,n} \is \frac{2}{\pi} c_{3,n}^f \,,
\nonum
\tilde{c}_{2,n} \is c_{2,n}^f + c_{3,n}^f \frac{4}{\pi} \ln( e^{\gamma_E} |n|/2) \,,
\nonum
\tilde{c}_{1,n} \is \frac{\pi}{2} c_{1,n}^f  + c_{2,n}^f \ln( e^{\gamma_E} |n|/2)
+ c_{3,n}^f \frac{2}{\pi} \ln^2( e^{\gamma_E} |n|/2) \,.
\ea
For $n=0$ the SLE parameters in (\ref{FHad10})
are not really defined and we omit them for now. For the induced
Bogoliubov parameters one has
\ba
\label{match21}
\tilde{\lb}_n^f \is \lb_n^f -
\frac{i}{\pi}(\lb_n^f\! -\! \mu_n^f) \ln \gamma_n(0)\,,
\nonum
\tilde{\mu}_n^f \is \mu_n^f -
\frac{i}{\pi}(\lb_n^f\! -\! \mu_n^f) \ln \gamma_n(0)\,
\ea 
and the velocity dominated initial values are
(up to a shared phase that may be needed to have the correct
large $n$ asymptotic expansion) one has 
\ba
\label{match22}
\tilde{z}_n^f(\tau_0) \is \frac{\sqrt{\pi}}{2} (\lb_n^f + \mu_n^f)
- \frac{i}{\sqrt{\pi}}\ln \gamma_n(\tau_0) (\lb_n^f - \mu_n^f)\,,
\nonum
\tilde{w}_n^f(\tau_0) \is - \frac{i}{\sqrt{\pi}} (\lb_n^f - \mu_n^f)\,.
\ea

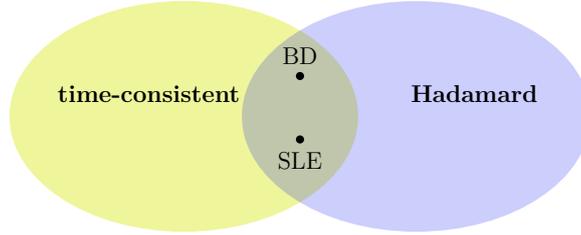
\begin{figure}[h]
  \centering
\resizebox{0.5\textwidth}{!}{%
  \begin{tikzpicture}
    \definecolor{yellowish}{RGB}{215,230,5}
    \definecolor{bluish}{RGB}{0,10,240}
    \fill[yellowish, opacity=0.4] (-2,0) ellipse (3 and 2);
    \fill[bluish,   opacity=0.2] ( 2,0) ellipse (3 and 2);
    \node at (-2.6,0.4) {\textbf{time‐consistent}};
    \node at (3.0,0.4)  {\textbf{Hadamard}};
    \fill (0,0.7) circle (2pt) node[above=2pt] {BD};
    \fill (0,-0.4) circle (2pt) node[below=2pt] {SLE};
  \end{tikzpicture}
}
  \caption{\label{Fig4} Intersection of time‐consistent and Hadamard states.}
\end{figure}

There are of course also Hadamard states which are not time consistent
simply because the Bogoliubov parameters depend on the
initial value time $\tau_0$. For example
\be
\label{comp4}
\lb_n(\tau_0) = \big(1 + e^{- \frac{n^2}{2\tau_0}}\big)^{1/2}\,, \quad
\mu_n(\tau_0) = e^{- \frac{n^2}{2\tau_0}}\,, 
\ee
differs from the BD parameters $\lb_n^{\rm BD} =1, \mu_n^{\rm BD}=0$
by $\tau_0$-dependent terms that do not have an expansion in
powers of $1/n$. It is therefore a Hadamard state but not
time consistent. The interplay between time consistent states and
Hadamard states is schematically summarized in Figure \ref{Fig4}. 

Using the notion of time consistent states as introduced
in Subsection \ref{section3.2} the AVD Results \ref{AVDexpansion},
\ref{AVDbounds}  
can be paraphrased as follows:
{\it all} time consistent states give rise to (matrix-) two-point
functions that are asymptotically velocity dominated in the sense
of (\ref{avdresult}). These states include many Hadamard states, in
particular the Bunch-Davies vacuum and the States of Low Energy. 


\subsection{Quantum $\sigma$‑field}

So far, our analysis has focused on the Gowdy scalar $\phi$,
canonically quantized as in (\ref{ModePP}). The gravitational origin
of the classical system is encoded in the constraints (\ref{constrL}).
As noted in (\ref{Fdef}) they can be solved for the gradient of
$\sigma$. In proper time gauge and $(t = e^{\tau}, \zeta)$
coordinates the gradient in (\ref{evolpt}) simplifies to 
\ba
\label{sigmaclass2} 
F_0^{\rm pt}(e^{\tau}, \phi) \is \frac{1}{2} [ (\dd_{\tau} \phi)^2 +
  e^{2 \tau} (\dd_{\zeta} \phi)^2 ] =: T_{00} = T_{11}\,.
\nonum
F_1^{\rm pt}(e^{\tau}, \phi) \is \dd_{\tau} \phi \dd_{\zeta} \phi =:
T_{01} = T_{10}\,.
\ea
Here, we introduced the "would-be energy–momentum tensor"
$T_{\mu\nu}$ of $\phi$ in the time dependent background. For
on-shell $\phi$ it obeys the partial conservation equation
$\partial_\tau T_{01}= \partial_\zeta T_{11}$. This ensures that 
the gradient formulas can consistently be integrated to
give \cite{Torre}
\begin{equation}
\label{sigmaclass1}
\sigma(\tau, \zeta) = \sigma(\tau_0,\zeta_0)
+ \int_{\tau_0}^{\tau} \! d\tau' \, T_{00}(\tau',\zeta) +
\int_{\zeta_0}^{\zeta} \! d\zeta' \, T_{01}(\tau_0,\zeta')\,.
\end{equation}
Since $T_{00} = T_{11}$ is spatially periodic
the last relation implies that $C_0$ below is conserved, $\dd_{\tau} C_0 =0$.
On the other hand, one sees from (\ref{sigmaclass1}) that in order
for $\sigma(\tau_0, \zeta) = \sigma(\tau_0, \zeta + 2\pi)$ to hold
one needs $C_0=0$ at $\tau = \tau_0$. Together,
\begin{equation}
\label{C0def} 
C_0 \stackrel{\displaystyle{!}}{=} 0\,, \quad
C_0 := \int_0^{2\pi} \frac{d \zeta}{2\pi} \,T_{01}\,.
\end{equation}
This is the remnant of the original constraints that still needs to be
incorporated into the solution for $\phi$. 
In the quantum theory, one must define the field $\sigma$—that is,
the components $T_{00}$ and $T_{01}$—as composite operators 
satisfying the partial conservation law and the constraint (\ref{C0def}).
An adiabatic renormalization \cite{PT} of $T_{00}$ has been considered early on
in \cite{Berger}.

For later use we prepare the mode expansions
\ba
\label{sigmaclass3} 
T_{00} \is \sum_{n \in \Z} e^{ i n \zeta} (T_{00})_n(\tau) \,,\quad
T_{01} = \sum_{n \in \Z} e^{ i n \zeta} (T_{01})_n(\tau) \,,
\nonum
(T_{00})_n(\tau) \is \frac{1}{2} \sum_m \big[
\dot{\phi}_{n-m} \dot{\phi}_m - (n-m) m \,e^{2 \tau} \phi_{n-m} \phi_m \big]  
\nonum
(T_{01})_n(\tau) \is i \sum_m m \,\dot{\phi}_{n-m} \phi_m\,.
\ea 
Here, the Fourier modes $\phi_m = \phi_m(\tau)$ are functions
of $\tau$ only and we often write $\dot{\phi}_m = \dd_{\tau} \phi_m$. 
The consistency condition needed is
\begin{equation}
\label{sigmaclass4} 
\dd_{\tau} (T_{01})_n = i n (T_{00})_n\,, \quad n \in \Z\,,
\end{equation} 
which indeed holds on account of $\ddot{\phi}_n = - n^2 \phi_n$. 

These relations remain valid upon canonical quantization,
provided that the ordering of the Fock space operators is respected and
the operator sums are treated as formal. For example,
the time dependent Hamilton operator is  
\ba
\label{Hamdef} 
\mathbb{H}(\tau) \is \frac{1}{2} \int_0^{2\pi} \!\frac{d \zeta}{2\pi}
T_{00}(\tau,\zeta) = \frac{1}{2} (T_{00})_0(\tau) = 
\nonum
\is \frac{1}{2} \sum_{n \in \Z} \big\{ \varepsilon_n(\tau) (a_n a_n^* +
a_{-n}^* a_{-n}) + d_n(\tau) a_n a_{-n} +
d_n(\tau)^* a_{-n}^* a_n^* \big\}\,,
\ea 
with $\varepsilon_n(\tau) = |\dd_{\tau} T_n|^2 + n^2 e^{2 \tau} |T_n(\tau)|^2$,
$d_n(\tau) = (\dd_{\tau} T_n)^2 + n^2 e^{2 \tau} T_n(\tau)^2$.
The zero-mode part of the $\sigma$ operator is by (\ref{sigmaclass1})
then formally given by integrating 
$\dd_{\tau} \int\! \frac{d\zeta}{2\pi} \sigma(\tau,\zeta)
= 2 \mathbb{H}(\tau)$. Of course, these formal sums cannot act on the
Fock space and some regularization and/or renormalization is needed.

Hadamard states are tailored towards renormalization of
quantum field theories in curved spacetimes \cite{KMreview, Hackbook}.
In the standard setting with a spatial slices diffeomorphic to $\R^d$
one can render the two‐point function finite by subtracting its singular
“Hadamard parametrix” in position space—i.e.~the part singular on the
lightcone—from the full two‐point function. For spatially homogeneous
backgrounds subtraction of a truncated Hadamard parametrix is consistent
with the older adiabatic renormalization, see e.g. \cite{PlaWinst}.
The precise relation has been elucidated in \cite{LR}, see also Appendix A
of \cite{BianchiSLE}.

In the polarized Gowdy model the spatial slices are circles, so the
light‐cone singularity acquires infinitely many $2\pi$–periodic images,
see (\ref{FHad4}).
Therefore, it is more convenient to work in Fourier space and
parameterize the two‐point function using its initial data $Z_n(\tau_0)$.
Since Hadamard states are singled out by the universal large $n$
asymptotics (\ref{FHad3}) of $|T_n(\tau)|^2$,
it follows that the initial–value matrix $Z_n(\tau_0)$ itself admits a
universal large $n$ expansion for every Hadamard state. Explicitly,
\be
\label{Tsplit4} 
Z_n(\tau_0) \asymp
\left.
\begin{pmatrix}
  \frac{2}{x} - \frac{1}{4 x^3} & - \frac{1}{x} + \frac{3}{8} \frac{1}{x^3}
  \\[2mm]
  - \frac{1}{x} + \frac{3}{8} \frac{1}{x^3} &
  \frac{x}{2} + \frac{9}{16} \frac{1}{x} - \frac{105}{256} \frac{1}{x^3}
\end{pmatrix} \right|_{x = |n| e^{\tau_0}} +O(n^{-5})\,.
\ee
We denote by $Z_n^{\rm Hr}(\tau_0)$ the matrix obtained by truncating the
large $n$ expansion of $Z_n(\tau_0)$ up to and including terms of order
$\mathcal{O}(n^{-r})$; the superscript “{\rm Hr}” stands for “Hadamard
expansion truncated
at order $r$.” Next, we consider point-split definitions of the vacuum
expectation values of \(T_{00}\) and \(T_{01}\) from \eqref{sigmaclass2},
aimed at:
\ba
\label{Tsplit1} 
&& T_{00}^W(\tau,\tau', \zeta\! - \!\zeta') := 
\frac{1}{4} \big[ \dd_{\tau} \dd_{\tau'} +
  e^{\tau + \tau'} \dd_{\zeta} \dd_{\zeta'} \big]
W^s(\tau,\tau', \zeta\! - \!\zeta')
\nonum
&& \quad = \frac{1}{4} W_0^s(\tau,\tau') + \frac{1}{2}
\sum_{n \geq 1} \cos n(\zeta\! - \!\zeta')
\big[ \dd_{\tau} \dd_{\tau'} W_n^s(\tau,\tau')
+ e^{\tau + \tau'} n^2 W_n^s(\tau,\tau') \big]\,, 
\nonum
&& T_{01}^W(\tau,\tau', \zeta\! - \!\zeta') := 
\frac{1}{2} \dd_{\tau} \dd_{\zeta'} W^s(\tau,\tau', \zeta \!- \!\zeta') 
\nonum
&& \quad = \sum_{n \geq 1} n \sin n (\zeta\! - \!\zeta')
\dd_{\tau} W_n^s(\tau,\tau')\,, 
\ea 
where $W_n^s(\tau,\tau') := T_n(\tau) T_n(\tau')^* +
T_n(\tau)^* T_n(\tau')$, $n \in \Z$. They obey the
quasi-conservation equation
\be
\label{Tsplit2}
\dd_{\zeta}  T_{00}^W(\tau,\tau', \zeta\! - \!\zeta')
= \frac{1}{2} \big(\!\!- \dd_{\tau'} + e^{-\tau + \tau'} \dd_{\tau} \big)\, 
T_{01}^W(\tau,\tau', \zeta\! - \!\zeta')\,,
\ee
which we seek to preserve after renormalization. 

To proceed, we recall the initial data realization of the
symmetric two point function
\be
\label{Tsplit3} 
W_n^s(\tau,\tau') = \big(\!\! - \dd_{\tau_0} \Delta_n(\tau,\tau_0) ,
\Delta_n(\tau,\tau_0) \big) Z_n(\tau_0)
\begin{pmatrix} - \dd_{\tau_0} \Delta_n(\tau',\tau_0) \\[2mm]
\Delta_n(\tau',\tau_0)
\end{pmatrix} \,.
\ee
Used in the Fourier kernels $\dd_{\tau} \dd_{\tau'} W_n^s(\tau,\tau')
+ e^{\tau + \tau'} n^2 W_n^s(\tau,\tau')$ and $\dd_{\tau} W_n^s(\tau,\tau')$,
one obtains realizations of $T^W_{00}$ and $T^W_{01}$ in terms of the
commutator function and the initial value matrix $Z_n(\tau_0)$. Importantly,
subtractions that
are carried out on the level of the initial value matrix will not affect
the time dependence and in particular preserve the validity of
(\ref{Tsplit2}). Our proposal for a ``truncated Hadamard
renormalization'' in this context is to replace $Z_n(\tau_0)$ in
(\ref{Tsplit3}) with 
\be
\label{Tsplit5}
Z_n^{\rm subHr}(\tau_0) := Z_n(\tau_0) - Z_n^{\rm Hr}(\tau_0)\,,
\ee
for a suitable order $r \in \N$. From (\ref{PPtwop3}) one recalls
that pointwise in $\tau,\tau_0$ the leading large $n$ behavior
of $\Delta_n(\tau,\tau_0)$ is $O(1/n)$, that of 
$\dd_{\tau_0} \Delta_n(\tau,\tau_0)$ is $O(1)$, and
that of $\dd_{\tau_0} \dd_{\tau} \Delta_n(\tau,\tau_0)$
is $O(n)$. For the three relevant Fourier kernels
$W_n^s(\tau,\tau')$, $n \dd_{\tau} W_n^s(\tau,\tau')$,
and $\dd_{\tau} \dd_{\tau'} W_n^s(\tau,\tau')
+ e^{\tau + \tau'} n^2 W_n^s(\tau,\tau')$, we need
a subtraction that produces a leading decay of $O(1/n^2)$
in order to ensure absolute convergence of the sums. 
Inspection of the matrix multiplications in (\ref{Tsplit3})
and its derivatives shows that in order to obtain 
coincidence limits represented by absolutely convergent 
series the following minimal subtractions are needed:  
\begin{equation}
\label{Tsplit6}
\phi^2:\;\; r=1\,, \quad T_{01}:\;\; r=2\,, \quad T_{00}:\;\; r=2\,.   
\end{equation}
We denote the subtracted Fourier kernels by
$[{\rm kernel}]^{\rm subHr}$. After the subtraction they
decay at least like $1/n^2$, pointwise in $\tau,\tau'$,
including $\tau'=\tau$. The $\zeta' \ra \zeta$ limit is therefore
well-defined and by common abuse of notation we
interpret the limit as the matrix element
$\bra 0| :\!{\rm operator}\!:_{\rm Hr} |0\ket$
of the composite operator aimed at (even if the existence of the
operator has not yet fully been justified, see \cite{KMreview}). In this
notation one has
\ba
\label{Tsplit7}
\bra 0 | :\!\phi(\tau,\zeta)^2 \!:_{H1} |0\ket \is
W_0^s(\tau,\tau) + 2 \sum_{n \geq 1} [W_n^s]^{\rm subH1}(\tau,\tau)\,.
\nonum
\bra 0 | :\!T_{00}(\tau,\zeta) \!:_{H2} |0\ket \is
\frac{1}{4} \dd_{\tau} \dd_{\tau'} W_0^s(\tau,\tau')\big|_{\tau' = \tau} 
+ \sum_{n \geq 1} [ \dd_{\tau} \dd_{\tau'} W_n^s
  + e^{\tau + \tau'}n^2 W_n^s]^{\rm subH2}(\tau,\tau) \,.
\nonum
\bra 0 | :\!T_{01}(\tau,\zeta) \!:_{H2} |0\ket \is 0\,.
\ea 
The first expression defines the counterpart of the ``power spectrum'',
the second is independent of $\zeta$ on account of translation invariance,
and the third is consistent with the extension of (\ref{Tsplit2})
to the coincidence limit. Alternatively, the matrix element
of the $n=0$ version of (\ref{sigmaclass4}) should at least be constant,
$\dd_{\tau} \bra 0 | \int_0^{ 2\pi} \!
\frac{d\zeta}{2\pi} :\!T_{01}(\tau,\zeta) \!:_{H2} |0\ket = 0$. 

The vanishing of this constant extents to higher order subtractions
\begin{equation}
\label{Tsplit8}
\bra 0 | :\!T_{01}(\tau,\zeta) \!:_{Hr} |0\ket = 0\,, \quad r \geq 2\,.
\end{equation}
This is because, once $[n \dd_{\tau} W_n^s]^{\rm subHr}(\tau,\tau)$ is
absolutely summable, its Fourier-sine series will vanish in the
$\zeta \ra \zeta'$ coincidence limit. Normally, the Fourier subtractions
(\ref{Tsplit5}) cannot be pushed to arbitrarily high orders,
as the coefficients in (\ref{Tsplit4}) are rapidly increasing
beyond $r=10$, say, reflecting the merely asymptotic nature of the
expansion. The case (\ref{Tsplit8}) is exceptional in that the
large coefficients are rendered irrelevant by the $\zeta' \ra \zeta$
limit of the convergent sine-Fourier series. For (\ref{Tsplit8}) we can
therefore conclude that the all order subtraction, representing the
actual Hadamard subtraction, also vanishes,
$\bra 0 |\! :\!T_{01}(\tau,\zeta) \!:_{\rm H} \!|0\ket = 0$.
Recall that the Hadamard normal product would be defined
(after stripping off test functions) by the coincidence limit of
\begin{equation}
\label{Tsplit9}
:\!\dd_{\tau} \phi(\tau,\zeta) \dd_{\zeta'} \phi(\tau', \zeta')\!:_H
\,= \dd_{\tau} \phi(\tau,\zeta) \dd_{\zeta'} \phi(\tau', \zeta') -
T_{01}^{W^{\rm subH}}(\tau,\tau', \zeta-\zeta') \,,
\end{equation}
where $T_{01}^{W^{\rm subH}}(\tau,\tau', \zeta-\zeta')$ derives from
$T_{01}^W(\tau,\tau', \zeta\!-\!\zeta') = \frac{1}{2} \dd_{\tau} \dd_{\zeta'}
W^s(\tau,\tau',\zeta\! -\! \zeta') $ by subtracting the full
position space Hadamard parametrix from $W^s$; see e.g.~\cite{KMreview}. In the coincidence limit the subtraction is
$\bra 0 |\! :\!T_{01}(\tau,\zeta) \!:_{\rm H} \!|0\ket$, so its
vanishing leads to the ``non-renormalization'' result
\ba
\label{Tsplit10} 
&& \lim_{\tau' \ra \tau, \zeta' \ra \zeta} 
\big\{ :\!\dd_{\tau} \phi(\tau,\zeta) \dd_{\zeta'} \phi(\tau', \zeta')
+ \dd_{\zeta'} \phi(\tau',\zeta') \dd_{\tau} \phi(\tau, \zeta) \!:_H
\big\} 
\nonum
&& \quad 
= \lim_{\tau' \ra \tau, \zeta' \ra \zeta}
\big\{ \dd_{\tau} \phi(\tau,\zeta) \dd_{\zeta'} \phi(\tau', \zeta')
+ \dd_{\zeta'} \phi(\tau',\zeta') \dd_{\tau} \phi(\tau, \zeta)
\big\}\,.
\ea
As a heuristic check one may consider the naive quantum
version of the Fourier modes $(T_{01})_n$ from
(\ref{sigmaclass3}), written as $(T_{01})_n = \frac{1}{2} \sum_m
m [ \dot{\phi}_{n-m} \phi_m + \phi_m \dot{\phi}_{n-m}]$,
with $\phi_n(\tau) = T_n(\tau) a_n + T_n(\tau)^* a_{-n}^*$,
$\dot{\phi}_n(\tau) = \dot{T_n}(\tau) a_n +\dot{T_n}(\tau)^* a_{-n}^*$.
Clearly,
for $n \neq 0$ there are no ordering ambiguities and the termwise
interpreted expectation value (being linear in $a_{n-m}, a_{-(n-m)}^*$
and $a_m, a_{-m}^*, n \neq 0)$ in the Fock vacuum simply
vanishes. For $n=0$ it is the antisymmetry in the $\sum_m m$   
sum (corresponding to the $\sin m(\zeta - \zeta')$ structure in
the point-split version) that removes the symmetric terms
containing $a_m a_{-m}$, $a_{-m}^* a_{m}^*$, as well as terms
arising by reordering $a_m^* a_m$. The upshot is that the
quantum version of the conserved charge $C_0$ can reasonably
be interpreted as 
\begin{equation}
\label{Tsplit11} 
C_0 =\, :\!C_0\!:_H \;= - i \lim_{N \ra \infty} \sum_{n=-N}^N n \,a_n^* a_n \,.
\end{equation}
Its vacuum expectation value vanishes and so does therefore that of
$:\!\!T_{01}(\tau,\zeta)\!\!:_H =  T_{01}(\tau, \zeta)= \sum_n e^{ i n \zeta}
(T_{01})_n$. Further, one may verify that
\begin{equation}
\label{Tsplit12} 
[i C_0, \phi(\tau, \zeta)] = \dd_{\zeta} \phi(\tau, \zeta)\,,
   \quad [i C_0, \mathbb{H}(\tau)] =0\,, 
\end{equation}
where $\mathbb{H}(\tau)$ is the above Fock space Hamiltonian.
The operator $C_0$ can be shown to be
selfadjoint on a dense (finite occupation number) domain of the
Fock space and by Stone's theorem therefore generates a
unitary group $\R \ni \zeta \mapsto e^{ i \zeta C_0}$.    
It acts on the Fourier modes in the expected way
\begin{equation}
\label{Tsplit13}
e^{ i \zeta C_0} \phi_n(\tau) e^{- i \zeta C_0} =
\sum_{n \geq 0} \frac{\zeta^n}{n!} {\rm ad}^n (i C_0) \phi_n(\tau)
  = e^{ in \zeta} \phi_n(\tau)\,,
\end{equation}
and similarly for $\pi_n(\tau)$. Here 
${\rm ad}(iC_0) A = [i C_0, A]$,
${\rm ad}^n(iC_0) A = [i C_0, {\rm ad}^{n-1}(iC_0) A]$, $n \geq 1$. 
In particular, for $\zeta = 2\pi$ the operator $e^{ i 2\pi C_0}$
acts like the identity on the field algebra generated by
(the $2\pi$-periodic) $\phi(\tau,\zeta), \pi(\tau, \zeta)$.
Finally, $e^{ i \zeta C_0} |0\ket = |0\ket$,
expresses the translation invariance of the Fock vacuum,
and accounts for the dependence of the two-point functions
considered on $\zeta\! - \!\zeta'$ only.


\section{Conclusions}

Gowdy cosmologies have played an important role in shaping
our understanding of the Big Bang singularity, specifically
through proofs of Asymptotic Velocity Domination (AVD) in
the polarized \cite{IsenMonc} and unpolarized case \cite{Ringstroemproof}.
Here we investigated for the polarized system a quantum version of
the AVD property, formulated in terms of a matrix two-point function
$\cW^s$ of the basic Gowdy scalar. Since the latter governs via
(\ref{diractwopt2}) also the two-point functions of the integrands of
Dirac observables the main question is whether $\cW^s$ exhibits some
form of AVD. This was answered in the affirmative through the results
(\ref{grad20}), (\ref{avdresult}). We take this as a `proof of principle'
for the viability of a quantum AVD scenario. Hopefully
it will turn out to admit as significant generalizations as the
seminal classical result in \cite{IsenMonc}. 

The immediate extension is of course to the non-polarized Gowdy system.
A formal expansion of classical solutions around those of the
velocity dominated system has been obtained in \cite{GMonc}. The
construction of classical Dirac observables is feasible by
taking advantage of an underlying Lax pair \cite{DiracGowdy}. The
required spatial
periodicity (for the $T^3$ topology) however presents considerable
complications compared to other two-Killing vector reductions. 
Through a Riemannian sigma-model formulation the quantum theory
is amenable to an all-order perturbative (weak Newton coupling)
analysis \cite{2Kren1,2Kren2,reducedquant}. The quantum AVD
property, on the other hand, is likely to be related to an
expansion in inverse powers of Newton's constant
(strong coupling) and requires a different methodology.  

\bigskip

{\bf Acknowledgments.} We would like to thank R.~Banerjee and R.~Penna
for discussions and valuable comments on the manuscript.  We also
appreciate W.~Musk's and C.~Ives' participation in other aspects of
this project. 
This work is supported in part by the PittPACC initiative.
\newpage

\appendix

\section{The Hamiltonian action and its VD counterpart} 
\label{appendixA}

It is a general property of two-Killing vector reductions that
entering with (\ref{2Kmetric}) into the Einstein-Hilbert action
will produce a valid reduced action principle, i.e.~one whose
variations produce field equation coinciding 
with the ones obtained by directly specializing the Einstein
field equations to metrics of the form (\ref{2Kmetric}). We therefore
take the reduced Hamiltonian action of the polarized $T^3$ system as a
starting point, see \cite{QGowdy0, TorreObs}. The velocity dominated system is normally
only defined through its field equations, see \cite{IsenMonc,RingstroemLR}.  
In preparation for the quantum theory we develop here
also an off-shell formulation in terms of an Hamiltonian action
principle and its symmetries.

\subsection{Gowdy cosmologies: the Hamiltonian action and its symmetries} 

In some fiducial foliation the Hamiltonian action reads
\ba
\label{SH}
S^{\smallcap{h}} \is \int_{t_i}^{t_f}\!\! d x^0 \int_0^{2\pi} \!\! dx^1
\big\{ \pi^{\rho} e_0(\rho)
+ \pi^{\sigma} e_0(\sigma) + \pi^{\phi} e_0(\phi) - n \cH_0 \big\}\,,
\nonum
\cH_0 \is - \lbn \pi^{\sigma} \pi^{\rho} - \frac{1}{\lbn}
( \dd_1 \rho \dd_1 \sigma - 2 \dd_1^2 \rho) +
\frac{\lbn}{2 \rho} (\pi^{\phi})^2 + \frac{\rho}{2 \lbn} (\dd_1 \phi)^2\,,
\nonum
\cH_1 \is \pi^{\rho} \dd_1 \rho + \pi^{\sigma} \dd_1 \sigma
-2 \dd_1 \pi^{\sigma} + \pi^{\phi} \dd_1 \phi\,,
\ea 
where $e_0 = \dd_0 - \cL_s$, with $\cL_s$ the Lie derivative along
the one-dimensional shift $s$. Upon integration by parts one can
render the $s$ dependence explicit to find $-s \cH_1$, with
$\cH_1$ as given. For definiteness we identify $S^1$ with
$\R/(2\pi \Z)$ and assume all fields in (\ref{SH}) to be $2\pi$-periodic
in $x^1$.  Further, $\lbn>0$ is the dimensionless reduced
Newton constant and $\sigma := \tilde{\sigma} + \frac{1}{2} \ln \rho$.  
The phase space is equipped with a Poisson structure, $\{\,\cdot\,,
\,\cdot\,\}$ determined by the basic brackets, 
$\{\rho(x^1), \pi^{\rho}(y^1) \} = \delta(x^1\!-\!y^1)$, 
$\{\sigma(x^1), \pi^{\sigma}(y^1) \} = \delta(x^1\!-\!y^1)$, 
$\{\phi(x^1), \pi^{\phi}(y^1) \} = \delta(x^1\!-\!y^1)$,
where the equal $x^0$ arguments are suppressed and  
all $\delta$ distributions are spatial $+1$ densities.

The constraints obey a closed Poisson algebra 
\ba
\label{constr0} 
\{ \cH_0(x), \cH_1(y)\} \is \dd_1 \cH_0 \delta(x-y) + 2 \cH_0(x)
\delta'(x-y)\,,
\nonum
\{ \cH_1(x), \cH_1(y)\} \is \dd_1 \cH_1 \delta(x-y) + 2 \cH_1(x)
\delta'(x-y)\,,
\nonum
\{ \cH_0(x), \cH_0(y)\} \is \delta'(x-y)[\cH_1(x) + \cH_1(y)]\,.
\ea 
The first two relations just express that the density weights
of $\cH_0, \cH_1$ are $+2$. By specialization
of the general (model independent) gravitational constraint algebra 
one would expect the last relation to contain a factor explicitly
dependent on the spatial metric. Specifically,
with the present density conventions one would for the $1\!+\!1$
dimensional constraint algebra expect an additional $\gamma(\gamma_{11})^{-1}$,
$\gamma := - \det \gamma_{\mu\nu}$, 
term on the right hand side of the last relation in (\ref{constr0}).
Using (\ref{2Dmetric}) this evaluates to $\gamma (\gamma_{11})^{-1} =
n^2 e^{\tilde{\sigma}}$, which is a spatial scalar. A short computation
shows that the redefinition $\cH_0^{\smallcap{ADM}} :=
n e^{\tilde{\sigma}/2}\cH_0$, $\cH_1^{\smallcap{ADM}} := \cH_1$,
maps (\ref{constr0}) into the expected ADM type constraint algebra.

On general grounds
$\cH_0(\eps) + \cH_1(\eps^1)$ is the generator of Hamiltonian gauge
variations of any (not explicitly time dependent) functional
$F$ on phase space built from the canonical variables, here
$\phi, \rho, \sigma$ and their canonical momenta.
The descriptors $(\eps, \eps^1)$ play the role of $(n,s)$ and
can both be viewed as spatial $-1$ densities. 
The variations $\delta_{\eps}^H F = \{ F, \cH_0(\eps) + \cH_1(\eps^1)\}$
have to be augmented by gauge variations of lapse and shift
in order to obtain an invariance of the Hamiltonian action
(\ref{SH}). The result is:
\be
\delta^{\smallcap{h}}_{\eps} S^{\smallcap{h}} = 0 \,, \quad \mbox{if} \quad
\eps|_{t_i} = 0 = \eps|_{t_f}\,,
\ee
where the restriction on the temporal gauge descriptors has a well
understood origin and carries over from full Einstein gravity \cite{Pons}.
For convenient reference we display the complete set of
Hamiltonian gauge variations: 
\begin{subequations}
\label{gtrans1}
\ba
\delta^{\smallcap{h}}_{\epsilon} n &=& \partial_{0} \epsilon
- (s \partial_{1} \epsilon - \epsilon \partial_{1} s)
+ \epsilon^{1} \partial_{1} n - n\partial_{1} \epsilon^{1}\,,
\nonum
\delta^{\smallcap{h}}_{\epsilon} s &=& \partial_{0} \epsilon^{1}
+(\epsilon^{1} \partial_{1} s - s\partial_{1} \epsilon^{1})
+ \epsilon \partial_{1} n - n \partial_{1} \epsilon\,,
\\[2mm]
\delta^{\smallcap{h}}_{\epsilon} \sigma &=& - \lbn \eps \pi^{\rho} 
+ \epsilon^{1} \partial_{1} \sigma + 2\partial_{1} \epsilon^{1}\,,
\nonum
\delta^{\smallcap{h}}_{\epsilon} \rho &=& - \lbn \eps \pi^{\sigma} + \eps^1 \dd_1 \rho\,, 
\nonum
\delta^{\smallcap{h}}_{\epsilon} \phi &=& \lbn \frac{\eps}{\rho} \pi^{\phi} 
+ \epsilon^{1} \partial_{1} \phi\,,
\\[2mm]
\delta^{\smallcap{h}}_{\epsilon} \pi^\sigma &=&
\dd_1\big(\!-\frac{\epsilon}{\lbn}  \dd_1 \rho +\eps^1 \pi^\sigma\big)\,,
\nonum
\delta^{\smallcap{h}}_{\epsilon} \pi^\rho &=&
\dd_1 \big(\!-\frac{\eps}{\lbn} \dd_1 \sigma + \eps^1 \pi^\rho \big)
- \frac{2}{\lbn} \dd_1^2 \eps 
-\frac{\eps}{2\lbn} (\dd_1 \phi)^2 + \frac{\eps \lbn}{2 \rho^2} (\pi^\phi)^2\,,
\nonum
\delta^{\smallcap{h}}_{\epsilon} \pi^\phi &=&
\dd_1\big(\frac{\epsilon \rho}{\lbn} \dd_1 \phi +\eps^1 \pi^\phi\big)\,.
\ea
\end{subequations} 
Unless stated otherwise we take the descriptors $\eps,\eps^1$ to be
spatially periodic.
The field equations of the system are readily obtained by varying
$S^{\smallcap{h}}$. In addition to the constraints
$\delta S^{\smallcap{h}}/\delta n = -\cH_0$,
$\delta S^{\smallcap{h}}/\delta s = -\cH_1$,
there are the velocity-momentum relations
$e_0(\rho) + \lbn n \pi^{\sigma} =0$,
$e_0(\sigma) + \lbn n \pi^{\rho} =0$,
$\rho e_0(\phi) - \lbn n \pi^{\phi} =0$, 
and the evolution equations 
\ba
\label{evol1} 
0= \frac{\delta S^{\smallcap{h}}}{\delta \rho} \is
- e_0(\pi^{\rho}) - \frac{1}{\lbn} \dd_1( n \dd_1 \sigma)
- \frac{2}{\lbn} \dd_1^2 n + \frac{\lbn n }{2 \rho^2} (\pi^{\phi})^2
- \frac{n}{2 \lbn} (\dd_1 \phi)^2\,,
\nonum
0 =\frac{\delta S^{\smallcap{h}}}{\delta \sigma} \is - e_0(\pi^{\sigma})
- \frac{1}{\lbn} \dd_1( n \dd_1 \rho)\,,
\nonum
0 = \frac{\delta S^{\smallcap{h}}}{\delta \phi} \is - e_0(\pi^{\phi})
+ \frac{1}{\lbn} \dd_1 (\rho n \dd_1 \phi)\,.
\ea 
These are such that the constraints are preserved under time evolution
\be 
\label{evol2}
e_0(\cH_0) - \frac{1}{n} \dd_1(n^2 \cH_1) =0\,, \quad
e_0(\cH_1) - \frac{1}{n} \dd_1(n^2 \cH_0) =0\,. 
\ee
Generally, let $F$ be a (not explicitly time dependent) functional $F$
on phase space that is also a spatial tensor. Then $\{ F, \cH_0(n)\} = e_0(F)$
encodes the evolution equations.

\subsection{VD Gowdy cosmologies: the Hamiltonian action and its symmetries} 

The Hamiltonian action of the velocity dominated system is 
\ba
\label{SHVD}
S^{\smallcap{hvd}} \is \int_{t_i}^{t_f}\!\! d x^0 \int_0^{2\pi} \!\! dx^1
\big\{ \pi^{\varrho} e_0(\varrho)
+ \pi^{\varsigma} e_0(\varsigma) + \pi^{\vp} e_0(\vp) -
n \cH_0^{\smallcap{vd}} \big\}\,,
\nonum
\cH_0^{\smallcap{vd}} \is - \lbn \pi^{\varsigma} \pi^{\varrho}  +
\frac{\lbn}{2 \varrho} (\pi^{\vp})^2 \,,
\nonum
\cH_1^{\smallcap{vd}} \is \pi^{\varrho} \dd_1 \varrho +
\pi^{\varsigma} \dd_1 \varsigma
-2 \dd_1 \pi^{\varsigma} + \pi^{\vp} \dd_1 \vp\,.
\ea
Here we write $\varrho,\varsigma,\vp$ for the counterparts of
$\rho,\sigma, \phi$ in the VD system. Strictly speaking, different
symbols should be used also for the VD versions of $n$, $s$, but
for readability's sake we keep the original ones. 
The phase space is again equipped with the natural Poisson
structure. The constraint algebra simplifies to
\ba
\label{constr0VD} 
\{ \cH_0^{\smallcap{vd}}(x), \cH_1^{\smallcap{vd}}(y)\} \is \dd_1 \cH_0^{\smallcap{vd}} \delta(x-y) + 2 \cH_0^{\smallcap{vd}}(x)
\delta'(x-y)\,,
\nonum
\{ \cH_1^{\smallcap{vd}}(x), \cH_1^{\smallcap{vd}}(y)\} \is \dd_1 \cH_1^{\smallcap{vd}} \delta(x-y) + 2 \cH_1^{\smallcap{vd}}(x)
\delta'(x-y)\,,
\nonum
\{ \cH_0^{\smallcap{vd}}(x), \cH_0^{\smallcap{vd}}(y)\} \is 0\,.
\ea 
The Poisson commutativity of the Hamiltonian constraint is characteristic
for Carroll type gravity systems, see e.g.~\cite{HenneauxC,GRstrongI},
and accounts for the dynamical decoupling
of spatial points. Nevertheless, the action (\ref{SHVD}) is still invariant
under diffeomorphism type gauge symmetries parameterized by two independent
functions of two variables. Explicitly,
\be
\delta^{\smallcap{hvd}}_{\eps} S^{\smallcap{hvd}} = 0 \,,
\quad \mbox{if} \quad \eps|_{t_i} = 0 = \eps|_{t_f}\,,
\ee
holds with the modified gauge variations 
\begin{subequations}
\label{gtransVD}
\ba
\delta^{\smallcap{hvd}}_{\epsilon} n &=& \partial_{0} \epsilon
- (s \partial_{1} \epsilon - \epsilon \partial_{1} s)
+ \epsilon^{1} \partial_{1} n - n\partial_{1} \epsilon^{1}\,,
\nonum
\delta^{\smallcap{hvd}}_{\epsilon} s &=& \partial_{0} \epsilon^{1}
+(\epsilon^{1} \partial_{1} s - s\partial_{1} \epsilon^{1})
\\[2mm]
\delta^{\smallcap{hvd}}_{\epsilon} \varsigma &=& - \lbn \eps \pi^{\varrho} 
+ \epsilon^{1} \partial_{1} \varsigma + 2\partial_{1} \epsilon^{1}\,,
\nonum
\delta^{\smallcap{hvd}}_{\epsilon} \varrho &=& - \lbn \eps \pi^{\varsigma}
+ \eps^1 \dd_1 \varrho\,, 
\nonum
\delta^{\smallcap{hvd}}_{\epsilon} \vp &=& \lbn \frac{\eps}{\varrho} \pi^{\vp} 
+ \epsilon^{1} \partial_{1} \vp\,,
\\[2mm]
\delta^{\smallcap{hvd}}_{\epsilon} \pi^{\varsigma} &=&
\dd_1(\eps^1 \pi^{\varsigma})\,,
\nonum
\delta^{\smallcap{hvd}}_{\epsilon} \pi^\varrho &=&
\dd_1 (\eps^1 \pi^\varrho) + \frac{\eps \lbn}{2 \varrho^2} (\pi^\vp)^2 \,,
\nonum
\delta^{\smallcap{hvd}}_{\epsilon} \pi^\vp &=& \dd_1(\eps^1 \pi^\vp)\,.
\ea
\end{subequations} 
So far we treated the VD system as a Carroll-type gravitational theory 
in its own right. The asymptotic velocity domination property posits
its dynamical relevance for the full Gowdy system, initially classically
and on-shell. There is, however, also a kinematic relation in that the
VD system arises as a (classical) scaling limit from the original one.
For Carroll systems this is usually done via a (coordinate dependent) 
``speed of light to zero'' limit. In the present context, we adopt a
merely foliation dependent scaling defined on phase space as follows:  
\be
\label{scale1}
\begin{array}{lll} 
n  \mapsto \ell^{-1} n\,, \quad\;\;  & s \mapsto s\,,
\quad  & \lbn \mapsto \ell^3 \lbn\,,
\\[2mm]
\rho \mapsto  \ell^2 \rho\,, \quad\;\;
& \pi^{\rho} \mapsto \ell^{-2} \pi^{\rho},
&
\\[2mm] 
\sigma  \mapsto \sigma + 3 \ln \ell\,,\quad\;\; 
&\pi^{\sigma}  \mapsto \pi^{\sigma}, &
\\[2mm] 
\phi \mapsto  \phi\,, \quad\;\;  & \pi^{\phi}  \mapsto \pi^{\phi}\,. &
\end{array}
\ee
In the line element (\ref{2Kmetric}) this enhances for $\ell \gg 1$
spacelike distances compared to timelike ones. Neighboring worldlines
are harder to communicate with and the light cone structure appears
anti-Newtonian. The scaling of the reduced Newton constant $\lbn$ is
adjusted such that the rescaled action (\ref{SH}) has a well defined
limit as $\ell \ra \infty$. Indeed, the limit is precisely
$S^{\smallcap{hvd}}$ from (\ref{SHVD}), upon renaming of the fields.
Likewise, the constraint algebra (\ref{constr0VD}) is a contraction
of (\ref{constr0}), and the gauge variations (\ref{gtransVD})
are limiting versions of (\ref{gtrans1}). 

\section{Classical Dirac observables}
\label{appendixB}

Two Killing vector reductions of Einstein gravity often admit
an infinite set of nonlocal conserved charges whose existence is
related to that of a Lax pair. On general grounds such conserved
charges are expected to give rise to (normally elusive) Dirac observables,
Poisson commuting with the constraints. For the polarized Gowdy system
the conserved charges become linear functionals of the basic Gowdy
scalar. A systematic characterization of Dirac observables
for the polarized Gowdy system has been given by Torre
\cite{TorreObs}, using an extended phase space formulation.
Due to the use of a polar decomposition in the last step it falls
however slightly short of providing fully explicit expressions for
them. Here we provide such a construction on the original rather
than an extended phase space in a way that makes contact to the
velocity dominated limit and in principle generalizes to the
non-polarized case.

A Dirac observable $\cO$ is by definition a functional of the phase
space variables that weakly Poisson commutes with the constraints,
see e.g.~\cite{Pons}. In the situation at hand, we aim at the strong
vanishing 
\begin{equation}
\label{obs0}
\{\cH_0, \cO \} =0 = \{\cH_1, \cO \}\,.
\end{equation} 
In addition to the local gauge invariance (\ref{gtrans1}) the
action $S^{\smallcap{h}}$ also has a trivial global invariance under constant
shifts $\phi(x) \mapsto \phi(x) + \alpha$, $\alpha \in \R$.   
It is generated by the Noether charge $Q = (2\pi)^{-1} \int_0^{2\pi}
\! dx^1 \pi^{\phi}(x)$. Since the constraints $\cH_0,\cH_1$ Poisson
commute with $Q$ the iterated images $\{ Q, \ldots \{ Q, \cO\} ...\}$ 
are Dirac observables as well. To avoid such rather trivial chains
it is natural to require that the Dirac observables Poisson commute
with $Q$ as well 
\begin{equation}
\label{obs1}
\{ Q, \cO\} =0\,.
\end{equation}
The goal in the following is to construct an infinite set of Dirac
observables which satisfy (\ref{obs0}), (\ref{obs1}) {\it strongly, off-shell,
  and without gauge fixing}. See e.g.~\cite{Pons} for the significance
of the latter. 
Since $\cH_1$ generates infinitesimal spatial translations
spatial integrals of spatial $+1$ densities will satisfy the
second condition in (\ref{obs0}). A solution of the first
condition, on the other hand, necessitates to the construction of
(nonlocal) conserved charges, which is normally elusive as it requires a
constructive approach to the solution of the initial value problem. 
As mentioned, this is often feasible for the two Killing vector
reductions of Einstein gravity, but for Gowdy cosmologies
has not been accomplished beyond the polarized sector.
In the following we describe a construction of off-shell Dirac
observables for the polarized Gowdy system that generalizes to
the unpolarized sector.

{\bf Gauge variation of the generating current.} Consider the basic
currents in the Hamiltonian formalism
\be
\label{ahvar2} 
\jmath_0^\h := \pi^{\phi}\,, \quad
\jmath_1^\h := \frac{\rho n}{\lbn} \dd_1 \phi\,,
\ee
whose conservation is one of the Hamiltonian evolution equations
$e_0(\pi^{\phi}) = \dd_1( \frac{\rho n}{\lbn} \dd_1 \phi)$. Using the
Hamiltonian
gauge variations (\ref{gtrans1}) one can check
that 
\begin{equation}
\label{ahvar3} 
\delta_{\eps}^\h \jmath_0^\h = \dd_1\Big(
\frac{\eps}{n} \jmath_1^\h + \eps^1 \jmath_0^\h\Big)\,,
\end{equation}
holds {\it without} using any equations of motion. 

Next, we aim at a one-parameter generalization of the form%
\footnote{The origin of these expressions lies in the $\dd/\dd \th$
derivatives of a Lax pair for the 2-Killing vector reductions, see \cite{DiracGowdy}. Use of
the Lax pair is overkill for the polarized case but (\ref{ahvar4})
turns out to generalize to the non-polarized Gowdy system.} 
\ba
\label{ahvar4} 
\jmath_0^\h(\th) &:=& \frac{\lbn^2}{[ \lbn^2 (\th + \tilde{\rho})^2 - \rho^2]^{3/2}}
  [ \lbn (\th + \tilde{\rho}) \pi^{\phi} - \frac{\rho^2}{\lbn} \dd_1\phi]\,,
  \nonum
\jmath_1^\h(\th) &:=& \frac{\lbn^2}{[ \lbn^2 (\th + \tilde{\rho})^2 - \rho^2]^{3/2}}
  [ (\th + \tilde{\rho}) \rho n \dd_1 \phi - \rho n \pi^{\phi}]\,,
\ea
for complex $\th \in \C$, ${\rm Im} \th \neq 0$.
Here we define 
\begin{equation} 
\label{ahvar5} 
\tilde{\rho}(x^0,x^1) := - \int_{y^1}^{x^1} \! dx \,\pi^{\sigma}(x^0, x) +
\tilde{\rho}(x^0,y^1)\,,
\end{equation}
for some reference point $y^1$. For its Hamiltonian gauge variation one
can self-consistently take 
\be
\label{ahvar6}
\delta_{\eps}^\h \tilde{\rho} = - \eps^1 \pi^{\sigma}
+ \frac{\eps}{\lbn} \dd_1 \rho\,.
\ee
This comes about as follows: since $\tilde{\rho}$ is not spatially
periodic (see below) the natural class of  gauge-type spatial
diffeomorphisms for its are those of the real line vanishing outside a
compact region. Evaluating the gauge variation of the integral in
(\ref{ahvar6}) one finds that that contribution from the $y^1$  
endpoint vanishes if $y^1 \ra - \infty$. One can therefore consistently
take $\tilde{\rho}(x^0,- \infty) =0$ and obtains (\ref{ahvar6}). 
Once (\ref{ahvar6}) is in place we treat $\tilde{\rho}$ as spatially
quasi-periodic via (\ref{rhotHqperiodic}) and for the fundamental
domain $[0,2\pi]$ use gauge descriptors of the same type as in
(\ref{gtrans1}).

Using only (\ref{gtrans1}) and (\ref{ahvar6}) one finds
\be
\label{alvar7}
\delta_{\eps}^\h \jmath_0^\h(\th) = \dd_1\Big(\frac{\eps}{n}
\jmath_1^\h(\th)
+ \eps^1 \jmath_0^\h(\th) \Big)\,,
\ee 
holds, again {\it without} using any equations of motion.
For noncompact spatial sections and with suitable fall-off
conditions the relation (\ref{alvar7}) would directly
identify the spatial integral of $\jmath_0^\h(\th)$ as
a one-parameter family of Dirac observables for the polarized
Gowdy system.

{\bf Periodic extension.} 
For the $T^3$ Gowdy cosmologies of prime interest in the context of
AVD, the identity (\ref{alvar7}) does not directly provide
Dirac observables by integration: while all the basic
fields and hence $\jmath_0^\h,\jmath_1^\h$ are spatially periodic,
the off-shell currents $\jmath_0^\h(\th), \jmath_1^\h(\th)$ are not.
This is because $\tilde{\rho}$ is not spatially periodic (neither
on- nor off-shell). From (\ref{ahvar5}) one has 
\begin{equation}
\label{rhotHqperiodic} 
\tilde{\rho}(x^0,x^1\!+\! 2\pi) - \tilde{\rho}(x^0,x^1)
= - \!\int_0^{2\pi} \! dx \,\pi^{\sigma}(x^0,x) =: -\pi^{\sigma}_0\,,
\end{equation}
using the spatial periodicity of $\pi^{\sigma}$.  
Importantly, $\pi_0^{\sigma}$ itself is gauge invariant,
\begin{equation}
\label{rhotHshift1} 
\delta^H_{\eps} \pi^{\sigma}_0 =0\,,
\end{equation} 
as can be seen from (\ref{gtrans1}). Here and from now on we
take the descriptors $\eps,\eps^1$ to be spatially periodic. 
An alternative way of looking at (\ref{rhotHshift1}) is by noting that
$\{\sigma_0, \pi_0^{\sigma} \} =1$, where $\sigma_0 = (2\pi)^{-1}
\int_0^{2\pi} \! dx \, \sigma(x)$ is the zero mode of $\sigma$. 
This zero mode simply does not occur in $\cH_0, \cH_1$, which
accounts for (\ref{rhotHshift1}). Taking into account the
nonperiodicity of (\ref{ahvar4}) one merely has 
\ba
&& \big\{ \int_0^{2\pi} \!\! dx \jmath_0^\h(x;\th),
\cH_0(\eps)+\cH_1(\eps^1)\big\}
\nonum
&& = \Big(\frac{\eps}{n}\Big)(0) \big[
\jmath_1^\h(2\pi;\th) - \jmath_1^\h(0;\th)\big]
+ \eps^1(0) \big[
\jmath_0^\h(2\pi;\th) - \jmath_0^\h(0;\th)\big]\,.
\ea 
Here we suppressed the shared time argument and assumed as
before that the descriptors $\eps,\eps^1$ are spatially periodic. 
The basic idea now is to replace $\jmath_0^\h(x,\th)$ on the
left hand side with its $2\pi$-periodic extension, i.e.~with
$\sum_n \jmath_0^\h(x \!+ \!2\pi n, \th)$. Assuming that this sum converges,
it will give rise on the right hand side to a telescopic sum which
vanishes.  In principle, the plain periodic extension can be modified
by multiplying
the $n$-th term by a function $s_n$ that is itself gauge invariant and
obeys $s_n|_{x^1 + 2\pi} = s_{n+1}$ as a function of $x^1$. Since the only
source of non-periodicity in our context is the combination
$\th + \tilde{\rho}$ it is natural to take $s_n$ a function of
$\th + \tilde{\rho}$, in which case $\delta^\h_{\eps} s_n =0$ is a
stringent requirement.

Proceeding along these lines, we consider for ${\rm Im}\th \neq 0$
the $s_n$-modified periodic extension of (\ref{ahvar4}). It leads
to sums of the form
\ba
\label{csums1} 
c_0(\tilde{\rho}+\th,\rho) &:=&
\sum_{n \in \Z} \frac{s_n(\th\! + \!\tilde{\rho})
  \lbn^3 (\th  + \tilde{\rho} - n \pi_0^{\sigma})}%
{[\lbn^2 (\th  + \tilde{\rho} - n \pi_0^{\sigma})^2 - \rho^2]^{3/2}}\,,
\nonum
c_1(\tilde{\rho}+\th,\rho) &:=&
\sum_{n \in \Z} \frac{\lbn^3 s_n(\th\! + \!\tilde{\rho})}%
    {[\lbn^2(\th  + \tilde{\rho} - n \pi_0^{\sigma})^2 - \rho^2]^{3/2}}\,.
\ea    
For reasons that will become clear shortly we take for $s_n$ the
generalized sign function $s_n(z-in \pi_0^{\sigma}) =
\exp\{3 i {\rm Arg}(z - i n \pi_0^{\sigma})\}
(\exp\{ 2 i {\rm Arg}(z -i n \pi_0^{\sigma})\})^{-3/2}$, ${\rm Im} z \neq 0$.
The sums (\ref{csums1}) then converge absolutely and for
${\rm Im}\th \neq 0$ define spatially periodic functions of $x^0,x^1$.
The summed expansions (\ref{csums2}) below also show them to be
smooth in $\rho,\tilde{\rho}$ for ${\rm Im}(\th) \neq 0$.  
Inserting (\ref{csums1}) into (\ref{ahvar4}) gives
\ba
\label{jasummed} 
\jmath_0^\smallcap{p}(\th):=
\sum_{n \in \Z} s_n(\th + \tilde{\rho}) \jmath^\h_0(\th - n \pi_0^{\sigma})
= c_0(\tilde{\rho} + \th, \rho) \pi^{\phi} -
c_1(\tilde{\rho} + \th, \rho) \frac{\rho^2}{\lbn^2} \dd_1 \phi\,,
\nonum
\jmath_1^{\smallcap{p}}(\th):=
\sum_{n \in \Z} s_n(\th + \tilde{\rho}) \jmath^\h_1(\th - n \pi_0^{\sigma})
= c_0(\tilde{\rho} + \th, \rho) \frac{\rho}{\lbn} n\dd_1 \phi
-c_1(\tilde{\rho} + \th, \rho) \frac{\rho}{\lbn} n \pi^{\phi}\,. 
\ea
The condition for the summed versions to
obey equation (\ref{alvar7}) amounts to a
simple pair of differential equations for $c_0,c_1$,
\be
\label{csums0}
\dd_{\tilde{\rho}} (\rho^2 c_1) = - \lbn^2 \rho \dd_{\rho} c_0\,, \quad
\rho \dd_{\tilde{\rho}} c_0 = - \dd_{\rho} (\rho^2 c_1)\,.
\ee
These imply that $c_0$ and $\rho^2 c_1$ are spatially periodic
solutions of $\lbn^{-2} \dd_{\tilde{\rho}}^2 f =
\dd_{\rho}^2f \pm \rho^{-1} \dd_{\rho} f$, respectively.

The spatial periodicity can be rendered explicit by expanding in powers of
$\rho$ using
\ba
\label{csums2} 
&\nspace &\nspace 
\frac{\lbn^3(\th \! + \!\tilde{\rho} \!- \!n \pi_0^{\sigma})}%
     {[\lbn^2 (\th  \!+\! \tilde{\rho}\! - \!n \pi_0^{\sigma})^2 - \rho^2]^{3/2}} =
     \sum_{l \geq 0} { -3/2 \choose l} (-\rho^2)^l
     s_n(\th \!+\! \tilde{\rho})
     \lbn^2[\lbn(\th \!+\! \tilde{\rho} \!-\! n \pi_0^{\sigma})]^{-2 l -2}, 
\nonum
& \nspace & \nspace \frac{\lbn^3}%
     {[\lbn^2(\th \! + \! \tilde{\rho}\! -\! n \pi_0^{\sigma})^2 - \rho^2]^{3/2}} =
     \sum_{l \geq 0} { -3/2 \choose l} (-\rho^2)^l
     s_n(\th \!+ \!\tilde{\rho})
     \lbn^3[\lbn(\th  \!+\! \tilde{\rho} \!-\! n \pi_0^{\sigma})]^{-2 l -3}, 
\ea 
The point of the $s_n$ modulated sums in (\ref{csums1}) is simply to
cancel the corresponding $s_n$'s in the expansions (\ref{csums2}). The
resulting sums over $n$ can then readily be performed. For 
$2 \leq l \in \N$ one has 
\be
\label{csums3} 
\sum_{n \in \Z} \frac{1}{(a\! + \!b n)^l} = \frac{(-)^l}{(l\!-\!1)!}
\Big( \frac{\dd}{\dd a} \Big)^{l-2} \sum_{n \in \Z} \frac{1}{(a\! + \!b n)^2}
= \frac{\pi^2}{b^2} \frac{(-)^l}{(l\!-\!1)!}
\Big( \frac{\dd}{\dd a} \Big)^{l-2} \frac{1}{\sin^2(\pi a/b)}\,.
\ee
With $a = \lbn(\th + \tilde{\rho})$, $b = - \lbn\pi_0^{\sigma}$ this
results in 
\ba
\label{csums4}
\nspace c_0(\th \!+\! \tilde{\rho}, \rho) \is
\Big( \frac{\pi}{\pi_0^{\sigma}} \Big)^2
\sum_{l \geq 0} { -3/2 \choose l} \frac{(- \rho^2/\lbn^2)^l}{(2 l + 1)!}
\Big( \frac{\dd}{\dd \th} \Big)^{2l} \!\frac{1}{\sin^2\big(
  \frac{\pi}{\pi_0^{\sigma}}(\th \!+\! \tilde{\rho}) \big)} ,
\nonum
\nspace c_1(\th \!+\! \tilde{\rho}, \rho) \is
- \Big(\frac{\pi}{ \pi_0^{\sigma}} \Big)^2
\sum_{l \geq 0} { -3/2 \choose l} \frac{(- \rho^2/\lbn^2)^l}{(2 l + 2)!}
\Big( \frac{\dd}{\dd \th} \Big)^{2l+1} \!\frac{1}{\sin^2\big(
  \frac{\pi}{\pi_0^{\sigma}}(\th \!+\! \tilde{\rho}) \big)} .
\ea
As a check on the innocuous nature of the $s_n$ modulation in (\ref{jasummed})
one can verify that the expressions (\ref{csums4}) indeed
solve the differential equations (\ref{csums0}).
This yields an expansion of $\jmath_0^{\smallcap{p}}(\theta)$ of the form
$\jmath_0^{\smallcap{p}}(\theta) = \sum_{j\geq 0}
\jmath_0^{\smallcap{p}}(\theta)_{2j}$, where the $j$-th term has scaling
weight $\ell^{-2j}$ under (\ref{scale1}). Explicitly,
\ba
\label{csums8}
\jmath_0^{\smallcap{p}}(\th)_0 &:=& \Big( \frac{\pi}{ \pi_0^{\sigma}} \Big)^2  
    \frac{ \pi^{\phi}}%
{\sin^2\big(\frac{\pi}{\pi_0^{\sigma}}(\th + \tilde{\rho}) \big)} \,,
\\[2mm]
\jmath_0^{\smallcap{p}}(\th)_{2j} &:=& \Big( \frac{\pi}{ \pi_0^{\sigma}} \Big)^2 
      { -3/2 \choose j} \frac{(-1)^j}{(2 j\! +\!1)!}
      \Big( \frac{\rho^2}{\lbn^2} \Big)^j
    \nonum
    &&\times \Big(\frac{\dd}{\dd \th} \Big)^{2 j}
    \Big[ \frac{ \pi^{\phi}}%
      {\sin^2\big(\frac{\pi}{\pi_0^{\sigma}}(\th + \tilde{\rho}) \big)}
      +\frac{1}{(2j\!+\!2)} \frac{\dd}{\dd \th} \frac{ \dd_1 \phi}%
{\sin^2\big(\frac{\pi}{\pi_0^{\sigma}}(\th + \tilde{\rho}) \big)} \Big]\,.
\nonumber
\ea

Finally, we note a third rewriting of the $c_0,c_1$. Using
\ba
\label{csums6} 
&& \frac{1}{\sin^2(z)} = - 4 \sum_{n \geq 1} n e^{\pm 2 i n z} \,,\quad
\pm {\rm Im} z >0\,,
\nonum
&& \sum_{l \geq 0} { -3/2 \choose l} \frac{x^{2 l}}{(2l+1)!} = J_0(x)\,,
\quad 
\sum_{l \geq 0} { -3/2 \choose l} \frac{x^{2 l}}{(2l+2)!} =
\frac{J_1(x)}{x}\,,
\ea
to manipulate (\ref{csums4}) one finds%
\footnote{Inserting for the left hand side the original sums
(\ref{csums1}) produces Bessel function identities which appear to be new.} 
\ba
\label{csums7}
c_0(\th + \tilde{\rho},\rho) \is -
\Big( \frac{2 \pi}{ \pi_0^{\sigma}} \Big)^2
\sum_{n \geq 1} n e^{ \pm 2 i \pi n (\th + \tilde{\rho})/\pi_0^{\sigma}}
J_0\Big( \frac{2 n \rho \pi}{\lbn \pi_0^{\sigma}} \Big)\,,
\nonum
c_1(\th + \tilde{\rho},\rho) \is \pm i
\Big( \frac{2 \pi}{\pi_0^{\sigma} } \Big)^2
\sum_{n \geq 1} n e^{ \pm 2 i \pi n (\th + \tilde{\rho})/\pi_0^{\sigma}}
\;\frac{\lbn}{\rho}J_1\Big( \frac{2 n \rho \pi}{\lbn \pi_0^{\sigma}} \Big)\,,
\ea
for $\pm {\rm Im} \th >0$. The occurrence of Bessel functions
is explained by the comment after (\ref{csums0}), where regularity
at $\rho =0$ excludes the Bessel $Y$-type solutions. Conversely, it
follows from (\ref{csums7}) that the Dirac observables
(\ref{jaDirac}) are a complete set in the sector regular
for $\rho/\lbn \ra 0$. Let us stress that (\ref{csums7}) holds off-shell,   
conceptually the Bessel functions entering here are unrelated to those in
(\ref{Tsol2}).

\begin{result}
Set 
\be
\label{jaDirac}
\cO(\th) := \int_0^{2\pi} \! \frac{dx^1}{2\pi}
\big\{c_0(\tilde{\rho} + \th, \rho) \pi^{\phi} -
c_1(\tilde{\rho} + \th, \rho) \frac{\rho^2}{\lbn^2} \dd_1 \phi\big\}\,,
\quad {\rm Im} \th \neq 0\,,
\ee
with $c_0$, $c_1$ given alternatively by (\ref{csums1}), (\ref{csums4}), (\ref{csums7}). Then $\cO(\th)$ is a one-parameter family of strong, off-shell Dirac observables for the polarized $T^3$ Gowdy cosmologies. The integrand $\jmath_0^{\smallcap{p}}(\theta)$ has an `anti-Newtonian' expansion in even powers of $\rho/\lbn$ given in (\ref{csums8}). The leading term gives rise to one-parameter family of strong, off-shell Dirac observables in the VD system via
\be
\label{VDDirac}
\cO^{\smallcap{vd}}(\theta) = \int_0^{2\pi}\!\!\frac{dx^1}{2\pi} \, \jmath_0^{\smallcap{p}}(\theta)_0\big|_{ \pi^{\phi} \mapsto \pi^{\vp},
  \pi^{\sigma} \mapsto \pi^{\varsigma}}\,.
\ee
\end{result}

Recall that $\rho>0$ is a temporal function whose $\rho =0$ level
set can be identified with the Big Bang. The Dirac observables (\ref{jaDirac})
therefore remain regular at the Big Bang and (\ref{csums8}) provides a
systematic expansion away from it, notably in {\it inverse} powers of the
reduced Newton constant $\lbn$. Upon substitution of the dynamical variables
of the VD system the $\rho=0$ term defines Dirac observables in the VD
system, as indicated in (\ref{VDDirac}). The invariance
$\delta_\eps^{\smallcap{hvd}}\cO^{\smallcap{vd}}(\theta) = 0$ can be
verified by direct computation from (\ref{gtransVD}).   
\medskip

{\bf On-shell specialization.} Subject to the evolution equations
the left hand side of the defining relation
$\delta_{\eps}^\h \jmath_0^\smallcap{p}(\th) = \dd_1[ (\eps/n)
  \jmath_1^\smallcap{p}(\th)
  + \eps^1 \jmath_0^\smallcap{p}(\th)]$ must for $\eps =n, \eps^1=0$
produce the infinitesimal time evolution $e_0(\jmath_0^{\smallcap{p}}(\th))$.
The Dirac observables thereby
give rise to a one-parameter family of on-shell conserved currents  
\be
\label{diracon1} 
e_0(\jmath^{\smallcap{p}}_0(\th)) = \dd_1 \jmath_1^{\smallcap{p}}(\th)\,.
\ee
Using the evolution equations (\ref{evol1}) and (\ref{csums0}) this
can indeed be verified directly. The on-shell version of $\tilde{\rho}$
obeys $e_0(\rho) = \lbn n \dd_1 \tilde{\rho}$,
$\lbn e_0(\tilde{\rho}) = n \dd_1 \rho$,
and the periodicity constant is $\pi_0^{\sigma} = \lbn^{-1} \int_0^{2\pi}
dx e_0(\rho)/n$. This holds without gauge fixing and/or the subsequent
coordinate specifications (\ref{gf2}), (\ref{folimatch}). In particular,
the scaling transformations (\ref{scale1}) remain intact and can be used
to introduce an $\ell$-grading as before. After the scaling weights have
been identified we proceed to fix the proper time gauge and impose the
coordinate choices (\ref{gf2}), (\ref{folimatch}). The latter
spoil the scaling as $t$ and $\zeta$ would have to be assigned weights
$\ell^2$ and $\ell$, respectively, but the limit $\ell \ra \infty$ is
not supposed to act on the coordinates. We therefore also set $\lbn =1$
and identify $\tilde{\rho}$ with $\zeta$, consistent with the specialization
to $\pi_0^{\sigma} =2 \pi$ and the comment after (\ref{gf2}). 
We denote the so-specialized coefficient functions by 
$c_0(\th + \zeta,e^{\tau})$, $c_1(\th + \zeta,e^{\tau})$ (with
$\lbn =1, \pi_0^{\sigma} =2\pi$ understood). They can be expressed
alternatively as the specializations of (\ref{csums1}),(\ref{csums4})
or (\ref{csums7}). For the Dirac observables (\ref{jaDirac}) this
yields an on-shell specialization of the form
\ba
\label{diracon2}
\cO^{\rm on}(\th) \is \int_0^{2\pi} \!\frac{d\zeta}{2\pi}
\big\{
c_0(\th\! + \!\zeta, e^{\tau}) \dd_{\tau} \phi^{\rm on} -  
c_1(\th\! + \!\zeta, e^{\tau}) e^{2\tau} \dd_{\zeta} \phi^{\rm on} \big\}\,,
\nonum
\phi^{\rm on} \is \sum_{n \in \Z} \big\{ T_n(\tau) e^{ i n \zeta} a_n +
T_n(\tau)^* e^{ -i n \zeta} a_n^* \big\}\,,
\ea
with $T_n$ a Wronskian normalized solution of $[\dd_{\tau}^2 +e^{2 \tau} n^2]
T_n(\tau) =0$. By construction $\dd_{\tau} \cO^{\rm on}(\th) =0$, which can
also be verified directly from (\ref{csums0}). The value of $\cO^{\rm on}(\th)$
can therefore be found from the $\tau \rightarrow - \infty$ limit of the
integrand. Since $c_0$, $c_1$ are regular at $\rho=0$ and AVD implies
$\dd_\tau \phi \rightarrow v$ for $\tau \rightarrow -\infty$ (see Section 2.2)
one infers $\cO^{\rm on}(\th) = \cO_{\smallcap{vd}}^{\rm on}(\th)$. In fact,
a fairly explicit expression for the on-shell value can be found as follows:
inserting $\phi^{\rm on}$ into the first line of (\ref{diracon2}) one finds 
\ba
\label{diracon3}
\cO^{\rm on}(\th) \is \sum_{ n \neq 0}
\big\{ \cO_n(\th) a_n + \cO_n(\th)^* a_n^* \big\} \,,
\nonum
\cO_n(\th) \is \frac{i}{\sqrt{\pi}}(\lb_n\! - \!\mu_n) \th(\mp n)
|n| e^{ \pm i |n| \th}\,, \quad \pm {\rm Im}\th >0\,.
\ea
We outline the derivation. In a first step, the coefficients $\cO_n(\th)$
come out as a linear combination of $\dd_{\tau} T_n(\tau)$ and
$i n e^{2\tau} T_n(\tau)$ with coefficients that are Fourier coefficients
of the $c_0, c_1$. The latter are conveniently evaluated using
(\ref{csums7}) and come out as
\ba
&& \int_0^{2\pi} \! \frac{d \zeta}{2 \pi}
e^{ i n \zeta} c_0(\zeta + \th, e^{\tau})
= - J_0(e^{\tau} |n|) e^{ \pm i |n| \th} \th( - |n|)\,,
\nonum
&& \int_0^{2\pi} \! \frac{d \zeta}{2 \pi}
e^{ i n \zeta} c_1(\zeta + \th, e^{\tau})
= \pm i |n| J_1(e^{\tau} |n|) e^{ \pm i |n| \th} \th( - |n|)\,,
\ea
where $\th(\mp n)$ is the step function, imposing $\mp n \in \N$.  
Expressing $\dd_{\tau} T_n(\tau)$ and $i n e^{2 \tau} T_n(\tau)$
in terms of Bessel $J_1, Y_1$ and $J_0,Y_0$ functions via (\ref{Tsol2}) 
many terms cancel and use of the Wronskian identity 
$Y_1(x) J_0(x) - Y_0(x) J_1(x) = -2/(\pi x)$ yields the above $\cO_n(\th)$.
Importantly, for ${\rm Im} \th \neq 0$ the
$\cO_n(\th)$ decay faster than any power in $|n|$. Indeed, for Hadamard
states we saw in Section \ref{sec4.2} that $\lb_n\ra 1$, $\mu_n \ra 0$,
for $|n| \ra \infty$; the rapid decay would however persist for any
$\lb_n,\mu_n$ polynomially bounded in $|n|$.

An analogous computation can be done for the VD observables
(\ref{VDDirac}). Here $\vp^{\rm on} = \sum_{n} \{
\mathfrak{t}_n(\tau) e^{ i n \zeta} a_n + \mathfrak{t}_n(\tau)^*
e^{ - n \zeta} a_n^*\}$, $\dd_{\tau}^2 \mathfrak{t}_n =0$, needs
to be inserted into the much simpler (specialized) on-shell form of
$\cO_0(\th)|_{ \pi^{\phi} \mapsto \pi^{\vp},
  \pi^{\sigma} \mapsto \pi^{\varsigma}}$. The integrals encountered are
for $\pm {\rm Im} \th>0$ readily evaluated by contour deformation,
while $\dd_{\tau} \mathfrak{t}_n = (-i/\sqrt{\pi})(\tilde{\lb}_n
\!-\!\tilde{\mu}_n)$ is already $\tau$ independent. The result is
\ba
\label{diracon4}
\cO_{\smallcap{vd}}^{\rm on}(\th) \is \sum_{ n \neq 0}
\big\{ \widetilde{\cO}_n(\th) a_n +
\widetilde{\cO}_n(\th)^* a_n^* \big\} \,,
\nonum
\widetilde{\cO}_n(\th) \is
\frac{i}{\sqrt{\pi}}(\tilde{\lb}_n\! - \!\tilde{\mu}_n)
\th(\mp n) |n| e^{ \pm i |n| \th}\,, \quad \pm {\rm Im}\th >0\,.
\ea
Since by (\ref{match8}) $\tilde{\lb}_n\! - \!\tilde{\mu}_n
=\lb_n\! - \!\mu_n$ one confirms 
\be
\label{diracon5}
\cO^{\rm on}(\th) = \cO_{\smallcap{vd}}^{\rm on}(\th) \,,
\ee
which is another manifestation of AVD.


\end{document}